\documentclass{elsart}
\usepackage{axodraw}
\usepackage{epsfig}
\usepackage{supertabular}
\usepackage{amssymb}

\newcommand{\lsim}
{\mathrel{\raisebox{-.3em}{$\stackrel{\displaystyle <}{\sim}$}}}

\def\asymp#1%
{\mathrel{\raisebox{-.4em}{$\widetilde{\scriptstyle #1}$}}}

\def\Nequal#1%
{\mathrel{\raisebox{-.5em}{$\stackrel{=}{\scriptstyle\rm#1}$}}}
\newcommand{\dsl}[1]{\not \hspace{-0.7mm}#1}
\def\dsl{\mathpalette\make@slash}
\def\make@slash#1#2{\setbox\z@\hbox{$#1#2$}%
  \hbox to 0pt{\hss$#1/$\hss\kern-\wd0}\box0}

\def\beq{\begin{equation}}
\def\eeq{\end{equation}}
\def\beqar{\begin{eqnarray}}
\def\eeqar{\end{eqnarray}}
\def\barr#1{\begin{array}{#1}}
\def\earr{\end{array}}
\def\bfi{\begin{figure}}
\def\efi{\end{figure}}
\def\btab{\begin{table}}
\def\etab{\end{table}}
\def\bce{\begin{center}}
\def\ece{\end{center}}
\def\nn{\nonumber}
\def\nl{\nonumber\\}
\def\nln{\nl[-1ex]}
\def\disp{\displaystyle}
\def\text{\textstyle}

\def\al{\alpha}

\def\Ga{\Gamma}
\def\ga{\gamma}
\def\de{\delta}
\def\De{\Delta}

\def\si{\sigma}

\def\refeq#1{\mbox{Eq.~(\ref{#1})}}
\def\refeqs#1{\mbox{Eqs.~(\ref{#1})}}
\def\refeqf#1{\mbox{(\ref{#1})}}
\def\reffi#1{\mbox{Figure~\ref{#1}}}

\def\refta#1{\mbox{Table~\ref{#1}}}
\def\reftas#1{\mbox{Tables~\ref{#1}}}
\def\refse#1{\mbox{Section~\ref{#1}}}
\def\refses#1{\mbox{Sections~\ref{#1}}}

\def\citere#1{\mbox{Ref.~\cite{#1}}}
\def\citeres#1{\mbox{Refs.~\cite{#1}}}

\newcommand{\TeV}{\unskip\,\mathrm{TeV}}
\newcommand{\GeV}{\unskip\,\mathrm{GeV}}
\newcommand{\MeV}{\unskip\,\mathrm{MeV}}

\newcommand{\ri}{{\mathrm{i}}}
\newcommand{\rd}{{\mathrm{d}}}

\newcommand{\Oa}{\mathswitch{{\mathcal{O}}(\alpha)}}

\newcommand{\M}{{\mathcal{M}}}

\def\mathswitchr#1{\relax\ifmmode{\mathrm{#1}}\else$\mathrm{#1}$\fi}

\newcommand{\PW}{\mathswitchr W}
\newcommand{\Pw}{\mathswitchr w}
\newcommand{\PZ}{\mathswitchr Z}

\newcommand{\Pp}{\mathswitchr p}
\newcommand{\Pe}{\mathswitchr e}

\newcommand{\Pdbar}{\bar{\mathswitchr d}}
\newcommand{\Pu}{\mathswitchr u}

\newcommand{\Pt}{\mathswitchr t}

\newcommand{\Pep}{\mathswitchr {e^+}}
\newcommand{\Pem}{\mathswitchr {e^-}}
\newcommand{\PWp}{\mathswitchr {W^+}}
\newcommand{\PWm}{\mathswitchr {W^-}}
\newcommand{\PWpm}{\mathswitchr {W^\pm}}

\def\mathswitch#1{\relax\ifmmode#1\else$#1$\fi}

\newcommand{\MW}{\mathswitch {M_\PW}}

\newcommand{\MZ}{\mathswitch {M_\PZ}}

\newcommand{\Me}{\mathswitch {m_\Pe}}

\newcommand{\GW}{\Gamma_{\PW}}

\newcommand{\sw}{\mathswitch {s_\Pw}}

\newcommand{\GF}{\mathswitch {G_\mu}}

\def\ie{i.e.\ }

\def\cf{cf.\ }

\newcommand{\sing}{{\mathrm{sing}}}
\newcommand{\finite}{{\mathrm{finite}}}

\newcommand{\DPA}{{\mathrm{DPA}}}

\newcommand{\Born}{{\mathrm{Born}}}
\newcommand{\born}{{\mathrm{Born}}}

\newcommand{\virt}{{\mathrm{virt}}}
\newcommand{\real}{{\mathrm{real}}}

\newcommand{\facto}{{\mathrm{fact}}}
\newcommand{\nonfact}{{\mathrm{nfact}}}
\renewcommand{\min}{{\mathrm{min}}}
\renewcommand{\max}{{\mathrm{max}}}

\newcommand{\Coul}{{\mathrm{Coul}}}

\newcommand{\U}{\mathrm{U}}
\newcommand{\SU}{\mathrm{SU}}

\def\Re{\mathop{\mathrm{Re}}\nolimits}
\def\Im{\mathop{\mathrm{Im}}\nolimits}

\newcommand{\RacoonWW}{{\sc RacoonWW}} 
\newcommand{\WRAP}{{\sc WRAP}} 
\newcommand{\PHEGAS}{{\sc PHEGAS}} 
\newcommand{\YFSWW}{{\sc YFSWW3}} 
\newcommand{\GENTLE}{{\sc GENTLE}}

\newcommand{\eeWWffff}{\Pep\Pem\to\PW\PW\to 4f}
\newcommand{\eeWWffffg}{\Pep\Pem\to\PW\PW(\gamma)\to 4f\gamma}
\newcommand{\eeffff}{\Pep\Pem\to 4f}
\newcommand{\eeffffg}{\Pep\Pem\to 4f\ga}

\renewcommand{\O}{{\mathcal O}}

\newcommand{\kon}{\hat{k}}

\newcommand{\ton}{\hat{t}}

\newcommand{\bare}{{\mathswitchr{bare}}}

\def\draftdate{\relax}
\def\mda{\relax}
\def\mua{\relax}
\def\mla{\relax}
\def\Mda{\relax}
\def\Mua{\relax}
\def\Mla{\relax}
\def\mpar#1{\relax}
\def\draft{
\def\thtystars{******************************}
\def\sixtystars{\thtystars\thtystars}
\typeout{}
\typeout{\sixtystars**}
\typeout{* Draft mode!
         For final version remove \protect\draft\space in source file *}
\typeout{\sixtystars**}
\typeout{}
\def\draftdate{\today}
\renewcommand{\mpar}[1]{{\marginpar{\hbadness10000\raggedright%
                      \sloppy\hfuzz10pt\boldmath\bf##1}}%
                      \typeout{marginpar: ##1}\ignorespaces}
\def\mua{\marginpar[\boldmath\hfil$\uparrow$]%
                   {\boldmath$\uparrow$\hfil}%
                    \typeout{marginpar: $\uparrow$}\ignorespaces}
\def\mda{\marginpar[\boldmath\hfil$\downarrow$]%
                   {\boldmath$\downarrow$\hfil}%
                    \typeout{marginpar: $\downarrow$}\ignorespaces}
\def\mla{\marginpar[\boldmath\hfil$\rightarrow$]%
                   {\boldmath$\leftarrow $\hfil}%
                    \typeout{marginpar: $\leftrightarrow$}\ignorespaces}
\def\Mua{\marginpar[\boldmath\hfil$\Uparrow$]%
                   {\boldmath$\Uparrow$\hfil}%
                    \typeout{marginpar: $\uparrow$}\ignorespaces}
\def\Mda{\marginpar[\boldmath\hfil$\Downarrow$]%
                   {\boldmath$\Downarrow$\hfil}%
                    \typeout{marginpar: $\downarrow$}\ignorespaces}
\def\Mla{\marginpar[\boldmath\hfil$\Rightarrow$]%
                   {\boldmath$\Leftarrow $\hfil}%
                    \typeout{marginpar: $\leftrightarrow$}\ignorespaces}
\overfullrule 5pt
\oddsidemargin -15mm
\marginparwidth 29mm
}

\def\stars{\strut\leaders\hbox{*}\hfill\strut}
\def\starline{\hfil\strut\hfil\hbox to \textwidth {\stars}\hfil}
\newlength{\parwidth}\newlength{\colonewidth}%
\newlength{\restpageheight}

\makeatletter
\newcommand{\cpcsuptable}[2]
{\settowidth{\colonewidth}{#1}\setlength{\parwidth}{\textwidth}%
\addtolength{\parwidth}{-\colonewidth}\addtolength{\parwidth}{-2em}%
\bce
\setlength{\restpageheight}{\@colroom}\addtolength{\restpageheight}{-\pagetotal}
\ifdim \restpageheight<20pt \pagebreak\fi
\begin{supertabular}[l]{p{\colonewidth}@{ }c@{ }p{\parwidth}}
#2
\end{supertabular}%
\ece
}%
\makeatother

\newcommand{\cpcsubtable}[2]
{%
\settowidth{\colonewidth}{#1}
\addtolength{\parwidth}{-\colonewidth}\addtolength{\parwidth}{-1em}%
\begin{tabular}[t]{@{}p{\colonewidth}@{ }c@{ }p{\parwidth}@{}}
#2
\end{tabular}
}%

\newcommand{\cpcitemtable}[2]
{\settowidth{\colonewidth}{#1}\setlength{\parwidth}{\textwidth}%
\addtolength{\parwidth}{-\leftmargin}%
\addtolength{\parwidth}{-\colonewidth}\addtolength{\parwidth}{-2em}%
\nobreak
\begin{flushleft}%
\begin{tabular}[l]{@{}p{\colonewidth}@{ }c@{ }p{\parwidth}}%
#2 
\end{tabular}%
\end{flushleft}%
}

\newcommand{\cpcframetable}[2]
{%
\settowidth{\colonewidth}{#1}\setlength{\parwidth}{\textwidth}%
\addtolength{\parwidth}{-\colonewidth}\addtolength{\parwidth}{-3em}%
\smallskip
\begin{tabular}{@{}|p{\colonewidth}@{ }c@{ }p{\parwidth}|@{}}
\hline #2\\\hline
\end{tabular}
\smallskip
}%

\newcommand{\cpcframtable}[2]
{%
\settowidth{\colonewidth}{#1}\setlength{\parwidth}{\textwidth}%
\addtolength{\parwidth}{-\colonewidth}\addtolength{\parwidth}{-3em}%
\smallskip
\begin{tabular}{@{}|p{\colonewidth}|p{\parwidth}|@{}}
\hline #2\\\hline
\end{tabular}
\smallskip
}%

\newenvironment{cpcdescription}
   {\begin{description}%
   \setlength{\itemsep}{2ex}}%
   {\end{description}}

\begin{document}
\begin{frontmatter}
% Title, authors and addresses
% use the thanksref command within \title, \author or \address for footnotes:
% \title{Title\thanksref{label1}}
% \thanks[label1]{}
% \author{Name\thanksref{label2}}
% \thanks[label2]{}
% \address{Address\thanksref{label3}}
% \thanks[label3]{}
% including your email address:
% \address{Address\thanksref{email}}
% \thanks[email]{E-mail: }

\strut\hfill DESY 02-154 \\
\strut\hfill KA-TP-13-2002\\
\strut\hfill PSI-PR-02-10\\
\strut\hfill UB-HET-02-04\\

\title{\RacoonWW1.3: A Monte Carlo program for four-fermion production
at $\mathbf{e}^+ \mathbf{e}^-$ colliders}%\thanksref{label0}}

% use optional labels to link authors explicitly to addresses:
% \author[label1,label2]{}
% \address[label1]{}
% \address[label2]{}

\author[label1]{A. Denner\thanksref{email}},
\author[label2]{S. Dittmaier}, %\thanksref{label5}},
\author[label3]{M. Roth},
\author[label4]{D. Wackeroth}

\address[label1]{Paul Scherrer Institut, CH-5232 Villigen PSI, Switzerland}
\address[label2]{Deutsches Electronen-Synchroton DESY, 
D-22603 Hamburg, Germany}
\address[label3]{Institut f\"ur Theoretische Physik, Universit\"at Karlsruhe, 
D-76128 Karslruhe, Germany}
\address[label4]{Department of Physics, 
SUNY at Buffalo, Buffalo, NY 14260, USA}
\thanks[email]{\parbox[t]{15cm}{Corresponding author\\
E-mail: Ansgar.Denner@psi.ch (A.~Denner)}}
%\thanks[email]{Corresponding author\\
%E-mail: Ansgar.Denner@psi.ch (A.~Denner),
%Stefan.Dittmaier@desy.de (S.~Dittmaier), 
%roth@particle.uni-karlsruhe.de (M.~Roth), 
%dow@ubpheno.physics.buffalo.edu (D.~Wackeroth)}
%\thanks[label5]{Heisenberg Fellow of the Deutsche Forschungsgemeinschaft}
\begin{abstract}
We present the Monte Carlo generator {\RacoonWW} that computes cross
sections to all processes $\eeffff$ and $\eeffffg$ and calculates the
complete $\Oa$ electroweak radiative corrections to $\eeWWffff$ in the
electroweak Standard Model in 
double-pole approximation.
The calculation of the tree-level processes $\eeffff$ and $\eeffffg$
is based on the full matrix elements for massless (polarized)
fermions.  When calculating radiative corrections to $\eeWWffff$, the
complete virtual doubly-resonant electroweak corrections are included, 
\ie the factorizable and non-factorizable virtual corrections in double-pole 
approximation, and the real corrections are based on the full matrix elements 
for  $\eeffffg$.
The matching of soft and collinear
singularities between virtual and real corrections is done
alternatively in two different ways, namely by using a subtraction
method or by applying phase-space slicing.  Higher-order
initial-state photon radiation and naive QCD corrections are taken 
into account.
{\RacoonWW} also provides anomalous triple gauge-boson couplings 
for all processes $\eeffff$ and anomalous quartic gauge-boson couplings
for all processes $\eeffffg$.
\end{abstract}

\begin{keyword}
% keywords here, in the form: keyword \sep keyword
radiative corrections \sep \PW-pair production \sep four-fermion production
\sep triple gauge-boson couplings \sep quartic gauge-boson couplings 
\sep Monte Carlo integration \sep event generation 
%\sep 
% PACS codes here, in the form: \PACS code \sep code
\PACS 12.15Lk \sep 12.15Ji \sep 14.70Fm 
\end{keyword}
\end{frontmatter}

\pagebreak[3]

{\bf PROGRAM SUMMARY}\nobreak

\noindent{\sl Title of program}: 
\RacoonWW, version 1.3

\noindent{\sl Program obtainable from}: 
CPC Program Library, Queen's University of Belfast, N.~Ireland

\noindent{\sl Computer}: 
any computer with FORTRAN 77 compiler. 
The program has been tested on 
DEC/ALPHA workstations and Linux PCs. %(Pentium III, 500 MHz)

\noindent{\sl Operating systems}:
UNIX, Linux % (Red Hat Linux 6.x, SUSE Linux 6.x)

\noindent{\sl Programming language used}: 
FORTRAN 77

\noindent{\sl Memory required to execute with typical data}: 
about 4\,MB

\noindent{\sl No.\ of bytes in distributed program, including test data, etc.}:
about 2\,MB

\noindent{\sl Distribution format}: uuencoded compressed tar file

\noindent{\sl Keywords}: 
radiative corrections, \PW-pair production, four-fermion production,
triple gauge-boson couplings, quartic gauge-boson couplings, 
Monte Carlo integration, event generation 

\noindent{\sl Nature of physical problem}\\
Precision calculations for W-pair production, four-fermion production,
and four-fermion plus photon production in $\Pep\Pem$ annihilation.

\noindent{\sl Method of solution}\\
Multi-channel Monte Carlo integration of perturbatively calculated
multidifferential cross sections including radiative corrections to
$\eeffff$.  The code can generate weighted and unweighted events. For
hadronization an interface to {\sc Pythia} is provided.

\noindent{\sl Restrictions on the complexity of the problem}\\
Virtual $\Oa$ radiative corrections are included in double-pole
approximation (DPA) only.  Higher-order initial-state radiation is
included via collinear structure functions in leading-logarithmic
approximation.  QCD corrections are included as a global
multiplicative factor $(1+\alpha_s/ \pi)$ for each hadronically
decaying W~boson.  The external fermions are assumed to be massless.
Below about 170 GeV the DPA is not applicable, and above 500 GeV
contributions become numerically important that are not yet
implemented in {\RacoonWW}.

\noindent{\sl Typical running time}\\
On a DEC/ALPHA workstation the computing time for calculating $10^7$
weighted events for tree-level predictions is about 2 hours.  When
including radiative corrections the computing time for $5\times 10^7$
weighted events amounts to about $30$ hours.  \pagebreak[4]

{\bf LONG WRITE-UP}
\def\thefootnote{\arabic{footnote}}%
\setcounter{footnote}{0}%
\section{Introduction}%
\label{se:intro}%
Experimental investigations of four-fermion production processes,
$\eeffff(\gamma)$, are phenomenologically very interesting for a
variety of reasons. The most prominent subprocess proceeds via
$\eeWWffff$, which is analyzed in the LEP2 experiment at the per-cent
level and can be studied with even higher precision at future
$\Pep\Pem$ linear colliders, where centre-of-mass (CM) energies go far
beyond the $210\GeV$ reached by LEP2, most probably up to the TeV
range. Upon reconstructing the W~bosons from their decay products, the
W-boson mass is determined at LEP2 with an accuracy of presently $\sim
40\MeV$, close to the finally aimed precision of $\sim30\MeV$.
Moreover, the triple gauge-boson couplings $\PZ\PW\PW$ and
$\gamma\PW\PW$ are directly constrained within a few per cent upon
analyzing angular distributions.  Recently the LEP2 collaborations
started to put bounds on specific deviations from the Standard-Model
(SM) quartic gauge-boson couplings, mainly by investigating cross
sections and photon-energy distributions in the processes
$\eeWWffffg$.  Of course, also other subchannels in four-fermion
production, such as Z-pair or single-W production, deliver important
information that is complementary to results obtained from W pairs,
although the corresponding accuracy at LEP2 is significantly lower.
Details about the present situation of the LEP2 data analysis can be
found in \citere{LEP2000} and about future prospects at linear
colliders in \citere{Accomando:1998wt,Abe:2001wn}.

To account for the high experimental accuracy on the theoretical side
is a great challenge: the W~bosons have to be treated as resonances in
the full four-fermion processes $\eeffff(\gamma)$, and radiative
corrections need to be included.  From the variety of the
above-mentioned physical issues it is also clear that semianalytical
predictions are not sufficient; the data analyses require realistic
studies with event generators that meet these theoretical
requirements.  In this article we describe the structure and the use
of the Monte Carlo (MC) event generator {\RacoonWW}, which is designed
for this task. Numerical results of {\RacoonWW} as well as the
underlying calculation of the cross sections and corrections have
already been published in
\citeres{Denner:1999gp,Denner:2000kn,Denner:2000bj,Denner:2001zp,Denner:2001vr}.
A general review of the state-of-the-art for four-fermion production
can be found in \citeres{Beenakker:1996kt} and
\cite{Grunewald:2000ju}.  The latter reference, in particular,
contains detailed tuned comparisons of {\RacoonWW} results to related
calculations, such as the semi-analytical calculation of
\citere{Beenakker:1998gr} and the MC generator \YFSWW\ 
\cite{Jadach:1995sp} for $\eeWWffff$ including $\O(\alpha)$
corrections in double-pole approximation (DPA), and the MC generators
\WRAP\ \cite{Montagna:2001uk} and \PHEGAS\ \cite{Papadopoulos:2000tt}
for tree-level predictions of $\eeffffg$.

{\RacoonWW} operates in three different modes: the {\it tree-level
  mode}, the {\em improved Born approximation (IBA) mode}, or the {\it
  radiative-correction (RC) mode}.  The tree-level mode provides
predictions for $\eeffff$ or $\eeffffg$ for all possible final states
with massless fermions in lowest order, based on the full set of
Feynman diagrams or on selected subsets \cite{Denner:1999gp}.
Gauge-boson decay widths are implemented as constant or running
widths, or within the {\it complex-mass scheme}. Since the latter
fully respects gauge invariance, gauge-invariance violations in the
other schemes can be controlled.  The initial-state radiation (ISR) in
leading-logarithmic (LL) approximation via the structure-function
method and the Coulomb singularity for slowly moving W~bosons can be
optionally included \cite{Denner:2001zp,Denner:2001vr}.  Anomalous
triple gauge-boson couplings are supported for all processes $\eeffff$
\cite{Denner:2001bd}, and for the process class $\eeffffg$ anomalous
quartic gauge-boson couplings are also available \cite{Denner:2001vr}.

The IBA mode calculates the leading universal corrections to the
$\eeWWffff$ processes \cite{Denner:2001zp} or to the process class
$\eeffffg$ \cite{Denner:2001vr}.  The IBA for $\eeWWffff$, which is
similar to the approach followed in the semi-analytical program
\GENTLE\ \cite{Bardin:1993bh}, is particularly interesting near the
W-pair production threshold where it is still applicable in contrast
to the full one-loop calculation in DPA.

The RC mode, which is designed for $\eeffff(\gamma)$, uses the $4f$
lowest-order contribution of the tree-level mode and further includes
the full electroweak $\Oa$ corrections in the DPA as defined in
\citere{Denner:2000bj}.  More precisely, only the virtual corrections
are treated in the DPA leading to factorizable and non-factorizable
corrections \cite{Denner:1998ia}. The former comprise the corrections
to the production and decay subprocesses, while the latter are due to
soft-photon exchange between these subprocesses. The real $\Oa$
corrections are based on the smallest gauge-invariant subset of the
full $\eeffffg$ matrix element that includes \PW-pair production.
They are evaluated analogously to the tree-level mode for $4f\gamma$
production.  Since the full calculation is consistently based on
massless external fermions, photons that are collinear to any external
charged fermion have to be recombined with this fermion, as described
in \citere{Denner:2000bj} in more detail. {\RacoonWW} provides such
recombination algorithms, which can, however, be replaced by
user-defined alternatives.  Finally, higher-order ISR at the LL level
is included up to $\O(\alpha^3)$ using the structure-function approach
\cite{sf}.  In this context, it should be emphasized that the
combination of the virtual corrections in DPA with the real
corrections based on full matrix elements is non-trivial. The
singularity structures of both parts are closely related, and a
mismatch has to be carefully avoided. {\RacoonWW} provides two
versions for the redistribution of the soft and collinear
singularities, resulting in two different branches, called {\it
  slicing} and {\it subtraction} branches. The former employs the
well-known phase-space slicing, the latter the subtraction procedures
of \citeres{Dittmaier:2000mb,Roth:1999kk}. Since even the phase-space
generation in the two branches is different, a comparison of results
obtained with the different branches is a highly non-trivial
consistency check.

Finally, {\RacoonWW} offers various further options, such as the
choice between different input-parameter schemes, the inclusion of
naive QCD corrections, a switch for the imaginary parts of loop
diagrams, and various settings to estimate the intrinsic uncertainty
of the DPA.  {\RacoonWW} is able to generate unweighted events by a
hit-and-miss algorithm.  In the RC mode unweighted events can only be
generated when using the slicing branch.  In order to account for the
complex peaking behaviour of the matrix elements {\RacoonWW} uses a
multi-channel integration technique \cite{Berends:1985gf} and adaptive
weight optimization \cite{Kleiss:1994qy}.

{\RacoonWW} has been tested up to a CM energy of $\sqrt{s}=500 \GeV$
in the RC mode.  The theoretical uncertainty of the total cross
section has been estimated to be $0.4\%$ for energies between $200
\GeV$ and $500 \GeV$, $0.5\%$ at $180 \GeV$, and $0.7\%$ for $170
\GeV$ \cite{Denner:2000bj,Grunewald:2000ju}.  The estimates result
from investigations of intrinsic ambiguities in the definition of the
DPA and from a detailed comparison with results obtained from the MC
generator \YFSWW\ \cite{Jadach:1995sp}.  In the RC mode of {\RacoonWW}
results have also been obtained for a CM energy of $1\TeV$, but no
detailed study of the theoretical uncertainties has been performed.
Around the \PW-pair production threshold at $160 \GeV$ the DPA is not
applicable, and above $500 \GeV$ higher-order radiative corrections
that are not yet implemented in {\RacoonWW}, e.g.\ large logarithmic
electroweak corrections beyond $\O(\alpha)$, may become numerically
sizeable.  In the tree-level mode {\RacoonWW} is tested up to CM
energies of $10 \TeV$ \cite{Denner:1999gp}, and IBA results have been
obtained for CM energies up to $1 \TeV$ \cite{Denner:2001zp}. The
tree-level and IBA modes are also applicable in the \PW-pair
production threshold region.

The paper is organized as follows: in \refse{sec:structure}, we sketch
the general structure of the program.  Section \ref{sec:details} is
devoted to the description of important details of the code. In
\refse{sec:usage}, we describe in detail the usage of {\RacoonWW}
including all important options. Section \ref{sec:summary} summarizes
the paper, and the appendix contains input and output of two sample
runs.

\section{Program structure}
\label{sec:structure}

{\RacoonWW} provides predictions for the cross section to 
the class of processes $\Pep \Pem \to 4f(\gamma)$ as given by
\begin{equation}
\label{eq:crosssection}
\int \rd \sigma = \frac{(2 \pi)^{4-3 n}}{2 s} \; \int \rd \Phi^{(n)}
\; | \M(p_+,p_-,k_3,\ldots ,k_{n+2})|^2,
\end{equation}
where $p_\pm$ and $k_i$ are the momenta of the incoming and outgoing
particles, respectively, and $n=4,5$ is the number of outgoing
particles.  The total energy of the scattering process in the CM
frame, where $\mathbf{p}_+ + \mathbf{p}_-=\mathbf{0}$, is denoted by
$\sqrt{s}=p_+^0+p_-^0$. The $n$-particle phase space is defined by
\begin{equation}
\rd \Phi^{(n)}=\prod\limits_{i=3}^{n+2} 
\frac{\rd^3 \mathbf{k}_i}{2 k_i^0} \,
\delta^{(4)} \left.\left( p_+ + p_- -\sum_{i=3}^{n+2} k_i \right)\right|_{k_i^0=\sqrt{\mathbf{k}_i^2+m_i^2}} \; .
\end{equation}
Results for the matrix element squared $|\M|^2$ in terms of helicity
amplitudes for the tree-level processes $\Pep\Pem\to 4f(\gamma)$ with
massless fermions are given in \citere{Denner:1999gp}, and the
calculation of radiative corrections is described in
\citere{Denner:2000bj}.  The phase-space integration is performed with
the help of a MC technique, which allows to calculate a variety of
observables simultaneously and to easily implement separation cuts and
photon recombination schemes in order to account for the experimental
situation.  The matrix element squared exhibits a complicated peaking
behaviour in various regions of the integration domain. In order to
obtain a numerically stable result and to reduce the MC integration
error, {\RacoonWW} is based on a multi-channel MC method
\cite{Berends:1985gf} with adaptive weight optimization
\cite{Kleiss:1994qy}.

The general structure of the program is determined by the generation
of the $n$-particle phase space, $\Phi^{(n)}$, and the evaluation of
the matrix element squared, $|\M|^2$. The program run starts with some
initializations and ends with the output of the integration result in
form of an output file for the total cross section and a number of
data files for various distributions.

The initialization of the matrix-element evaluation consists to a
large extent of specifying the input parameters and the different
options provided by \RacoonWW. The initialization of the phase-space
generation in the multi-channel MC integration approach essentially
amounts to the identification of kinematical channels for the given
final state.  While the user is free to choose the parameters and
flags for options in the input file (and to a lesser extent also in
{\tt racoonww.f} and {\tt public.f}), the initialization of the
multi-channel phase-space generators is done automatically by
\RacoonWW.

{\RacoonWW} consists of 11 fortran files and a makefile.  The usage of
the program is described in the file {\tt README}.  {\RacoonWW} also
includes 5 default input files with corresponding output and data
files. They represent the recommended choices for input parameters and
flags of the different modes provided by \RacoonWW.  The general
structure of the program is reflected by a series of subroutine calls
in the main program in {\tt racoonww.f}:
\cpcsuptable{\tt CALL EVENTGENERATION}{%
{\tt CALL READINPUT} &:& the input file is read. \\
{\tt CALL INITIALIZE} &:& 
parameters and flags are set, and the MC integration is 
initialized (specification of channels for multi-channel integration). \\
{\tt CALL INTEGRATION} &:& 
the MC integration and the generation of {\tt NEVENTSW} weighted events
are performed, the 
hit-and-miss algorithm for generation of unweighted events is initialized. \\
{\tt CALL EVENTGENERATION} &:& 
in the main program it is only called if ${\tt NEVENTSUNW}>0$; 
then {\tt NEVENTSUNW} unweighted events are generated by using 
a hit-and-miss algorithm (see \refse{subsec:unweighting}).         
This subroutine is also called in the subroutine
{\tt INTEGRATION} where it generates weighted events. \\
{\tt CALL PY4FRM} &:& 
the interface to {\sc Pythia} \cite{Sjostrand:2001yu} to include parton 
showering and hadronization is called. 
The subroutine {\tt PY4FRM} is provided by 
{\sc Pythia} (see also \refse{subsec:unweighting}). \\
{\tt CALL WRITEOUTPUT} &:& the results are written to the output file and 
data files.\\
}

Besides the choices made for the input parameters and flags in the
input file, the user can change the default values for the remaining
parameters and flags of {\RacoonWW} in the subroutine {\tt INITIALIZE}
in {\tt racoonww.f} and in the subroutine {\tt PARAMETER} in {\tt
  public.f}.  The specification of the input in {\RacoonWW} is
described in detail in \refse{subsec:input}. After the initialization,
the MC integration is performed in
\cpcsuptable{\tt AVPREL}{%
\multicolumn{3}{l}{{\tt SUBROUTINE INTEGRATION(AVPREL,SIPREL)}}\\ 
{\tt AVPREL} &:& total cross section calculated with 
{\tt NEVENTSW} weighted events,\\ 
{\tt SIPREL} &:& corresponding statistical error \\
}%
by calling the MC kernels in
\cpcsuptable{\tt UNWEIGHTING}{%
\multicolumn{3}{l}{{\tt SUBROUTINE
EVENTGENERATION(ATOTSQR,A1SQR,A2SQR,UNWEIGHTING)}}\\ 
{\tt ATOTSQR} &:& 
total matrix element squared for the event (input for {\tt PY4FRM}),\\
{\tt A1SQR} &:& 
matrix element squared for the configuration with fermions 
$f_3,\bar{f}_4$ and $f_5, \bar{f}_6$ as the 
two colour singlets (input for {\tt PY4FRM}),\\
{\tt A2SQR} &:& matrix element squared for the 
configuration with fermions $f_3,\bar{f}_6$ and $f_5,\bar{f}_4$ as the 
two colour singlets (input for {\tt PY4FRM}), \\
{\tt UNWEIGHTING} &:& flag for generating weighted 
(${\tt UNWEIGHTING}=0$) or unweighted (${\tt UNWEIGHTING}=1$) events; 
this flag is automatically controlled by {\tt NEVENTSUNW}.\\
}
{\RacoonWW} provides an unweighting procedure based on a hit-and-miss
algorithm. It is activated when the user sets ${\tt NEVENTSUNW}>0$ in
the main program, where {\tt NEVENTSUNW} is the number of unweighted
events.  When generating unweighted events, parton showering and
hadronization for coloured final states can be taken into account by
activating the interface to {\sc Pythia} \cite{Sjostrand:2001yu} in
the main program. The information about the (unweighted) event is
stored in the HEP standard format~\cite{hepevt} in the common block
{\tt HEPEVT} (see \refse{subsec:unweighting}). The variables {\tt
  ATOTSQR}, {\tt A1SQR}, and {\tt A2SQR} are provided for the
generation of unweighted events.
%, ${\tt UNWEIGHTING}=1$. 
A detailed discussion of the unweighting procedure is given in
\refse{subsec:unweighting}.

By default {\RacoonWW} generates only weighted events (${\tt
  NEVENTSUNW}=0$).  The number of weighted events, {\tt NEVENTSW}, is
set in the input file. The events are represented by the four-momenta
of the final-state particles and their corresponding weights.  A
weight, $w(r_0^j,\mathbf{r}^{\,j})$, assigned to the MC point $(r_0^j,
\mathbf{r}^{\,j})$ of the $j$th event, is composed of the total
density of the event, $g_{{\rm tot}}$, (\cf\refse{sec:multichannel})
and the matrix element squared as
\begin{equation}
\label{eq:evweight}
w(r_0^j,{\mathbf{r}}^{j})=\frac{f(k_i)}{g_{{\rm tot}}(k_i)}
\quad \mbox{ with } \quad
f(k_i)=\frac{(2 \pi)^{4-3 n}}{2 s} \; | \M(k_i)|^2,
\end{equation}
where $r_0^j,{\mathbf{r}}^{j}$ is a set of pseudo-random numbers.
As usual, the MC estimate 
$\bar I$ of the total cross section is obtained as
($N\equiv{\tt NEVENTSW}$)
\begin{equation}
\label{eq:iestimate}
\bar I=\frac{1}{N} \sum_{j=1}^N w(r_0^j,{\mathbf{r}}^{\,j}).
\end{equation}
In the RC mode of {\RacoonWW} more than one weight per (weighted)
event is generated corresponding to the kinematics of the
subcontributions, \ie whether the weight is a $4f$ or $4f\gamma$ event
with or without ISR, or corresponding to the DPA part of the virtual
corrections, or to one of the subtraction terms.  The weights for each
weighted event (see \reftas{tab:wslicing1}--\ref{tab:wslicing2})
together with the corresponding four-momenta are provided in the
common block {\tt EVENT} in {\tt racoonww.f}:
\cpcsuptable{\tt WEIGHT(1:29)}{%
\multicolumn{3}{l}{{\tt COMMON/EVENT/WEIGHT,PBEAM,P,WEIGHTBORN,PBORN,NSUBEVENT}} \\
{\tt WEIGHT(1:29)} &:& all weights generated for one weighted event, \\
{\tt PBEAM(I,J)} &:& four-momenta of incoming particles (before ISR), \\
&& ${\tt PBEAM(1,0)+PBEAM(2,0)} = \sqrt{s}$, \\
&& 
\begin{tabular}[l]{@{}l@{ }c@{ }l}
${\tt I}=1,2$ &:& incoming positron, electron, \\ 
${\tt J}=0,\ldots,3$ &:& four-vector components,
\end{tabular} 
\\ 
{\tt P(I,J,K)} &:& four-momenta (after possible ISR), \\ 
&& ${\tt P(1,0,K)+P(2,0,K)} \leq \sqrt{s}$, \\
&& 
\cpcsubtable{${\tt K}=1,\ldots,29$}{%
${\tt I}=1,\ldots,7$  &:& \sloppy 
numerates the external particles\\&&
($1=\mathrm{incoming~positron}$, $2=\mathrm{incoming~electron}$,\\&& 
3,\ldots,6${}=\mathrm{outgoing~fermions}$, $7=\mathrm{photon}$), \\
${\tt J}=0,\ldots,3$  &:& four-vector components,\\
${\tt K}=1,\ldots,29$ &:& numerates the weights,
}%
\\
{\tt WEIGHTBORN}&:& weight for tree-level four-fermion process, \\  
{\tt PBORN(I,J)}&:& momenta corresponding to {\tt WEIGHTBORN}, \\
&&
\begin{tabular}[l]{@{}l@{ }c@{ }l}
${\tt I}=1,\ldots,6$ &:& numerates the external particles, \\
${\tt J}=0,\ldots,3$ &:& four-vector components,
\end{tabular} \\
{\tt NSUBEVENT} &:& number of weights corresponding to one weighted event, \\
&& 
\begin{tabular}[l]{@{}l@{ }c@{ }l}
${\tt SMC}=1,3$ &:& 8 weights per event (slicing branch), \\
${\tt SMC}=2$   &:& 29 weights per event (subtraction branch).
\end{tabular}\\
}

These two key components of the event generation, phase-space
generation and matrix-element evaluation, are described in detail in
\refses{subsec:phasespace} and \ref{subsec:matrixelement},
respectively.

For a realistic simulation of $\eeffff(\gamma)$ processes separation
cuts and a photon recombination procedure have to be applied to take
into account the detector response. {\RacoonWW} provides several
experiment-inspired recombination procedures and sets of separation
cuts, which can be specified in the input file. They are defined in
{\tt public.f} in
\cpcsuptable{\tt PBEAM(1:2,0:3)}{%
\multicolumn{3}{l}{
{\tt SUBROUTINE RECOMBINATION(PBEAM,PIN,POUT,NOCUT)}} \\
{\tt PBEAM(1:2,0:3)} &:& four-momenta of incoming particles (before ISR), \\
{\tt PIN(1:7,0:3)} &:& four-momenta of external particles (input), \\
{\tt POUT(1:7,0:3)} &:& four-momenta after eventual recombination (output), \\
{\tt NOCUT}&:& indicates whether the event is rejected  
(${\tt NOCUT}=0$) or has passed all cuts (${\tt NOCUT}=1$)\\
}%
and 
\cpcsuptable{\tt PBEAM(1:2,0:3)}{%
\multicolumn{3}{l}{{\tt SUBROUTINE CUT(PBEAM,P,NPARTICLE,NOCUT)}}\\ 
{\tt PBEAM(1:2,0:3)} &:& four-momenta of incoming particles (before ISR), \\
{\tt P(1:7,0:3)} &:& four-momenta of external particles (after possible ISR),\\
{\tt NPARTICLE} &:& number of particles in the final state,\\
{\tt PBEAM,NOCUT} &:& see subroutine {\tt RECOMBINATION}.\\
}%
Finally, the data files for the histograms are filled in subroutine
{\tt SETTING} in {\tt public.f}, and the results for the total and
differential cross sections are written to the output and data files
in the subroutine {\tt WRITEOUTPUT}.

In the following we describe in detail the most important features of
the program and its usage.

\section{Details of the program}
\label{sec:details}

{\RacoonWW} is a merger of two MC programs, which use the same
routines for the evaluation of the matrix elements but different,
completely independent routines for the phase-space generation.
Moreover, they differ in the treatment of the soft and collinear
singularities arising in the calculation of the radiative corrections
to $\eeWWffff$. One employs the subtraction method of
\citeres{Dittmaier:2000mb,Roth:1999kk} and the other applies
phase-space slicing (for a review see e.g.\ \citere{Harris:2001sx}).
The two branches of {\RacoonWW} are named accordingly, the subtraction
and slicing branches, and the corresponding MC integration kernels are
\cpcsuptable{\tt XV(1:15)}{%
  \multicolumn{3}{l}{{\tt FUNCTION KERN(XV,WGT)}}\\
  {\tt XV(1:15)} &:& pseudo-random numbers for the generation of one
  weighted event, \\
{\tt WGT} &:& not used in this version of {\RacoonWW}\\
}%
for the slicing branch in {\tt kern.f} and
\cpcsuptable{\tt RANDOM(1:15)}{%
\multicolumn{3}{l}{{\tt SUBROUTINE SUBTRACTION(RANDOM)}}\\ 
{\tt RANDOM(1:15)}&:& pseudo-random numbers for the generation of one
weighted event\\
}%
for the subtraction branch in {\tt subtraction.f}, respectively. The
pseudo-random number generator {\tt RANS} can be found at the end of
{\tt racoonww.f}. The user chooses between these two branches by
setting the flag {\tt SMC} in the input file, ${\tt SMC}=1$ or 3 for
the slicing branch and ${\tt SMC}=2$ for the subtraction branch. The
value ${\tt SMC}=3$ has to be used when unweighted events are
generated in the RC mode.  The initialization of the event generation
and the setting of branch-specific options for the slicing and
subtraction branches of {\RacoonWW} is done in
\cpcsuptable{\tt NEVENTSW}{%
\multicolumn{3}{l}{{\tt SUBROUTINE INITSLICING(ENERGY,CONV,SCALE,NEVENTSW,NOUT,IOUT)}}\\ 
{\tt ENERGY} &:& CM energy in GeV (${\tt ENERGY}=\sqrt{s}$), \\
{\tt CONV}  &:& conversion factor from ${\rm GeV}^{-2}$ to fbarn, \\
{\tt SCALE} &:& QED splitting scale in the structure functions 
(see \citere{Denner:2000bj} and \refse{subsec:matrixelement}),\\
{\tt NEVENTSW} &:& number of weighted events,\\
{\tt NOUT}  &:& number assigned to the output file, \\
{\tt IOUT}  &:& number assigned to the file {\tt optimization.info}\\
}
in {\tt slicing.f} and 
\cpcsuptable{\tt ENERGY,CONV,SCALE,NOUT}{%
\multicolumn{3}{l}{{\tt SUBROUTINE INITSUBTRACTION(ENERGY,CONV,SCALE,NOUT)}}\\
{\tt ENERGY,CONV,SCALE,NOUT} &:& 
see subroutine {\tt INITSLICING}\\ 
}% 
in {\tt subtraction.f}. In particular, the specification of the
kinematical channels for the multi-channel MC integration for the
chosen four-fermion final state is done automatically in these
routines.

After the initialization, the MC kernels are called.  The phase-space
cuts are applied first, and if the event is not discarded, the total
phase-space weight, $1/g_{{\rm tot}}$, and the matrix element squared,
$|\M|^2$, of \refeq{eq:evweight} are calculated.  For the
matrix-element evaluation the user can choose between the tree-level,
IBA, and RC modes of {\RacoonWW} by setting the flag {\tt SRC} in the
input file. The implementation of the three modes is described in
detail in \refse{subsec:matrixelement}.

\subsection{The phase-space generation and multi-channel MC integration}
\label{subsec:phasespace}

The numerical integration of \refeq{eq:crosssection} is rather
complicated owing to the rich peaking structure of the integrand,
since the amplitudes to $\eeffff$ and, in particular, to $\eeffffg$
involve a huge number of propagators that become resonant or are
enhanced in various phase-space regions.  To obtain reliable numerical
results, events have to be sampled more frequently in the integration
domains where the integrand is large. This redistribution of events is
called {\it importance sampling}. In practice, this means that the
mapping of the pseudo-random numbers $r_i$, $0\le r_i \le 1$, into the
space of final-state momenta has to be chosen such that the
corresponding Jacobian $1/g({{\mathbf r}})$ cancels the propagators of
the differential cross section at least partially.

\subsubsection{Propagator mappings}
\label{se:propmap}

In order to smoothen a single propagator with momentum transfer $q$
in the square of a matrix element,
\beq
\label{eq:prop}
|\M|^2\propto \frac{1}{(q^2-M^2)^2+M^2 \Ga^2},
\eeq
the phase space is parametrized in such a way that the virtuality $q^2$
of the propagator is chosen as an integration variable. 
The integral over $q^2$ is transformed into an integral over the random 
number $r$:
\beq
\int^{q^2_{\max}}_{q^2_{\min}} \rd q^2 =\int_0^1 \rd r \, 
\frac{1}{g_{\mathrm{prop}}(q^2(r))}. 
\eeq
The dependence of $q^2$ on the pseudo-random number $r$ has to be
chosen such that the Jacobian $1/g_{\mathrm{prop}}$ smoothens the
propagator \refeqf{eq:prop}.  In {\sc RacoonWW} two types of mappings
are used according to the mass $M$ of the propagating particle. Since
all occurring massive particles possess a finite decay width $\Gamma$,
a suitable mapping is of Breit--Wigner type ($M\ne 0$, $\Gamma\ne 0$):
\beqar
\nn
\label{eq:maps1}
q^2(r)&=&M \Ga \tan [y_1 +(y_2 - y_1 ) r]+M^2,\\
g_{\mathrm{prop}}(q^2)&=&\frac{M \Ga}{(y_2 - y_1 ) [(q^2-M^2)^2+M^2 \Ga^2]}
\eeqar 
with
\beq
y_{1,2}=\arctan
\left(\frac{q^2_{\min,\max}-M^2}{M \Ga}\right).
\eeq
For massless particles the following mappings are appropriate
($M=\Gamma=0$):
\beqar
\nn
\label{eq:maps2}
q^2(r)&=&\left[r (q^2_{\max})^{1-\nu}+
(1-r) (q^2_{\min})^{1-\nu}\right]^{\frac{1}{1-\nu}},\\
g_{\mathrm{prop}}(q^2)&=&\frac{1-\nu}
{(q^2)^\nu\left[(q^2_{\max})^{1-\nu}-(q^2_{\min})^{1-\nu}\right]} \,;\\[1em]
\label{eq:maps3}
q^2(r)&=& (q^2_{\max})^r \, (q^2_{\min})^{1-r},\qquad
g_{\mathrm{prop}}(q^2)=
\frac{1}{q^2\ln (q^2_{\max}/q^2_{\min})}. 
\eeqar
The mapping of \refeq{eq:maps2} is only valid for $\nu\ne 1$, 
while \refeq{eq:maps3} is a substitute of \refeq{eq:maps2}
for $\nu=1$. The naive expectation $\nu=2$ from the squared matrix element 
is not necessarily the best choice because the propagator poles are 
partially cancelled in the collinear limit. It turns out that a proper 
value is $\nu\lsim 1$. 

In this way all propagators of a given Feynman diagram can be
smoothened by choosing an appropriate set of integration variables and
mappings as given in \refeqs{eq:maps1}--\refeqf{eq:maps3}.  The
integration variables that do not correspond to a specific propagator
mapping are sampled uniformly, i.e.\ 
\beq
\label{eq:nomaps}
q^2(r) = q^2_{\max} r+q^2_{\min} (1-r), \qquad
\phi(r) = 2 \pi r, \qquad \cos\theta(r) =2 r-1
\eeq
for invariant masses $q^2$, azimuthal angles $\phi$, and polar angles $\theta$,
respectively.

\subsubsection{Multi-channel approach}
\label{sec:multichannel}

Obviously, importance sampling of all propagators appearing in the
amplitude is not possible by a single phase-space mapping.  The {\it
  multi-channel approach} \cite{Berends:1985gf} provides a solution to
this problem. For each peaking structure a suitable set of invariants
is chosen, and accordingly a mapping of random numbers $\mathbf r$
into the final-state momenta $k_i$, so that the resulting Jacobian of
this mapping $1/g_k$ cancels this particular peaking behaviour of
$f(k_i)$. The phase-space integral of \refeq{eq:crosssection} reads
\beq
\label{eq:multichannel}
\int \rd \Phi^{(n)}=
\int_0^1 \rd r_0 \, \sum_{k=1}^{N_{\mathrm{ch}}} 
\theta(r_0-\beta_{k-1}) \theta(\beta_{k}-r_0) 
\int_0^1 \rd^{3n-4} {\mathbf r} \, \frac{1}{g_{\mathrm{tot}}(k_i({\mathbf r}))},
\eeq
where the $\beta_i$ define a partition of unity, \ie
$\beta_i=\sum_{j=1}^i \alpha_j$, $\beta_0=0$,
$\beta_{N_{\mathrm{ch}}}=\sum_{j=1}^{N_{\mathrm{ch}}}\alpha_j=1$, and
${N_{\mathrm{ch}}}$ denotes the number of all channels.  In
\refeq{eq:multichannel} an additional random number $r_0$ is
introduced in order to select a {\it channel} $k$ randomly with
probability $\alpha_k\ge 0$.  The numbers $\alpha_k$ are called {\it a
  priori weights}.  All {\it local densities} $g_k$ are combined into
the {\it total density} $g_{\mathrm {tot}}$,
\beq
g_{\mathrm{tot}}=
\sum_{k=1}^{N_{\mathrm{ch}}} \alpha_k \, g_k,
\eeq
which is expected to smoothen the integrand over the whole phase
space.  In this way, it is possible to simultaneously include
different phase-space mappings which are suitable for different parts
of the integrand.  To be specific, for each (tree-level) Feynman
diagram one channel exists, so that all propagators (except massive
$t$-channel propagators in the slicing branch) of the squared diagram
are smoothened by the corresponding local density $g_k$.  The local
densities are obtained as the product of the corresponding Jacobians,
as given in \refeqs{eq:maps1}--\refeqf{eq:maps3}.

The a priori weights $\alpha_k$ are adapted in the early phase of the
MC run several times to optimize the convergence behaviour of the
numerical integration.  This {\it adaptive weight optimization}
\cite{Kleiss:1994qy} increases the a priori weights $\alpha_k$ for
such channels that give large contributions to the total cross
section.  After a certain number of generated events a new set of a
priori weights $\alpha_k^{\mathrm{new}}$ is calculated according to
\begin{eqnarray}
\label{eq:adaptopt}
\alpha_k^{\mathrm{new}} &\propto& \alpha_k 
\sqrt{\frac{1}{N}\sum_{j=1}^N \, 
\frac{g_k(k_i({{\mathbf r}}^{\,j})) \, [w(r_0^j,{{\mathbf r}}^{\,j})]^2}
{g_{\mathrm{tot}}(k_i({{\mathbf r}}^{\,j}))}}, \qquad
\sum_{k=1}^{N_{\mathrm{ch}}} \alpha_k^{\mathrm{new}}=1.
\end{eqnarray}

The error of the MC integration can be estimated as
\begin{equation}\label{eq:MCerror}
\delta \bar I=\sqrt{\frac{W({\mathbf \alpha})-\bar{I}^2}{N}},
\qquad 
W({\mathbf \alpha})=
\frac{1}{N} \sum_{j=1}^N [w(r_0^j,{{\mathbf r}}^{\,j})]^2
\end{equation}
with the MC estimate $\bar I$ of \refeq{eq:iestimate}.

The phase-space integration is implemented in {\RacoonWW} as follows:
\begin{itemize}
\item First the phase-space generators are initialized by identifying
  all channels needed to account for the peaking behaviour of the
  process under consideration. For the processes $\eeffff$, there are
  between 6 and 128 different channels, for $\eeffffg$ between 14 and
  928 channels. Each channel smoothens a particular combination of
  propagators that results from a characteristic Feynman diagram. The
  identification of the channels in the slicing and subtraction
  branches of {\RacoonWW} is described in \refse{subsubsec:slicing}
  and \refse{subsubsec:subtraction}, respectively.
\item The channel that is used to generate the event is picked
  randomly with probability $\alpha_k$. Using the randomly chosen
  channel, the momentum configuration $k_i$ is determined from the
  corresponding phase-space generator.  Finally, using the momentum
  configuration $k_i$ the local densities of all channels $g_k$ are
  calculated. From those densities the total density
  $g_{\mathrm{tot}}$ of the event is determined. Apart from the
  evaluation of the matrix elements, the calculation of the local
  densities is the most time-consuming part of the numerical
  integration.
\item In order to reduce the MC error further, the {\em a priori
    weights} $\alpha_k$ are recalculated several times in the MC
  integration according to \refeq{eq:adaptopt}.
\end{itemize}

\subsubsection{The phase-space generators}

The actual calculation of the event kinematics is decomposed into
three steps: the calculation of {\it time-like invariants} $s_i$, of
{\it $2\to 2$ scattering processes}, and of {\it $1\to 2$ decays}.
The phase-space integration of \refeq{eq:crosssection} reads
\beqar
\int \rd \Phi^{(n)} &=& 
\prod_{i=1}^{n-2}  \, 
\int_{s_{i,\min}}^{s_{i,\max}} \rd s_i \, 
%\prod_{j=\tau_0}^\tau \int \rd \Phi^{2\to 2}_j \, 
\prod_{j=1}^\tau \int \rd \Phi^{2\to 2}_j \, 
\prod_{k=1}^{n-1-\tau} \int \rd \Phi^{1\to 2}_k,
\label{eq:PSsplit}
\eeqar
where the phase spaces of the $2 \to 2$ scattering processes and the
$1 \to 2$ decays are denoted by $\Phi^{2\to 2}_j$ and $\Phi^{1\to
  2}_k$, respectively.  The $n-2$ invariants $s_i$ result from the
phase-space factorization into scattering processes and decays.  The
number $\tau$ of $2 \to 2$ scattering processes can range from 1 to
$n-1$. If there is only one $2 \to 2$ scattering process ($\tau=1$),
this can involve either a $t$-channel or an $s$-channel propagator. If
there are more scattering processes ($\tau\ge2$), these involve only
$t$-channel propagators. If only one $t$-channel $2\to 2$ scattering
process is present, \ie for the first two topologies in the first row
of \reffi{fig:SpecificPS} and the first three topologies in the second
row of \reffi{fig:SpecificPS}, all the $s_i$ correspond to the
virtualities of propagators and are mapped as described in
\refse{se:propmap}. The angles and invariants appearing in the
parametrizations of the phase spaces are sampled as described below.
If several $t$-channel propagators are present, \ie for $\tau\ge2$,
some of the $s_i$ do not correspond to virtualities of propagators in
the diagram; for such variables no mappings are introduced, they are
generated uniformly.  Topologies with no $t$-channel propagator, are
obtained from those with only one $t$-channel propagator by replacing
the $t$-channel scattering process with an $s$-channel scattering
process.  The corresponding invariant is fixed to the CM energy
squared $s$, and the phase space is parametrized as for the $1\to 2$
decays in \refeq{eq:decays} or equivalently for the $2\to2$ scattering
processes in \refeq{eq:scattering} without propagator mapping
($\nu=M=\Gamma=0$).  Diagrams with four-particle vertices result from
the contraction of propagators in the diagrams of
\reffi{fig:SpecificPS}. Consequently, they also fit into the
decomposition \refeqf{eq:PSsplit}, and the invariants corresponding to
the contracted propagators are sampled uniformly.  More details on
phase-space parametrizations can be found in
\citeres{Roth:1999kk,By73}.

For each class of diagrams with the same propagator structure we adopt an 
own channel for the numerical integration. No special channels are provided 
for interference contributions. 
\bfi
\centerline{
%\framebox{
\setlength{\unitlength}{1pt}
\begin{picture}(220,200)(0,0)
\ArrowLine( 25, 25)(100, 25)
\Line(100, 25)(100,175)
\ArrowLine( 25,175)(100,175)
\ArrowLine(100, 25)(200, 25)
\Line(100,125)(150,125)
\ArrowLine(150,125)(200,110)
\ArrowLine(150,125)(200,140)
\ArrowLine(100, 75)(200, 75)
\ArrowLine(100,175)(200,175)
\Vertex(100, 25){3}
\Vertex(100, 75){3}
\Vertex(150,125){3}
\Vertex(100,125){3}
\Vertex(100,175){3}
\put(105, 45){$t_1$}
\put(105, 95){$t_3$}
\put(105,145){$t_2$}
\put(127,133){$s_1$}
\put( 35,  8){$p_+$}
\put( 35,185){$p_-$}
\put(175,  8){$k_6$}
\put(175,185){$k_3$}
\put(175, 80){$k_5$}
\put(175,140){$k_4$}
\put(175,102){$k_7$}
\Text(245,125)[l]{$s_3$}
\Text(215,107.5)[l]{$s_2$}
\CArc( 70,125  )(170,340, 20)
\CArc(100,107.5)(110,340, 20)
\end{picture}
}
%}
\vspace*{-.5em}
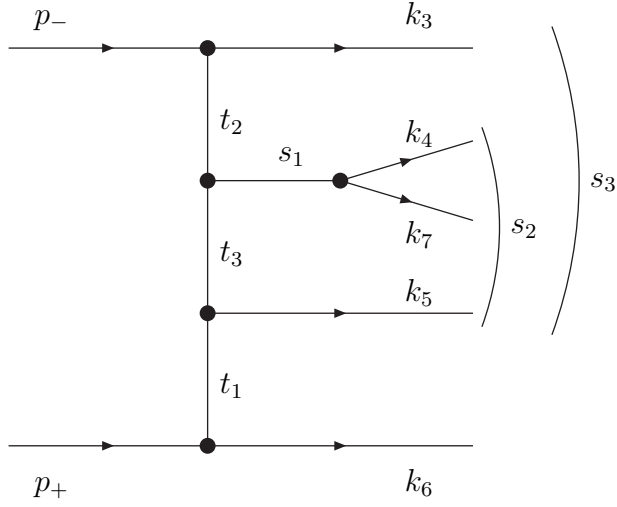
\caption{Generic diagram of a specific channel}
\label{fig:SpecificPS}
\efi 

We illustrate the general strategy by considering the specific
topology shown in \reffi{fig:SpecificPS}, where a particular choice for
the incoming and outgoing momenta $p_\pm$ and $k_i$ is made.
For this example the phase-space parametrization \refeq{eq:PSsplit} reads
\beqar
\int \rd \Phi^{(n)}\Big|_{\mathrm{Fig.\ref{fig:SpecificPS}}}&=& 
\prod_{i=1}^3 \int_{s_{i,\min}}^{s_{i,\max}} \rd s_i \, 
\int \rd \Phi^{2\to 2}(p_+,p_-;k_6,k_{3457}) \, 
\int \rd \Phi^{2\to 2}(p_-,p_+-k_6;k_3,k_{457}) \,
\nn\\ && {}\times
\int \rd \Phi^{2\to 2}(p_+-k_6,p_--k_3;k_{5},k_{47}) \,
\int \rd \Phi^{1\to 2}(k_4,k_{7}),
\hspace*{2em}
\eeqar
where sums of outgoing momenta are abbreviated by $k_{ij}=k_i+k_j$,
$k_{ijk}=k_i+k_j+k_k$, etc.  Note that only the time-like invariant
$s_1=k_{47}^2$ corresponds to a virtuality of a propagator in the
diagram, while $s_2=k_{457}^2$ and $s_3=k_{3457}^2$ correspond to
invariant masses of fictitious final-state particles within the first
two $2\to 2$ scattering processes, symbolized by the arches in
\reffi{fig:SpecificPS}.  More examples for the phase-space generation
can be found in \citeres{Denner:1999gp,Roth:1999kk}.

In detail we proceed as follows.
\subsubsection*{Calculation of time-like invariants}

First of all, the time-like invariants $s_i$ are determined.  The
invariants corresponding to $s$-channel propagators are calculated
from \refeqs{eq:maps1}--\refeqf{eq:maps3}. All other time-like
invariants are sampled uniformly, \ie calculated from
\refeq{eq:nomaps}.  The determination of the invariants $s_i$ is
ordered in such a way that the $s_i$ nearest to final-state particles
are calculated first, followed by the next-to-nearest $s$-channel
propagators, and so on.  The lower and upper limits $s_{i,\min/\max}$,
in general, are functions of the already determined invariants. To
enhance the efficiency of the numerical integration in the presence of
separation cuts, in some cases these cuts are taken into account in
the evaluation of $s_{i,\min/\max}$.
 
\subsubsection*{$2\to2$ scattering processes}

The phase spaces for the $2\to2$ scattering processes
of \refeq{eq:PSsplit} are parametrized as
\beq\label{eq:scattering}
\int \rd \Phi^{2\to2}(q_1,q_2;p_1,p_2)
=\frac{1}{8 \sqrt{(q_1 q_2)^2 -q_1^2 q_2^2}}
\int_0^{2\pi} \rd \phi^* \,
\int_{t_{\min}}^{t_{\max}} \rd t,
\eeq
where $q_{1,2}$ are the incoming and $p_{1,2}$ are the outgoing
momenta, and $\phi^*$ is the azimuthal angle defined by $p_1$ and
$q_1$ in the CM frame of the subprocess where ${\bf q}_1+{\bf q}_2=0$.
The argument $t=(p_1-q_1)^2$ of the $t$-channel propagator is
calculated according to \refeqs{eq:maps1}--\refeqf{eq:maps3} for
$t$-channel scattering processes (apart from massive $t$-channel
propagators in the slicing branch that are sampled uniformly) and
according to \refeq{eq:nomaps} for $s$-channel scattering processes.
The azimuthal angle $\phi^*$ is sampled uniformly.  Since the
corresponding polar angle $\cos\theta^*$ of this scattering process
depends on the invariant $t$ linearly,
\beq
t=\frac{q_1^2 p_2^2+q_2^2 p_1^2-2 (q_1 q_2)(p_1 p_2)
+2 \sqrt{(q_1 q_2)^2-q_1^2 q_2^2} \sqrt{(p_1 p_2)^2-p_1^2 p_2^2} \cos \theta^*}
{(q_1+q_2)^2},
\eeq
the calculation of the momenta and the Lorentz transformation  
into the laboratory (LAB) frame is simple and, thus, skipped here. 
Explicit formulas can, for instance, be found in \citeres{Roth:1999kk,By73}.

Cuts on angles between outgoing particles and the beams effectively
reduce the integration range of $t$.  For the $2\to2$ scattering
processes that are attached to the original incoming particles these
cuts are in some cases included in the determination of
$t_{\min/\max}$ to enhance the efficiency.

\subsubsection*{$1\to2$ particle decays}

The phase-space integration for the $1\to2$ particle decays in
\refeq{eq:PSsplit} reads
\beq\label{eq:decays}
\int \rd \Phi^{1\to 2}(p_1,p_2)
=\frac{\sqrt{(p_1 p_2)^2-p_1^2 p_2^2}}{4 (p_1+p_2)^2}
\int_0^{2 \pi} \rd \phi^* \, 
\int_{-1}^1 \rd \cos \theta^*,
\eeq
where $\phi^*$ and $\theta^*$ are the azimuthal and polar angles 
in the rest frame of the decaying particle, respectively.
The variables $\phi^*$ and $\cos \theta^*$ are sampled uniformly.
The momenta of the outgoing particles are $p_1$ and $p_2$.
As in the former step, the calculation of the momenta and the Lorentz 
transformation into the LAB frame is straightforward.

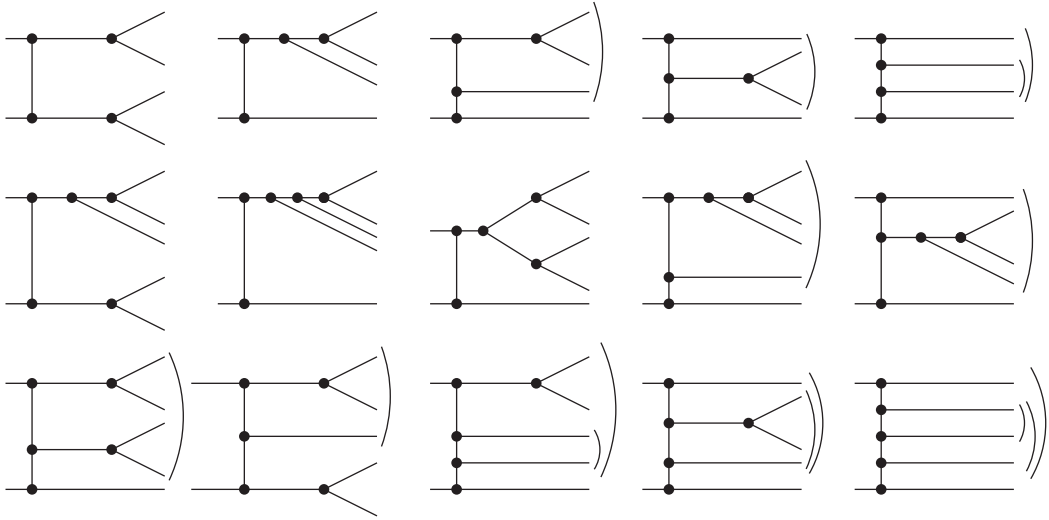
\begin{figure}
\begin{center}
%\framebox{
\begin{picture}(400,190)(0,-20)
\Line(10,160)(50,160)
\Line(10,130)(50,130)
\Line(20,130)(20,160)
\Line(50,130)(70,140)
\Line(50,130)(70,120)
\Line(50,160)(70,170)
\Line(50,160)(70,150)
\Vertex(50,160){2}
\Vertex(50,130){2}
\Vertex(20,160){2}
\Vertex(20,130){2}
\Line(90,160)(130,160)
\Line(90,130)(150,130)
\Line(100,130)(100,160)
\Line(115,160)(150,142.5)
\Line(130,160)(150,170)
\Line(130,160)(150,150)
\Vertex(130,160){2}
\Vertex(100,130){2}
\Vertex(100,160){2}
\Vertex(115,160){2}
\Line(170,160)(210,160)
\Line(170,130)(230,130)
\Line(180,130)(180,160)
\Line(180,140)(230,140)
\Line(210,160)(230,170)
\Line(210,160)(230,150)
\Vertex(210,160){2}
\Vertex(180,130){2}
\Vertex(180,160){2}
\Vertex(180,140){2}
\CArc(180,155)(55,340,20)
\Line(250,160)(310,160)
\Line(250,130)(310,130)
\Line(260,130)(260,160)
\Line(260,145)(290,145)
\Line(290,145)(310,155)
\Line(290,145)(310,135)
\Vertex(260,130){2}
\Vertex(260,145){2}
\Vertex(260,160){2}
\Vertex(290,145){2}
\CArc(280,147.5)(35,336,24)
\Line(330,160)(390,160)
\Line(330,130)(390,130)
\Line(340,130)(340,160)
\Line(340,150)(390,150)
\Line(340,140)(390,140)
\Vertex(340,130){2}
\Vertex(340,140){2}
\Vertex(340,150){2}
\Vertex(340,160){2}
\CArc(360,150)(37,338,22)
\CArc(380,145)(14,330,30)
\Line(10,100)(50,100)
\Line(10,60)(50,60)
\Line(20,60)(20,100)
\Line(50,60)(70,70)
\Line(50,60)(70,50)
\Line(50,100)(70,110)
\Line(50,100)(70,90)
\Line(35,100)(70,82.5)
\Vertex(50,100){2}
\Vertex(50,60){2}
\Vertex(20,100){2}
\Vertex(20,60){2}
\Vertex(35,100){2}
\Line(90,100)(130,100)
\Line(90,60)(150,60)
\Line(100,60)(100,100)
\Line(130,100)(150,110)
\Line(130,100)(150,90)
\Line(120,100)(150,85)
\Line(110,100)(150,80)
\Vertex(130,100){2}
\Vertex(100,60){2}
\Vertex(130,100){2}
\Vertex(100,100){2}
\Vertex(120,100){2}
\Vertex(110,100){2}
\Line(170,87.5)(190,87.5)
\Line(170,60)(230,60)
\Line(180,60)(180,87.5)
\Line(190,87.5)(210,100)
\Line(190,87.5)(210,75)
\Line(210,100)(230,110)
\Line(210,100)(230,90)
\Line(210,75)(230,85)
\Line(210,75)(230,65)
\Vertex(180,60){2}
\Vertex(180,87.5){2}
\Vertex(190,87.5){2}
\Vertex(210,100){2}
\Vertex(210,75){2}
\Line(250,100)(290,100)
\Line(250,60)(310,60)
\Line(260,60)(260,100)
\Line(290,100)(310,110)
\Line(290,100)(310,90)
\Line(275,100)(310,82.5)
\Line(260,70)(310,70)
\Vertex(290,100){2}
\Vertex(260,60){2}
\Vertex(290,100){2}
\Vertex(260,100){2}
\Vertex(275,100){2}
\Vertex(260,70){2}
\CArc(260,90)(57,335,25)
\Line(330,100)(390,100)
\Line(330,60)(390,60)
\Line(340,60)(340,100)
\Line(370,85)(390,95)
\Line(370,85)(390,75)
\Line(355,85)(390,67.5)
\Line(340,85)(370,85)
\Vertex(370,85){2}
\Vertex(340,60){2}
\Vertex(340,100){2}
\Vertex(370,85){2}
\Vertex(340,85){2}
\Vertex(355,85){2}
\CArc(340,83.75)(57,340,20)
\Line(10,30)(50,30)
\Line(10,-10)(70,-10)
\Line(20,-10)(20,30)
\Line(50,30)(70,40)
\Line(50,30)(70,20)
\Line(50,5)(70,15)
\Line(50,5)(70,-5)
\Line(20,5)(50,5)
\Vertex(50,30){2}
\Vertex(20,-10){2}
\Vertex(20,30){2}
\Vertex(20,5){2}
\Vertex(50,5){2}
\CArc(20,17.5)(57,335,25)
\Line(80,30)(130,30)
\Line(80,-10)(130,-10)
\Line(100,-10)(100,30)
\Line(130,-10)(150,0)
\Line(130,-10)(150,-20)
\Line(130,30)(150,40)
\Line(130,30)(150,20)
\Line(100,10)(150,10)
\Vertex(130,30){2}
\Vertex(130,-10){2}
\Vertex(100,30){2}
\Vertex(100,-10){2}
\Vertex(100,10){2}
\CArc(100,25)(55,340,20)
\Line(170,30)(210,30)
\Line(170,-10)(230,-10)
\Line(180,-10)(180,30)
\Line(210,30)(230,40)
\Line(210,30)(230,20)
\Line(180,10)(230,10)
\Line(180,0)(230,0)
\Vertex(210,30){2}
\Vertex(180,-10){2}
\Vertex(180,30){2}
\Vertex(180,0){2}
\Vertex(180,10){2}
\CArc(220,5)(14,328,32)
\CArc(180,20)(60,335,25)
\Line(250,30)(310,30)
\Line(250,-10)(310,-10)
\Line(260,-10)(260,30)
\Line(290,15)(310,25)
\Line(290,15)(310,5)
\Line(260,15)(290,15)
\Line(260,0)(310,0)
\Vertex(290,15){2}
\Vertex(260,-10){2}
\Vertex(260,30){2}
\Vertex(260,0){2}
\Vertex(260,15){2}
\CArc(280,12.5)(35,335,25)
\CArc(280,15)(38,330,30)
\Line(330,30)(390,30)
\Line(330,-10)(390,-10)
\Line(340,-10)(340,30)
\Line(340,20)(390,20)
\Line(340,0)(390,0)
\Line(340,10)(390,10)
\Vertex(340,20){2}
\Vertex(340,-10){2}
\Vertex(340,30){2}
\Vertex(340,0){2}
\Vertex(340,10){2}
\CArc(380,15)(14,330,30)
\CArc(370,10)(28,332,28)
\CArc(360,15)(42,330,30)
\end{picture}
%}
\end{center}
\caption{The generic phase-space generators symbolized by the topologies 
of the corresponding diagrams. The arcs indicate the intermediate $2\to2$ 
scattering processes used in the phase-space parametrizations. Topologies
with an $s$-channel $2\to 2$ scattering process are obtained from the ones
with only one $t$-channel scattering process by contracting
the $t$-channel propagator.}
\label{fig:topologies}
\end{figure}

\begin{sloppypar}
  All topologies corresponding to the generic phase-space generators
  for the processes $\eeffff$ and $\eeffffg$ are shown in
  \reffi{fig:topologies}.
\end{sloppypar}

\subsubsection{Phase-space generation in the slicing branch}
\label{subsubsec:slicing}

In the slicing branch, the $2\to 2$ scattering processes and the 
$1\to 2$ decays are combined in the function
\beq
R_2(s_i,{\tt MAPI}/{\tt MAPIG})=\left\{
\begin{array}{ll}
\disp\int\rd\Phi^{1\to 2}(p_1,p_2) \quad & \mbox{for} \quad {\tt MAPI},{\tt MAPIG}=0, \\
\disp\int\rd\Phi^{2\to 2}(q_1,q_2;p_1,p_2) \quad & \mbox{for} \quad 
{\tt MAPI},{\tt MAPIG}=1,2,3,
\end{array}
\right.
\eeq
where $s_i=(p_1+p_2)^2$. Using the function $R_2$,
the phase-space integration of \refeq{eq:PSsplit} reads
\beqar
\label{eq:four}
\int \rd \Phi^{(4)}&=& 
%\int_{s_{1,\min}}^{s_{1,\max}} \rd s_1 \, 
%\int_{s_{2,\min}}^{s_{2,\max}} \rd s_2
\int \rd s_1 \, \int \rd s_2 
 \, R_2(s_1,{\tt MAPI(5)}) \, 
R_2(s_2,{\tt MAPI(4)})\, R_2(s,{\tt MAPI(3)}), \\
\label{eq:five}
\int \rd \Phi^{(5)}&=& 
%\int_{s_{1,\min}}^{s_{1,\max}} \rd s_1 \, 
%\int_{s_{2,\min}}^{s_{2,\max}} \rd s_2 \, 
%\int_{s_{3,\min}}^{s_{3,\max}} \rd s_3
\int \rd s_1 \, \int \rd s_2 \, \int \rd s_3 
 \, R_2(s_1,{\tt MAPIG(5)}) \, 
\nonumber \\ && {} \times 
R_2(s_2,{\tt MAPIG(4)}) \, R_2(s_3,{\tt MAPIG(3)})\, R_2(s,{\tt MAPIG(8)}).
\eeqar
The propagator mappings for the functions $R_2$ and for the variables
$s_1,s_2,s_3$ are specified by the values of the variables {\tt MAPI}
and {\tt MAPIG} for the four- and five-particle phase spaces,
respectively. In this way all channels for the multi-channel
integration of the processes $\eeffff$ and $\eeffffg$ can be
constructed from the two and three basic phase-space decompositions,
respectively, shown in \reffi{fig:topo}.  Note that several topologies
of \reffi{fig:topologies} fit into single structural diagrams of
\reffi{fig:topo}.  The definitions of the variables $s_i$ and their
phase-space boundaries are given in \refta{ta:invariants}.
\begin{figure}
\centerline{
%\framebox{
\begin{picture}(190,100)(10,0)
\put(20,100){${\tt MAPI(6)}=1$:}
\Line(20,80)(50,50)
\Line(20,20)(50,50)
\Line(50,50)(120,80)
\Line(50,50)(120,60)
\Line(50,50)(120,40)
\Line(50,50)(120,20)
\GCirc(60,50){20}{.5}
\CArc( 70, 70)( 60,345,15)
\put(135, 68){$s_1$}
\CArc( 70, 30)( 60,345,15)
\put(135, 28){$s_2$}
\end{picture}
%}
%\framebox{
\begin{picture}(190,100)(10,0)
\put(20,100){${\tt MAPI(6)}=2$:}
\Line(20,80)(50,50)
\Line(20,20)(50,50)
\Line(50,50)(120,80)
\Line(50,50)(120,60)
\Line(50,50)(120,40)
\Line(50,50)(120,20)
\GCirc(60,50){20}{.5}
\CArc( 70, 70)( 60,345,15)
\put(133, 68){$s_1$}
\CArc( 90, 60)( 55,330,30)
\put(150, 58){$s_2$}
\end{picture}
%}
}
\centerline{
%\framebox{
\begin{picture}(140,130)(20,0)
\put(20,110){${\tt MAPIG(6)}=1$:}
\Line(20,80)(50,50)
\Line(20,20)(50,50)
\Line(50,50)(120,90)
\Line(50,50)(120,70)
\Line(50,50)(120,50)
\Line(50,50)(120,30)
\Line(50,50)(120,10)
\GCirc(60,50){20}{.5}
\CArc( 65, 80)( 60,345,15)
\put(130, 78){$s_1$}
\CArc( 65, 20)( 60,345,15)
\put(130, 18){$s_3$}
\CArc( 88, 70)( 55,330,30)
\put(148, 68){$s_2$}
\end{picture}
%}
%\framebox{
\begin{picture}(140,130)(20,0)
%\put(20,110){${\tt MAPIG(6)}=2$ (${\tt MAPIG(4)}=0$):}
\put(20,110){${\tt MAPIG(6)}=2$:}
\Line(20,80)(50,50)
\Line(20,20)(50,50)
\Line(50,50)(120,90)
\Line(50,50)(120,70)
\Line(50,50)(120,50)
\Line(50,50)(120,30)
\Line(50,50)(120,10)
\GCirc(60,50){20}{.5}
\CArc( 65, 80)( 60,345,15)
\put(130, 78){$s_1$}
\CArc( 65, 40)( 60,345,15)
\put(130, 38){$s_2$}
\CArc( 60, 60)( 85,335,25)
\put(150, 58){$s_3$}
\end{picture}
%}
%}
%\centerline{
%\framebox{
\begin{picture}(160,130)(20,0)
\put(20,110){${\tt MAPIG(6)}=3$:}
\Line(20,80)(50,50)
\Line(20,20)(50,50)
\Line(50,50)(120,90)
\Line(50,50)(120,70)
\Line(50,50)(120,50)
\Line(50,50)(120,30)
\Line(50,50)(120,10)
\GCirc(60,50){20}{.5}
\CArc( 65, 80)( 60,345,15)
\put(130, 78){$s_1$}
\CArc( 88, 70)( 55,330,30)
\put(148, 68){$s_2$}
\CArc( 78, 60)( 85,335,25)
\put(168, 58){$s_3$}
\end{picture}
}
%}
\caption{The basic phase-space decompositions from which 
  all channels for the processes $\eeffff$ and $\eeffffg$ can be
  constructed, as used in the slicing branch}
\label{fig:topo}
\end{figure}

\begin{table}
\begin{center}
\begin{tabular}{@{}|l|l|r@{}c@{}l|@{}}
\hline
topology & invariants & \multicolumn{3}{l|}{phase-space boundaries} \\
\hline
\hline
${\tt MAPI(6)}=1$ & $s_1=(k_{{\tt II(1)}}+k_{{\tt II(2)}})^2$ & 
$0$ & $\le s_1 $ & $\le s$ \\
& $s_2=(k_{{\tt II(3)}}+k_{{\tt II(4)}})^2$ & 
$0$ & $\le s_2 $ & $\le (\sqrt{s}-\sqrt{s_1})^2$  \\
\hline
${\tt MAPI(6)}=2$ & $s_1=(k_{{\tt II(1)}}+k_{{\tt II(2)}})^2$ & 
$0 $ & $\le s_1 $ & $\le s$ \\
& $s_2=(k_{{\tt II(1)}}+k_{{\tt II(2)}}+k_{{\tt II(3)}})^2$ & 
$s_1 $ & $\le s_2 $ & $\le s$ \\  
\hline
\hline
${\tt MAPIG(6)}=1$ & $s_1=(k_{{\tt IIG(1)}}+k_{{\tt IIG(2)}})^2$ & 
$0 $ & $\le s_1 $ & $\le s$ \\
& $s_2=(k_{{\tt IIG(1)}}+k_{{\tt IIG(2)}}+k_{{\tt IIG(3)}})^2$ & 
$s_1 $ & $\le s_2 $ & $\le s$ \\  
& $s_3=(k_{{\tt IIG(4)}}+k_{{\tt IIG(5)}})^2$ & 
$0 $ & $\le s_3 $ & $\le (\sqrt{s}-\sqrt{s_2})^2$ \\  
\hline
${\tt MAPIG(6)}=2$ & $s_1=(k_{{\tt IIG(1)}}+k_{{\tt IIG(2)}})^2$ & 
$0 $ & $\le s_1 $ & $\le s$ \\
& $s_2=(k_{{\tt IIG(3)}}+k_{{\tt IIG(4)}})^2$ & 
$0 $ & $\le s_2 $ & $\le (\sqrt{s}-\sqrt{s_1})^2$ \\  
& $s_3=(k_{{\tt IIG(1)}}+k_{{\tt IIG(2)}}+k_{{\tt IIG(3)}}+k_{{\tt IIG(4)}})^2$
& $(\sqrt{s_1}+\sqrt{s_2})^2 $ & $\le s_3 $ & $\le s$ \\  
\hline
${\tt MAPIG(6)}=3$ & $s_1=(k_{{\tt IIG(1)}}+k_{{\tt IIG(2)}})^2$ & 
$0 $ & $\le s_1 $ & $\le s$ \\
& $s_2=(k_{{\tt IIG(1)}}+k_{{\tt IIG(2)}}+k_{{\tt IIG(3)}})^2$ & 
$s_1 $ & $\le s_2 $ & $\le s$ \\  
& $s_3=(k_{{\tt IIG(1)}}+k_{{\tt IIG(2)}}+k_{{\tt IIG(3)}}+k_{{\tt IIG(4)}})^2$
& $s_2 $ & $\le s_3 $ & $\le s$ \\
\hline
\end{tabular}
\end{center}
\vspace*{.5em}
\caption{The definition of the invariants $s_i$ in \refeqs{eq:four} and
\refeqf{eq:five} and their phase-space boundaries for the different 
topologies in the slicing branch}
\label{ta:invariants}
\end{table}
The phase-space generators of individual channels are built
generically from these basic phase-space decompositions in {\tt
  kern.f} by calling subroutines {\tt PHSPGEN\_4} or {\tt PHSPGEN\_5}:
\cpcsuptable{\tt MAPI,MAPIG}{%
\multicolumn{3}{l}{
{\tt SUBROUTINE PHSPGEN\_4(IDENS,II,MAPI,VZI,NU,S,X,WPS,ICUT)}} \\
\multicolumn{3}{l}{
{\tt SUBROUTINE PHSPGEN\_5(IDENS,IIG,MAPIG,VZIG,NU,S,X,WPS,ICUT)}}  \\
{\tt IDENS}&:& determines whether the four-momenta $k_i$ (${\tt IDENS}=0$)
or the density $g_k$ is calculated (${\tt IDENS}=1$), \\
{\tt II,IIG} &:& 
{\tt II(1:4)},{\tt IIG(1:5)} specify the permutations of the final-state 
particles,
${\tt II,IIG}=1,\ldots,4$ denote the final-state fermions 
in the order of the input file, ${\tt IIG}=5$ denotes the photon,\\
{\tt MAPI,MAPIG}&:& {\tt MAPI(6)}$=1,2$ 
and {\tt MAPIG(6)}$=1,2,3$ numerate the basic topologies, \\
&& {\tt MAPI(1:2)}, {\tt MAPIG(1:2)}, and {\tt MAPIG(7)} specify the 
$s$-channel propagator mappings of the variables $s_1,s_2,s_3$ of 
\refeqs{eq:four} and \refeqf{eq:five}:\\
&& \cpcsubtable{${\tt MAPI,MAPIG}=0$}{
${\tt MAPI,MAPIG}=0$&:& no mapping, \\
${\tt MAPI,MAPIG}=1$&:& mapping for a massless  
propagator \refeq{eq:maps2} or \refeq{eq:maps3}, \\ 
${\tt MAPI,MAPIG}=2$&:& mapping for a W-boson propagator 
\refeq{eq:maps1},\\ 
${\tt MAPI,MAPIG}=3$&:& mapping for a Z-boson propagator 
\refeq{eq:maps1},\\
}\\
&&{\tt MAPI(3:5)}, {\tt MAPIG(3:5)}, and {\tt MAPIG(8)} specify the 
$t$-channel propagator mappings in the functions $R_2$ of
\refeqs{eq:four} and \refeqf{eq:five}:\\
&& \cpcsubtable{${\tt MAPI,MAPIG}=0$}{%
${\tt MAPI,MAPIG}=0$&:& no mapping ($s$-channel $2\to 2$ scattering), \\
${\tt MAPI,MAPIG}=1$&:& no mapping 
(massive particle in $t$-channel propagator), \\
${\tt MAPI,MAPIG}=2$&:& mapping for a massless $t$-channel propagator \\
& & using \refeq{eq:maps2} or \refeq{eq:maps3}, \\ 
}\\
{\tt VZI,VZIG} &:& {\tt VZI(1:2),VZIG(1:3)}$=\pm 1$,
the final-state momenta of the corresponding $2\to 2$ scattering process  
are interchanged if ${\tt VZI,VZIG}=-1$, \\
%(see Tables~\ref{tab:pabcdone}-\ref{tab:pabcdfive}),\\
{\tt NU}&:& parameter $\nu$ of \refeqs{eq:maps2}, \\
{\tt S}&:&CM energy squared,\\
{\tt X}&:& $3 n-4$ random numbers,\\
{\tt WPS}&:& local density $g_k$ of the channel, \\
{\tt ICUT}&:& the event is rejected if ${\tt ICUT}=1$. If separation
cuts are directly taken into account in the phase-space generation
(see e.g.\ Eq.~(A.4) of \citere{Denner:1999gp}), it can happen that
$s_{{\rm min}}>s_{{\rm max}}$ or
$t_{{\rm min}}>t_{{\rm max}}$. Then {\tt ICUT} is set to 1. \\
}

The channels for $4f$ and $4f\gamma$ production are obtained from the
basic topologies by specifying the parameters {\tt II,MAPI,VZI} and
{\tt IIG,MAPIG,VZIG}, respectively.  The parameters {\tt VZI,VZIG}
distinguish between $t$-channel processes $q_1+q_2 \to p_1+p_2$ with
invariants $t=q_1-p_1$ or $t=q_1-p_2$. For instance, the topology of
\reffi{fig:SpecificPS} where the final-state particle with momentum
$k_6$ is emitted from the electron instead of the positron is obtained
by changing the sign of {\tt VZIG(1)}.  The specifications of the
relevant parameters for each channel for a given final state is done
in the subroutine {\tt ADDMAP} in {\tt slicing.f}.  For the example of
the multi-peripheral diagram given in \citere{Denner:1999gp}, the
multi-channel parameters are {\tt MAPIG(1:8)}$=\{1,3,2,0,0,3,0,2\}$,
{\tt IIG(1:5)}$=\{3,5,4,2,1\}$, and {\tt VZIG(1:2)}$=\{1,-1\}$ ({\tt
  VZIG(3)} is not used for ${\tt MAPIG(4)}=0$ topologies).

The propagator mappings specified by the parameters {\tt MAPI} and
{\tt MAPIG} are performed according to
\refeqs{eq:maps1}--\refeqf{eq:maps3} in {\tt kern.f} by the
subroutines {\tt SMAP} and {\tt TMAP}, respectively.  The
corresponding Jacobians ({\tt WPSI}) are calculated in the subroutines
{\tt SDENS} and {\tt TDENS}.  The Lorentz transformations used in the
calculation of the outgoing momenta consist of a combination of boosts
along the $z$ direction and rotations described by the subroutines
{\tt BOOST\_Z} and {\tt ROTATION}
%
%\cpcsuptable{\tt ECM,PCM}{
%\multicolumn{3}{l}{{\tt 
%SUBROUTINE BOOST\_Z(ECM,PCM,P4,P3)}} \\
%{\tt ECM,PCM} &:& Lorentz $\gamma$ factor and  $v\gamma$ with $v$ being the
%relative four-velocity of the two frames, \\
%{\tt P4,P3} &:& energy and $z$ component,  \\
%}
%
%and 
%
%\cpcsuptable{\tt P1,P2,P3}{
%\multicolumn{3}{l}{{\tt
%SUBROUTINE ROTATION\_2(VZI,ST,CT,SP,CP,P1,P2,P3)}} \\
%{\tt ST,CT}&:& sine and cosine of $\theta_i$, \\
%{\tt ST,CT}&:& sine and cosine of $\phi_i$, \\
%{\tt P1,P2,P3} &:& $x,y,z$ components \\ 
%}
%
in {\tt kern.f}.

\subsubsection{Phase-space generation in the subtraction branch}
\label{subsubsec:subtraction} 

In the subtraction branch, the generation of the momenta of the 
final-state particles and the calculation of the local densities 
are done in 

\cpcsuptable{\tt RANDOM}{%
\multicolumn{3}{l}{
{\tt SUBROUTINE PHASESPACE(RANDOM,ROOTS,P,S,G,I,NG,SWITCH)}} \\
{\tt RANDOM}&:& %$3 n-4$ 
random numbers,\\
{\tt ROOTS} &:& total scattering energy (after possible ISR), \\ 
{\tt P(I,J)}&:& four-momenta of external particles (after possible ISR), \\ 
&& 
\begin{tabular}[l]{@{}l@{ }c@{ }l}
${\tt I}=1,\ldots,7$ &:& numerates the external particles 
($7=\mathrm{photon}$), \\
${\tt J}=0,\ldots,3$ &:& four-vector components,\\
\end{tabular} \\
{\tt S(I,K)}&:& invariants of pairs of the external particles 
(${\tt S(I,K)}=s_{ik}=2p_ip_k>0$), \\
&& 
\begin{tabular}[l]{@{}l@{ }c@{ }l}
${\tt I,K}=1,\ldots,7$ &:& numerate the external particles 
($7=\mathrm{photon}$), \\
\end{tabular} \\
{\tt G}&:& local density $g_i$ of the channel $I$, \\
{\tt I}&:& number of the channel, \\
{\tt NG}&:& number of the phase-space generator,\\
{\tt SWITCH}&:& decides between event generation, 
calculation of the local density, and initialization of the generator,\\
&&\cpcsubtable{${\tt SWITCH}=0$}{%
${\tt SWITCH}=0$&:& an error occurred (event will be rejected),\\
${\tt SWITCH}=1$&:& generation of momenta and invariants,\\
${\tt SWITCH}=2$&:& calculation of the local density of the given channel,\\
${\tt SWITCH}=3$&:& checks whether the diagram exists (initialization),\\
${\tt SWITCH}=4$&:& resulting flag if the diagram exists\\
}\\
}

in {\tt subtraction.f}. The subroutine {\tt PHASESPACE} is called
several times in the subroutine {\tt SUBTRACTION} where the weights of
the different subcontributions are calculated. Three different
phase-space generators are used:
\cpcsuptable{${\tt NG}=1$}{%
${\tt NG}=1$ &:& four-particle phase-space generator used for 
tree-level process $\eeffff$ (including optionally  higher-order LL 
corrections from ISR) or for the IBA to the processes $\eeWWffff$, \\
${\tt NG}=2$ &:& four-particle phase-space generator for the $4 f$ part  
of $\O(\alpha)$ radiative corrections, \\
${\tt NG}=3$ &:& five-particle phase-space generator used for the 
$4 f \gamma$ part of the $\O(\alpha)$ radiative corrections, 
for the tree-level process $\eeffffg$, or for the IBA to the 
processes $\eeffffg$.  \\
}

The subroutine {\tt PHASESPACE} is decomposed into 15 parts which
correspond to the different topologies contributing to the processes
$\eeffff$ and $\eeffffg$. There are 5 topologies for the processes
$\eeffff$ and 10 topologies for $\eeffffg$, as shown in
\reffi{fig:topologies}.
% or in the subroutine {\tt PHASESPACE} in {\tt subtraction.f}.

The initialization of the channels is done by calling
\cpcsuptable{\tt INPUTONSHELL}{%
\multicolumn{3}{l}{%
{\tt SUBROUTINE INITPHASESPACE(NUPS,Q,NG,NPARTICLE,INPUTONSHELL,SRC,}}\\
\multicolumn{3}{l}{\qquad {\tt SIGNAL,NOCUTS)}} \\
{\tt NUPS}&:& parameter $\nu$ used in the propagator mapping of
\refeq{eq:maps2},\\
{\tt Q(I)}&:& charges of the external fermions ($I=1,\ldots,6$),\\ 
{\tt NG}&:& number of the phase-space generator, \\ 
{\tt NPARTICLE}&:& number of final-state particles (${\tt NPARTICLE}=4,5$), \\
{\tt INPUTONSHELL}&:& 
if ${\tt INPUTONSHELL}=1$ the phase space is generated for on-shell W~bosons,
\\
{\tt SRC}&:& include special channels for the subtraction functions of the 
$4 f \gamma$ part of the radiative corrections for ${\tt SRC}=1$ 
and ${\tt NPARTICLE}=5$, \\
{\tt SIGNAL}&:& restricts to subsets of diagrams as given in 
\refta{tab:subsets},\\
{\tt NOCUTS}&:& decides whether cuts are included in the event generation 
(${\tt NOCUTS}=0$) or not (${\tt NOCUTS}=1$)\\
}
in {\tt subtraction.f}. The subroutine {\tt INITPHASESPACE} for the
three phase-space generators is called in {\tt INITSUBTRACTION} in
{\tt subtraction.f}, where the options for the phase-space generators
and for different branches can be found.

To initialize the phase-space generators, {\tt SEARCHGENERATOR} (which
is just a second entry of {\tt PHASESPACE}) is called in {\tt
  INITPHASESPACE} with ${\tt SWITCH}=3$. If the diagram of the tested
channel exists, {\tt SEARCHGENERATOR} yields ${\tt SWITCH}=4$, and the
channel is used in the phase-space generation.  The actual selection
of the channels is done by inserting the external particles ${\tt
  EP1},\ldots,{\tt EP7}$ (and the emitter {\tt EM} and spectator {\tt
  SP} for special channels related to subtraction functions) and the
virtual particles ${\tt IP1},\ldots,{\tt IP4}$ in all possible ways,
and testing whether the vertices of the topological diagram exist or
not. This is done in the functions {\tt V3} and {\tt V4} for the
three- and four-particle vertices, respectively. The vertices of the
SM are defined in the functions {\tt V3G} and {\tt V4G}. Some
additional criteria are added to select subsets of diagrams ({\tt
  CHECKSIGNAL}), and to avoid double-counting of equivalent channels,
\ie channels which have the same propagator mappings.  To speed up the
initialization, the virtual particles are inserted one after another,
and as soon as a vertex does not exist the diagram is discarded.  All
informations of the phase-space generators are stored in the common
block {\tt CHANNELS} in {\tt subtraction.f}.

For ${\tt SWITCH}=1$ the event is generated, \ie the momenta of the
final-state particles are calculated from the random numbers {\tt
  RANDOM}, the incoming momenta {\tt P(1:2,0:3)}, and the scattering
energy {\tt ROOTS}.  For ${\tt SWITCH}=2$ the local density {\tt G} of
the channel {\tt I} is calculated from momenta {\tt P}, invariants
{\tt S}, and from {\tt ROOTS}.  The invariants {\tt S} are calculated
in the subroutine {\tt SDET} once per event for all local densities.
The actual calculations are done by calling the subroutines {\tt
  SPROP}, {\tt DECAY}, and {\tt PROCESS} in {\tt subtraction.f}:
\cpcsuptable{\tt MASS,WIDTH}{%
\multicolumn{3}{l}{
{\tt SUBROUTINE SPROP(RANDOM,S,G,MASS,WIDTH,NU,SMIN,SMAX,ONSHELL,SWITCH)}}\\
\multicolumn{3}{l}{
{\tt SUBROUTINE DECAY(RANDOM,Q,P1,P2,G,S0,S1,S2,SWITCH)}} \\
\multicolumn{3}{l}{{\tt SUBROUTINE 
PROCESS(RANDOM,Q1,Q2,P1,P2,QT,G,MASS,WIDTH,NU,S0,S1,S2,T,T1,}}\\
\multicolumn{3}{l}{\qquad{\tt T2,CCUT,SWITCH)}} \\
{\tt RANDOM}&:& random numbers,\\
{\tt S}&:& invariant generated by {\tt SPROP}, \\
{\tt G}&:& local density of the channel, \\
{\tt MASS,WIDTH}&:& mass and width of the Breit--Wigner propagator 
mapping of \refeq{eq:maps1}, \\
{\tt NU}&:& parameter $\nu$ of the propagator mapping of \refeq{eq:maps2},\\
{\tt SMIN,SMAX}&:& minimal and maximal value allowed for the invariant {\tt S}
in {\tt SPROP}, \\
{\tt ONSHELL}&:& on-shell approximation for W~bosons used in {\tt SPROP} 
if ${\tt ONSHELL}=1$, \\
{\tt CCUT}&:& cut on cosine of scattering angle applied only for $2\to 2$
scattering processes ({\tt PROCESS}), \\
{\tt SWITCH}&:& see subroutine {\tt PHASESPACE}, \\
\multicolumn{3}{l}{the remaining arguments are given 
in \reffi{fig:processdecay}.} \\
}%
\bfi
\centerline{
%\framebox{
\setlength{\unitlength}{1pt}
\begin{picture}(200,180)(0,0)
\ArrowLine( 50, 20)( 50, 60)
\ArrowLine( 50, 60)(150, 60)
\ArrowLine(150, 20)(150, 60)
\ArrowLine( 50, 60)( 50,100)
\ArrowLine(150, 60)(150,100)
\Vertex( 50, 60){3}
\Vertex(150, 60){3}
\put(  0,145){{\tt SUBROUTINE PROCESS:}}
\put( 67, 66){${\tt T}=(q_1-p_1)^2$}
\put( 67, 49){${\tt QT}=q_1-p_1$}
\put(  5, 35){${\tt Q1}\equiv q_1$}
\put(157, 35){${\tt Q2}\equiv q_2$}
\put(  5, 75){${\tt P1}\equiv p_1$}
\put(157, 75){${\tt P2}\equiv p_2$}
\put( 30,  5){${\tt T1}=q_1^2$}
\put(130,  5){${\tt T2}=q_2^2$}
\put( 30,110){${\tt S1}=p_1^2$}
\put(130,110){${\tt S2}=p_2^2$}
\put( 70,125){${\tt S0}=(p_2+p_2)^2$}
\end{picture}
%}
%\framebox{
\setlength{\unitlength}{1pt}
\begin{picture}(160,160)(0,0)
\ArrowLine( 80, 20)( 80, 60)
\ArrowLine( 80, 60)(120,100)
\ArrowLine( 80, 60)( 40,100)
\Vertex( 80, 60){3}
\put(  0,145){{\tt SUBROUTINE DECAY:}}
\put( 45, 35){${\tt Q}\equiv q$}
\put( 10, 75){${\tt P1}\equiv p_1$}
\put(110, 75){${\tt P2}\equiv p_2$}
\put( 15,110){${\tt S1}=p_1^2$}
\put(105,110){${\tt S2}=p_2^2$}
\put( 65,  5){${\tt S0}=q^2$}
\end{picture}
%}
}
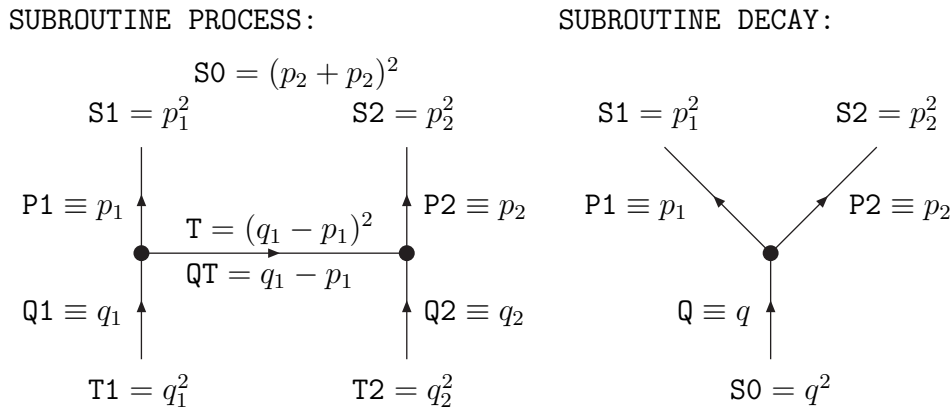
\caption{The generic diagrams of the $2\to 2$ scattering process and 
$1\to 2$ decay and the arguments of the corresponding subroutines 
{\tt PROCESS} and {\tt DECAY} in {\tt subtraction.f}}
\label{fig:processdecay}
\efi 

The propagator mappings of the massless propagators of \refeqs{eq:maps1} and 
\refeqf{eq:maps2} and the calculation of the corresponding Jacobians are 
done by calling the functions {\tt H} and {\tt JACOBIAN}, respectively, 
in the subroutines {\tt SPROP} and {\tt PROCESS} in {\tt subtraction.f}.
For the boosts and rotations of momenta the subroutines {\tt BOOST} and 
{\tt ROTATION} in {\tt subtraction.f} are used.

For the subtraction functions corresponding to the bremsstrahlung
process $\eeffffg$, special channels are included in the calculation
to improve the numerical stability. Since these involve the Born cross
section to $\eeffff$, a mapping of the $4 f \gamma$ phase space into
the $4 f$ phase space is needed.  The mapping from the five- to the
four-particle phase space and back is performed by calling the
subroutines {\tt MAP54} and {\tt MAP45} in {\tt PHASESPACE},
respectively.  The propagators of the Born process $\eeffff$ are then
smoothened by the four-particle phase-space generator. Additional
mappings implemented in {\tt MAP54} and {\tt MAP45} account for the
smoothening of the remaining peaking behaviour of the subtraction
functions, \ie of the splitting functions.

\subsection{The matrix-element evaluation}
\label{subsec:matrixelement} 

In this section, we describe the matrix-element evaluation in the
three different modes of {\RacoonWW}. Furthermore, we briefly sketch
the treatment of higher-order ISR and QCD corrections, and the
implementation of anomalous gauge-boson couplings.

\subsubsection{The tree-level mode (${\tt SRC}=0$)} 
\label{subsubsec:treelevel}

In the tree-level mode (${\tt SRC}=0$) {\RacoonWW} calculates either
the processes $\eeffff$ (${\tt SBORN4}=1,2,3$) or the radiative
processes $\eeffffg$ (${\tt SBORN5}=1$) at lowest order for all
possible four-fermion final states. The flags {\tt SBORN4} and {\tt
  SBORN5} are set in the input file.  The Coulomb singularity to the
tree-level process $\eeffff$ can be included by setting ${\tt
  SCOULTREE}=1$, which corresponds to the variant 1 of
\refse{subsubsec:iba}.  The treatment of the Coulomb singularity in
$\eeffffg$ processes is described in \refse{subsubsec:iba}.

The different process classes are characterized according to the
appearing intermediate vector bosons, \ie whether the reactions
proceed via charged-current (CC), or neutral-current (NC)
interactions, or via both interaction types (see
\refta{tab:classification}).  The classification can be performed by
considering the quantum numbers of the final-state fermion pairs.  As
shown in \refta{tab:classification}, we identified 11 classes of
processes. This classification does not change when a photon is
radiated.
\begin{table}
\bce
\tabcolsep 7pt
\begin{tabular}{|l|ccccccc|} \hline
\multicolumn{8}{|c|}{CC processes: {\tt CC} = 1, {\tt NC} = 0}\\ \hline
& $\bar{f}_1$ & $f_2$ & $\to $ & $f_3$ & $\bar{f}_4$ & $f_5$ & $\bar{f}_6$ \\ \hline
${\tt SYM}=0$ & $\Pep$ & $\Pem$ & $\to$ & $f$     & $\bar f'$   & $ F$     & $ \bar F'$ \\ 
${\tt SYM}=1$ & $\Pep$ & $\Pem$ & $\to$ & $\nu_e$ & $\Pep$      & $f$      & $\bar f'$ \\
${\tt SYM}=2$ & $\Pep$ & $\Pem$ & $\to$ & $f$     & $\bar f'$   & $\Pem$   & $\bar \nu_e$ \\ \hline
\multicolumn{8}{|c|}{NC processes: {\tt CC} = 0, {\tt NC} = 1}\\ \hline
& $\bar f_1$ & $f_2$ & $\to $ & $f_3$ & $\bar f_4$ & $f_5$ & $\bar f_6$ \\ \hline
${\tt SYM}=0$ & $\Pep$ & $\Pem$ & $\to$ & $f$     & $\bar f$    & $F$      & $\bar F$  \\
${\tt SYM}=1$ & $\Pep$ & $\Pem$ & $\to$ & $f$     & $\bar f$    & $f$      & $\bar f$  \\
${\tt SYM}=2$ & $\Pep$ & $\Pem$ & $\to$ & $\Pem$  & $\Pep$      & $f$      & $\bar f$  \\
${\tt SYM}=3$ & $\Pep$ & $\Pem$ & $\to$ & $\Pem$  & $\Pep$      & $\Pem$   & $\Pep$ \\ \hline
\multicolumn{8}{|c|}{Mixed (CC \& NC) processes: {\tt CC} = 1, {\tt NC} = 1}\\ \hline
& $\bar f_1$ & $f_2$ & $\to $ & $f_3$ & $\bar f_4$ & $f_5$ & $\bar f_6$ \\ \hline
${\tt SYM}=0$ & $\Pep$ & $\Pem$ & $\to$ & $f$     & $\bar f$    & $f'$     & $\bar f'$ \\
${\tt SYM}=1$ & $\Pep$ & $\Pem$ & $\to$ & $\nu_e$ & $\bar\nu_e$ & $f$      & $\bar f$  \\
${\tt SYM}=2$ & $\Pep$ & $\Pem$ & $\to$ & $\nu_e$ & $\bar\nu_e$ & $\nu_e$ & $\bar\nu_e$ \\
${\tt SYM}=3$ & $\Pep$ & $\Pem$ & $\to$ & $\nu_e$ & $\bar\nu_e$ & $\Pem$   & $\Pep$ \\ \hline
\end{tabular}
\ece
\caption{The classification of the four-fermion processes $\eeffff$.
Here, $f$ and $F$ denote different fermions ($f\ne F$) that are neither 
electrons nor electron neutrinos ($f,F \ne \Pem,\nu_\Pe$), and their 
weak-isospin partners are denoted by $f'$ and $F'$, respectively. 
The parameters {\tt SYM}, {\tt NC}, and {\tt CC} are determined in 
the subroutine {\tt INIT\_EPEM} in {\tt ee4fa\_amps.f}.}
\label{tab:classification}
\end{table}
In the following, we often use the shorthands CC11 and CC03. CC11
denotes the set of up to 11 Feynman diagrams contributing to the
simplest CC process in \refta{tab:classification} (${\tt SYM}=0$),
which is the smallest gauge-invariant subset of diagrams including
W-pair production. CC03 denotes the 3 diagrams that include 2 resonant
W~bosons, which is a non-gauge-invariant subset of CC11 (see also
\citere{Grunewald:2000ju}).

The calculation of the matrix elements to the tree-level processes
$\eeffff$ and $\eeffffg$ has been described in detail in
\citere{Denner:1999gp}.  For $\eeffff$ the matrix elements of all
Feynman diagrams can been reduced to two generic functions that are
related to the two basic graphs shown in \reffi{fig:ee4fdiags}. For
$\eeffff$, the calculation is similar to the one in \citere{Be94}.
For $\eeffffg$, all graphs that are obtained from one of the generic
graphs in \reffi{fig:ee4fdiags} by attaching one additional real
photon in all possible ways are combined into one generic function.
Finally, also all amplitudes to $\eeffffg$ can be constructed from
only two generic functions.
\bfi
\begin{center}
\setlength{\unitlength}{1pt}
\begin{picture}(420,150)(0,-20)
\Text(0,120)[lb]{a) Abelian graph}
\Text(210,120)[lb]{b) Non-abelian graph}
\put(20,-8){
\begin{picture}(150,100)(0,0)
\ArrowLine(35,70)( 5, 95)
\ArrowLine( 5, 5)(35, 30)
\ArrowLine(35,30)(35,70)
\Photon(35,30)(90,20){2}{6}
\Photon(35,70)(90,80){-2}{6}
\Vertex(35,70){2.0}
\Vertex(35,30){2.0}
\Vertex(90,80){2.0}
\Vertex(90,20){2.0}
\ArrowLine(90,80)(120, 95)
\ArrowLine(120,65)(90,80)
\ArrowLine(120, 5)( 90,20)
\ArrowLine( 90,20)(120,35)
\put(55,82){$V_1$}
\put(55,10){$V_2$}
\put( 20,50){$f_g$}
\put(-15,110){$\bar f_a(p_a,\si_a)$}
\put(-15,-10){$f_b(p_b,\si_b)$}
\put(125,90){$\bar f_c(p_c,\si_c)$}
\put(125,65){$f_d(p_d,\si_d)$}
\put(125,30){$\bar f_e(p_e,\si_e)$}
\put(125, 5){$f_f(p_f,\si_f)$}
\end{picture}
}
\put(245,-8){
\begin{picture}(150,100)(0,0)
\ArrowLine( 15,50)(-15, 65)
\ArrowLine(-15,35)( 15, 50)
\Photon(15,50)(60,50){2}{5}
\Photon(60,50)(90,20){-2}{5}
\Photon(60,50)(90,80){2}{5}
\Vertex(15,50){2.0}
\Vertex(60,50){2.0}
\Vertex(90,80){2.0}
\Vertex(90,20){2.0}
\ArrowLine(90,80)(120, 95)
\ArrowLine(120,65)(90,80)
\ArrowLine(120, 5)( 90,20)
\ArrowLine( 90,20)(120,35)
\put(30,58){$V$}
\put(62,70){$\PW$}
\put(62,18){$\PW$}
\put(-35,75){$\bar f_a(p_a,\si_a)$}
\put(-35,20){$f_b(p_b,\si_b)$}
\put(125,90){$\bar f_c(p_c,\si_c)$}
\put(125,65){$f_d(p_d,\si_d)$}
\put(125,30){$\bar f_e(p_e,\si_e)$}
\put(125, 5){$f_f(p_f,\si_f)$}
\end{picture}
}
\end{picture}
\end{center}
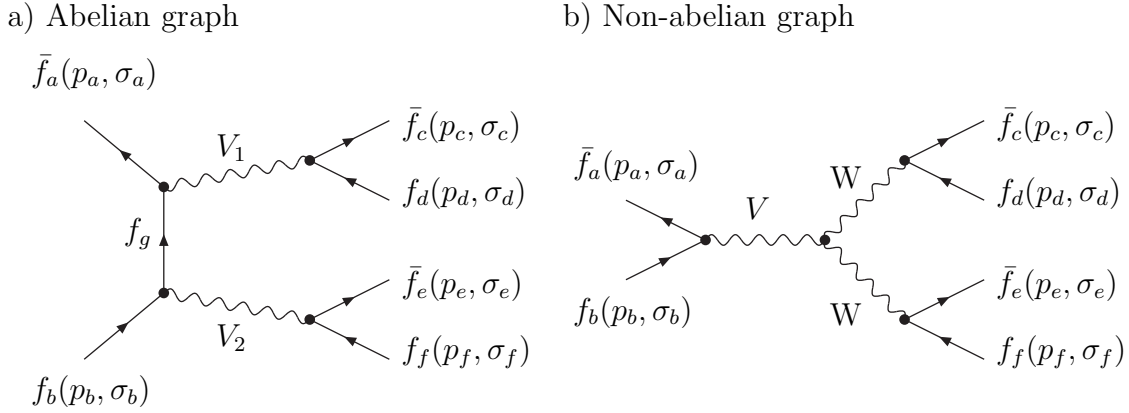
\caption{Generic diagrams for $\Pep\Pem\to 4f$}
\label{fig:ee4fdiags}
\efi
For the helicity amplitudes corresponding to the generic
contributions, concise results have been obtained by using the
Weyl--van~der~Waerden (WvdW) formalism (see \citere{wvdw} and
references therein).

The helicity amplitudes to $\eeffff$ are calculated in the function
{\tt M2\_EPEM\_4F} by calling either the subroutine {\tt M\_CC} or 
{\tt M\_NC} or both, depending on the process under consideration:

\cpcsuptable{{\tt M(-1:1,}\ldots{\tt ,-1:1)}}{%
\multicolumn{3}{l}{{\tt SUBROUTINE M\_CC(M,SGN,IF1,IF2,IF3,IF4,IF5,IF6)}} \\
\multicolumn{3}{l}{{\tt SUBROUTINE M\_NC(M,SGN,IF1,IF2,IF3,IF4,IF5,IF6)}} \\
{\tt M(-1:1,}\ldots{\tt ,-1:1)} &:& helicity amplitudes, \\
{\tt SGN} &:& relative sign of different diagrams, \\
{\tt IF1},\ldots,{\tt IF6} &:& specify the final-state fermions. \\
}

The helicity amplitudes are expressed in terms of WvdW spinor
products, which are calculated in the subroutine {\tt SETPRODS}. Note
that {\tt SETPRODS} must to be called before the matrix elements are
evaluated by the subroutines {\tt M2\_EPEM\_4F} or {\tt
  M2\_EPEM\_4FA}.
%, where {\tt DIR(7)} denotes the direction of the final-state particles 
%(${\tt dir(i)}=+1(-1)$ fir incoming(outgoing) particles). 
The helicity amplitudes for a four-fermion final state, {\tt M},
are built up by using two generic functions related to the graphs 
of \reffi{fig:ee4fdiags} as calculated in the subroutines {\tt ADDMVV} for 
the abelian diagram and {\tt ADDMVVV} for the non-abelian diagram of 
\reffi{fig:ee4fdiags}, respectively, with the appropriate choices for 
the parameters {\tt SGN}, {\tt IF1},\ldots,{\tt IF6}, and
by specifying the intermediate gauge bosons $V$, $V_1$, $V_2$ of
\reffi{fig:ee4fdiags} by the variables {\tt IV}, {\tt IV1}, {\tt IV2}, which
are arguments of  {\tt ADDMVV} and {\tt ADDMVVV}.

The generic functions for $\eeffffg$ can be constructed in a similar
way. The helicity amplitudes to $\eeffffg$ are built up in the
subroutines {\tt M\_CCA} and {\tt M\_NCA}.  Again the contributions of
all diagrams can be taken into account by calling the corresponding
subroutines {\tt ADDMAVV} and {\tt ADDMAVVV} and making the
appropriate choices for the parameters {\tt SGN}, {\tt
  IF1},\ldots,{\tt IF6}, {\tt IV}, {\tt IV1}, {\tt IV2}.

For the NC (${\tt SYM}=0,1$) and CC/NC (${\tt SYM}=0$) type of
processes with pure hadronic final states, one of the gauge bosons in
the diagram of \reffi{fig:ee4fdiags} can be a gluon. In {\RacoonWW}
these contributions are taken into account when choosing ${\tt
  SQCDEPEM}=1$ in the input file for both processes $\eeffff$ and
$\eeffffg$.  Then, the complete matrix elements for the fully hadronic
channels result from the sum of the purely electroweak and the
gluon-exchange contributions. The contribution of the gluon-exchange
diagrams are calculated in the subroutines {\tt MG\_NC} and {\tt
  MG\_NCA} for the process classes $\eeffff$ and $\eeffffg$,
respectively.

\subsubsection{The improved-Born approximation (IBA) mode (${\tt SRC}=2,3$)}
\label{subsubsec:iba}

{\RacoonWW} provides an IBA for the process $\eeWWffff$ (${\tt
  SRC}=2$) including only diagrams of the CC03 class, and to the
radiative processes $\eeffffg$ (${\tt SRC}=3$), as described in
\citeres{Denner:2001zp} and \cite{Denner:2001vr}, respectively.  For
both process classes universal leading electroweak corrections are
taken into account.  Universal radiative corrections are corrections
that are either connected to specific important subprocesses, such as
collinear photon emission, or that originate from large corrections
related to renormalization effects, like running couplings, and
involve characteristic enhancement factors. Owing to their
universality such corrections can be related in a simple way to the
lowest-order matrix element of the underlying process.

Near the W-pair-production threshold, a particularly important
correction is induced by the Coulomb singularity. This arises from
diagrams where a soft photon is exchanged between two nearly on-shell
W~bosons close to their kinematical production threshold and results
in a simple factor that depends on the momenta $k_\pm$ of the
\PWpm~bosons \cite{Denner:1998ia,coul},
\beqar 
\rd \sigma_{\Coul}&=& 
\left[1+\delta_{\Coul}(s,k_+^2,k_-^2)\,g(\bar\beta)\right] 
\rd \sigma_{\Born}, \nonumber \\[0.2em]
\delta_{\Coul}(s,k_+^2,k_-^2) &=& 
\frac{\alpha(0)}{\bar\beta} 
\Im\left[\ln\left(\frac{\beta-\bar\beta+\Delta_M} 
{\beta+\bar\beta+\Delta_M}\right)\right], \nonumber \\[0.1em]
\bar\beta &=& 
\frac{\sqrt{s^2+k_+^4+k_-^4-2sk_+^2-2sk_-^2-2k_+^2k_-^2}}{s},\nl 
\beta &=& \sqrt{1-\frac{4(\MW^2-\ri\MW\GW)}{s}}, \qquad \Delta_M = 
\frac{|k_+^2-k_-^2|}{s},
\eeqar  
with the fine-structure constant $\alpha(0)$, and $s=(k_++k_-)^2$.
The auxiliary function
\beq  
g(\bar\beta) = \left(1-\bar\beta^2\right)^2  
\eeq  
is introduced to restrict the impact of $\delta_{\Coul}$ to the
threshold region. 

A further ingredient in the IBA is the LL contribution induced by ISR.
This is included by using the structure-function approach \cite{sf},
where the full IBA matrix elements squared to $\eeWWffff$ and
$\eeffffg$ are multiplied by the ISR structure functions. Their scale
dependence can be used to adjust the IBA to the full correction, but
also to estimate the intrinsic uncertainty of the IBA by choosing
different values for $Q$.

When the IBA to $\eeWWffff$ is calculated, the (off-shell)
improved Born matrix element squared described by 
\cpcsuptable{\tt P(1:7,0:3)}{%
\multicolumn{3}{l}{{\tt FUNCTION M2\_BORN\_OFFSH(P,QCC03,QGZ)} }\\
{\tt P(1:7,0:3)}&:& external momenta (after possible ISR),\\
{\tt QCC03}&:& {chooses between the lowest-order CC03 
cross section ({\tt QCC03}=0) and the IBA for the CC03 cross section 
({\tt QCC03}=1),}\\ 
{\tt QGZ}&:& decides whether the Z-boson width, {\tt GZ}, is used
in the Z~propagator (${\tt QGZ}=1$) or not (${\tt QGZ}=0$) \\
}%
in {\tt eeWW4f\_amps.f} is convoluted with the ISR structure
functions.  The Born matrix element includes all contributions from
the CC03 class of diagrams (${\tt QCC03}=1$) which have been modified
in such a way that the universal renormalization effects induced by
the running of $\alpha$ and by $\Delta\rho$ are absorbed. This is
achieved \cite{bo92} by the replacements
\beq
\frac{e^2}{\sw^2} \;\to\; 4\sqrt{2}\GF\MW^2, \qquad e^2 \;\to\;
4\pi\alpha(s) 
\eeq 
in the amplitudes, which implies that weak-isospin exchange involves
the coupling $\GF\MW^2$ and pure photon exchange the coupling
$\alpha(s)$.  The running of the electromagnetic coupling is induced
by light (massless) charged fermions only, \ie we evaluate $\alpha(s)$
by
\beq 
\disp \alpha(s) = \frac{\alpha(\MZ^2)}
{1-\frac{\alpha(\MZ^2)}{3\pi}\ln(\frac{s}{\MZ^2})
\sum_{f\ne\Pt}N^{\mathrm{c}}_f Q_f^2},
\eeq 
where $\alpha(\MZ^2)={\tt ALPHAZ}$ is set in the subroutine {\tt
  PARAMETER} in {\tt public.f}.  Furthermore, the Coulomb singularity
is taken into account, and naive QCD corrections can be optionally
included.  The W-boson width is treated in the constant-width scheme.
The complex Z-boson mass is used (${\tt QGZ}=1$) in the
\PZ~propagators in order to regularize the Z~resonance below the
W-pair production threshold, which may be reached after ISR.

The IBA for $\eeffffg$ processes is based on the full matrix elements
of the tree-level mode in the \GF~scheme. Moreover, leading ISR
corrections are included via structure functions, and the naive QCD
correction can be optionally included. Again, the constant-width
scheme is used.  As the tree-level mode, the IBA is not restricted to
a certain class of processes, and the subset of diagrams included is
governed by the choice of the flags {\tt SSIGEPEM5} and {\tt SQCDEPEM}
in the input file.  For $\eeffffg$ the naive implementation of the
Coulomb singularity involves some freedom, since the momenta of the
W~bosons depend on the photon momentum in different ways for photon
radiation off initial-state and final-state particles. To control this
ambiguity the Coulomb factor can be evaluated in two different ways:
\begin{enumerate}
\item In the first variant (${\tt SCOULTREE}=1$), the complete matrix
  element squared is multiplied with the Coulomb correction factor
  using $k_+=k_3+k_4$ and $k_-=k_5+k_6$.  In this way the correct
  Coulomb correction is multiplied to all diagrams where the photon is
  radiated off initial-state particles.
\item In the second variant (${\tt SCOULTREE}=2$) this prescription is
  improved by choosing the correct Coulomb factor to all contributions
  involving the most resonant $\PWp$ and $\PWm$~bosons as determined
  from invariant masses in the final state.  To this end, the W-boson
  momenta entering the Coulomb correction factor are fixed as
\beqar 
(k_+,k_-) &=& \left\{\barr{lll} (k_3+k_4,k_5+k_6)
&\ \mathrm{for}\ & \De_{34}<\De_{347}, \De_{56}<\De_{567},\\
(k_3+k_4+k_7,k_5+k_6) &\ \mathrm{for}\ & \De_{34}>\De_{347},
\De_{56}<\De_{567}\quad \mathrm{or}\ \\ &&
\De_{34}>\De_{347},\De_{56}>\De_{567}, \De_{347}<\De_{567},\\
(k_3+k_4,k_5+k_6+k_7) &\ \mathrm{for}\ & \De_{34}<\De_{347},
\De_{56}>\De_{567}\quad \mathrm{or}\ \\ &&
\De_{34}>\De_{347},\De_{56}>\De_{567} , \De_{347}>\De_{567}, \earr
\right.\nln 
\eeqar 
where $\De_{ij} = |(k_i+k_j)^2-\MW^2|$ and
$\De_{ijl} = |(k_i+k_j+k_l)^2-\MW^2|$. Thus, the correct Coulomb
correction factor is applied to all dominating doubly-resonant
contributions.
\end{enumerate}

Finally, for the IBA the W-boson width $\GW$ is calculated in lowest
order using the \GF~scheme. This choice guarantees that the
``effective branching ratios'', which result after integrating out the
decay parts, add up to one when summing over all channels. Of course,
if naive QCD corrections are taken into account by multiplying with
$(1+\alpha_{\mathrm{s}}/\pi)$ for each hadronically decaying W~boson,
for consistency these QCD factors are also included in the calculation
of the total W-boson width.

\subsubsection{The radiative correction (RC) mode (${\tt SRC}=1$)}
\label{subsubsec:radcor}

In the RC mode (${\tt SRC}=1$) the complete
electroweak corrections to $\eeWWffff$ in DPA are calculated as
described in detail in \citere{Denner:2000bj}.  
Including $\Oa$ corrections, the cross section to $\eeffff(\gamma)$
is composed of the following contributions
\begin{equation}
\label{eq:crosssection0}
\rd \sigma =\rd \sigma_{\Born}^{\eeffff}+\rd \sigma_{\virt}^{\eeffff}
+ \rd\sigma^{\eeffffg}.
\end{equation}
Here $\rd\sigma_{\Born}^{\eeffff}$ and $\rd\sigma^{\eeffffg}$ are 
the lowest-order 
cross sections to $\eeffff$ and to $\eeffffg$, respectively,
and $\rd \sigma_{\virt}^{\eeffff}$ denotes the virtual one-loop
corrections.

The $\eeffffg$ process (${\tt SBORN5}=1$) represents the
bremsstrahlung contribution, where the soft and collinear
singularities are either subtracted from the matrix element squared
(subtraction branch) or avoided by imposing technical cuts on the
photon phase space (slicing branch). In the RC mode the calculation of
the $\eeffffg$ process is always restricted to the CC11 class of
diagrams (${\tt SSIGEPEM5}=5$).  Simultaneously, the tree-level
process $\eeffff$ (${\tt SBORN4}=1$) can be calculated including all
diagrams or only a subset of diagrams depending on the choice for {\tt
  SSIGEPEM4} in the input file.

Both the virtual and the real corrections involve soft and collinear
singularities. These singularities are extracted by separating the
cross sections into finite and singular parts:
\beqar\label{eq:finsing}
\rd\sigma_{\virt}^{\eeffff} &=&
\rd\sigma_{\virt,\finite}^{\eeffff} + 
\rd\sigma_{\virt,\sing}^{\eeffff},\nl
\rd\sigma^{\eeffffg} &=&
\rd\sigma_{\finite}^{\eeffffg} + 
\rd\sigma_{\sing}^{\eeffffg}.
\eeqar
In {\RacoonWW}, the DPA is only applied to the finite part of 
the virtual corrections,
\begin{eqnarray}
\label{eq:virtfinite}
\rd\sigma_{\virt,\finite}^{\eeffff} &\ \to\ & 
\rd\sigma_{\virt,\finite,\DPA}^{\eeWWffff}=
\rd\sigma_{\virt,\DPA}^{\eeWWffff}
-\rd\sigma_{\virt,\sing,\DPA}^{\eeWWffff}.
\end{eqnarray}
The doubly-resonant virtual corrections are composed as 
\begin{eqnarray}
\rd\sigma_{\virt,\DPA}^{\eeWWffff}&=&
\frac{1}{2 s} \rd\Phi_{4f} \biggl\{
2\Re \Bigl[\Bigl(\M^{\eeWWffff}_{\born,\DPA}\Bigr)^*
\de\M^{\eeWWffff}_{\virt,\facto,\DPA}\Bigr]
\nonumber\\
&&{}+ 
\Bigl|\M^{\eeWWffff}_{\born,\DPA}\Bigr|^2 \de^{\virt}_{\nonfact,\DPA} \biggr\},
\end{eqnarray}
where $\M^{\eeWWffff}_{\virt,\facto,\DPA}$ denotes the matrix element
for the factorizable virtual corrections and
$\de^{\virt}_{\nonfact,\DPA}$ is the factor describing the
non-factorizable virtual corrections.  The matrix element squared to
both contributions are calculated in {\tt eeWW4f\_DPA.f} by
\cpcsuptable{\tt PON(1:7,0:3)}{%
\multicolumn{3}{l}{{\tt FUNCTION M2(P,PON)}}\\
{\tt P(1:7,0:3)}&:&  four-momenta of external particles, \\
{\tt PON(1:7,0:3)}&:& four-momenta of external particles for on-shell W~bosons.\\
}

The full four-fermion phase space (with off-shell W~bosons) is
generated, but the matrix elements for the DPA part are evaluated with
kinematics for on-shell W bosons.
This necessitates a mapping of the momenta of the final-state fermions
such that the momenta of the W bosons become on shell.
This is done in {\tt eeWW4f\_DPA.f} by
\cpcsuptable{${\tt N}=1,\ldots,4$}{%
\multicolumn{3}{l}{{\tt SUBROUTINE OFTOON(P,PON,MW2,N)}}\\
{\tt P,PON}&:& see function {\tt M2}, \\
{\tt MW2}&:& W-boson mass squared, \\
${\tt N}=1,\ldots,4$&:& different versions of the on-shell mappings (see \refse{subsec:input}). \\
}
The factorizable contribution to the matrix element squared is
calculated in the function {\tt DM2(F)} in {\tt eeWW4f\_amps.f} in
terms of invariant coefficients {\tt F(1:10,-1:1)}
which multiply so-called standard matrix elements, $\M^\sigma_n$, as%
\footnote{In \citere{Denner:2000bj} the sum over $n$ runs from 1 to 7,
  since the terms for $n=8,9,10$ could be expressed in terms of the
  first 7 terms.  However, in {\RacoonWW} the version with 10
  invariant coefficients is used.}
\beqar
\M^{\eeWWffff,\si}_{\virt,\facto,\DPA}(\Pp_+,\Pp_-,\kon_+,\kon_-,k_+^2,k_-^2)
&=&\sum_{n=1}^{10} F^\sigma_n(s,\ton) 
\M^\sigma_n(\Pp_+,\Pp_-,\kon_+,\kon_-,k_+^2,k_-^2), \nl*
\eeqar
where $\hat{k}_\pm$ are the on-shell momenta of the \PWpm~bosons, and
$\ton=(p_+-\kon_+)^2$.  It is important to realize that the invariant
coefficients $F^\sigma_n(s,\ton)={\tt F}(n,\sigma)$ depend only on the
scalar products of the momenta of the on-shell-projected W~bosons, but
not on the momenta of their decay products.  For the default choice
${\tt QREAL}=1$, the real part of {\tt F} is taken, \ie only the real
part of the factorizable virtual corrections is included.  The form
factors to $\eeWWffff$ are calculated in the subroutine {\tt CALCFORM}
for ${\tt QFAST}=0$ and in {\tt FASTFORM} for ${\tt QFAST}=1$ in {\tt
  eeWW4f\_DPA.f}.  The standard matrix elements are calculated in the
subroutine {\tt CALCSME} in {\tt eeWW4f\_amps.f}.

The actual numerical evaluation of the invariant coefficients
$F^\sigma_n(s,\ton)$ can be rather CPU-time consuming, since the
expressions for the one-loop contributions to these coefficients are
rather involved (see \citeres{rcwprod1,rcwprod2}). Moreover, the
employed Passarino--Veltman reduction \cite{pa79} of tensor integrals
to scalar integrals breaks down at the boundary of the phase space,
necessarily leading to numerical instabilities in the very forward and
backward directions of the W-production angle $\theta$. In order to
solve these problems of CPU time and numerical instabilities,
{\RacoonWW} uses as default (${\tt QFAST}=1$) a numerical
reconstruction of the coefficients by making use of an expansion in
Legendre polynomials $P_l(\cos\theta)$:
\beq
F^\sigma_n(s,\ton) = 
\sum_{l=0}^\infty \, \frac{1}{\ton} \, c^\sigma_{n,l}(s) P_l(\cos\theta),
\label{eq:Fsuml}
\eeq
which involves only simple algebra. Thus, its evaluation is extremely
fast.  In practice, relatively few generalized Fourier coefficients
$c^\sigma_{n,l}(s)$ are needed; for example, taking $l=0,\ldots,20$
reproduces the full calculation of $F^\sigma_n(s,\ton)$ within roughly
six digits for moderate scattering angles and LEP2 energies.  The
generalized Fourier coefficients are calculated via Gaussian
integration during the initialization in the subroutine {\tt CFOURIER}
in {\tt eeWW4f\_DPA.f}.  Note also that the generalized Fourier series
remains stable in the forward and backward directions where the
original evaluation of $F^\sigma_n(s,\ton)$ breaks down.  The explicit
factor $\ton$ in \refeq{eq:Fsuml} was included in order to account for
a $t$-channel pole in some of the $F^\sigma_n(s,\ton)$; without this
factor the expansion is less efficient and requires much more terms in
the sum over $l$ in \refeq{eq:Fsuml}.  The Legendre polynomials
$P_l(\cos\theta)$ are calculated in {\tt eeWW4f\_DPA.f} by the
subroutine {\tt LEGP}.

As described in \citere{Denner:2000bj} (and references therein), the
virtual non-factorizable corrections can be written in terms of a
correction factor $\delta_{\nonfact,\DPA}^\virt$ to the lowest-order
cross section.  This factor is calculated in {\tt eeWW4f\_DPA.f} by
the function {\tt DNONF} with the possibility of choosing between
different implementations (${\tt QNF}=1,2,3$), as described in
\refse{subsec:input}.

The remaining part of the electroweak radiative corrections, \ie the
soft and collinear parts of virtual and real photon radiation and the
hard bremsstrahlung contribution are calculated exactly, \ie using
off-shell kinematics and based on the full matrix elements to
$\eeffff$ and $\eeffffg$ (${\tt SSIGEPEM5}=5$) of the tree-level mode.
The matching of the virtual photon corrections calculated in DPA with
the exact real photon radiation cannot be done unambiguously as far as
the IR-finite terms are concerned.  The definition of the virtual
singular cross section is a matter of convention, since finite terms
can be redistributed between singular and finite parts.  This freedom,
which is beyond DPA accuracy, is reflected by the different choices
provided by {\RacoonWW} in defining
$\rd\sigma_{\virt,\finite,\DPA}^{\eeWWffff}$ (${\tt QSOFT}=1,2,3$ and,
for the subtraction branch, ${\tt SSUB}=1,2$).

The singular part of the virtual corrections,
$\rd\sigma_{\virt,\sing}^{\eeffff}$, factorizes into the lowest-order
cross section and a simple correction factor.  The correction factor
of the part which is subtracted,
$\rd\sigma_{\virt,\sing,\DPA}^{\eeWWffff}$, is calculated in the
slicing branch by the function {\tt DSOFT1} in {\tt eeWW4f\_DPA.f} for
${\tt QSOFT}=1,2$ or by calling {\tt SOFTFAC} in {\tt kern.f} for
${\tt QSOFT}=3$.  In the subtraction branch, the part which is
subtracted is calculated by the subroutine {\tt SUB4} in {\tt
  subtraction.f} for ${\tt QSOFT}=1,2,3$.  The corresponding
lowest-order cross section to $\eeWWffff$ is calculated in the
subroutine {\tt CALCSME} in {\tt eeWW4f\_amps.f}.

The soft and collinear singularities in the real part,
$\rd\sigma_{\sing}^{\eeffffg}$, can be split off, and the integration
over the singular regions of the photon phase space can be performed
analytically.  The result can be written as a convolution of a
structure function with the lowest-order cross section
$\rd\sigma_{\Born}^{\eeffff}$.  When adding
$\rd\sigma_{\virt,\sing}^{\eeffff}$ and
$\rd\sigma_{\sing}^{\eeffffg}$, all soft singularities and all
collinear singularities associated with the final state cancel, and
the resulting ``singular'' cross section,
\beq
\rd\sigma_{\virt+\real,\sing}^{\eeffff}
=\rd\sigma_{\virt,\sing}^{\eeffff}+\rd\sigma_{\sing}^{\eeffffg},
\eeq
contains, apart from finite terms, only collinear singularities
associated with the initial state, \ie leading logarithms of the form
$\ln(s/\Me^2)$, at least for sufficiently inclusive observables.

Finally, we arrive at the master formula for the cross section:
\begin{equation}\label{eq:master}
\rd \sigma =
 \rd \sigma_{\Born}^{\eeffff}+
 \rd \sigma_{\virt,\finite,\DPA}^{\eeWWffff}
+ \rd\sigma_{\virt+\real,\sing}^{\eeffff}
+ \rd\sigma_{\finite}^{\eeffffg}.
\end{equation}
Since the contribution $\rd\sigma_{\virt+\real,\sing}^{\eeffff}$ is
not treated in DPA, the LL photonic corrections resulting from ISR are
treated exactly in our approach.
 
As a result, in the RC mode {\RacoonWW} calculates several weights
corresponding to processes with $4f$ and $4f\gamma$ kinematics.  The
cross sections $\rd \sigma_{\Born}^{\eeffff}$, $\rd
\sigma_{\virt,\finite,\DPA}^{\eeWWffff}$,
$\rd\sigma_{\virt+\real,\sing}^{\eeffff}$, and
$\rd\sigma_{\finite}^{\eeffffg}$ of \refeq{eq:master} are obtained
from the corresponding $4f$ weights and the $4f\gamma$ weight,
respectively.

\subsubsection{Higher-order leading-logarithmic (LL) corrections}

Higher-order LL corrections from ISR can be taken into account in all
three modes by setting ${\tt SISR}=1$ in the input file and are
implemented in LL approximation via the structure-function approach
(see Section~5 of \citere{Denner:2000bj}), \ie by convoluting the
tree-level or IBA cross section with the structure functions of
\citeres{Beenakker:1996kt,sf}.  The ISR structure functions depend on
the QED splitting scale, $Q={\tt SCALE}$, which is not fixed in LL
approximation and has to be set to a typical momentum scale of the
process.  In the RC mode $\Oa$ corrections from ISR are calculated
completely, and the $\Oa$ LL corrections are implicitly included.  In
order to avoid double counting of radiative corrections, the LL
$\O(\alpha)$ contributions from structure functions multiplied by the
CC11 Born cross section are subtracted from the full $\O(\alpha)$
corrections. Thus, for ${\tt SRC}=1$ and ${\tt SISR}=1$, the
dependence on the QED scale, $Q$, only enters in $\O(\alpha^2)$, \ie
at the two-loop level.

In the slicing branch of {\RacoonWW} the ISR structure
functions are calculated in {\tt kern.f} by 
\cpcsuptable{{\tt LLE},{\tt BET},{\tt ALP}}{%
\multicolumn{3}{l}{\tt FUNCTION STRUCI(XX,DXX,LLE,BET,ALP)}\\
{\tt XX},{\tt DXX} &:& random numbers, ${\tt DXX}=1-{\tt XX}$, \\
{\tt LLE},{\tt BET},{\tt ALP} &:& 
${\tt BET}={\tt ALP}({\tt LLE}-1)/ \pi$, 
${\tt LLE}=2 \ln({\tt SCALE}/ \Me)$,
${\tt ALP}=\alpha(0)$.\\
}

In the subtraction branch they are obtained by calling 
\cpcsuptable{\tt RANDOM(1:2)}{%
\multicolumn{3}{l}{\tt SUBROUTINE ISR(RANDOM,X1,X2,STFCT,WISR,SWITCH)} \\
{\tt RANDOM(1:2)} &:& random numbers, \\
{\tt X1,X2} &:& energy fraction of incoming particles (after ISR), \\
{\tt STFCT} &:& value of the structure functions only, \\
{\tt WISR} &:& total weight for generation of {\tt X1,X2} 
including structure functions, \\
{\tt SWITCH} &:& auxiliary flag (${\tt SWITCH}=0$ if an error occurs), \\
}%
in {\tt subtraction.f}.  By default, {\tt SCALE} is set to the CM
energy, ${\tt SCALE}=\sqrt{s}$, in the subroutine {\tt INITIALIZE} in
{\tt racoonww.f}.

\subsubsection{Naive QCD  corrections}

In all three modes of {\RacoonWW}, the so-called naive QCD corrections
can be included by setting ${\tt QQCD}=2$ in the input file. QCD
corrections enter the processes with hadronic final-states at two
different places.  On the one hand, the hadronic W-boson width
receives a QCD correction, namely a factor
$(1+\alpha_{\mathrm{s}}/\pi)$ \cite{rcwdecay2,rcwdecay1}, if the
W-boson width is calculated (${\tt QGW}=4$).  This affects the total
W-boson width $\GW$ in the W~propagators.  On the other hand, each
hadronically decaying W~boson receives a QCD correction to the $\PW
q\bar q'$ vertex. If the full phase space for gluon emission is
integrated over, this correction reduces to a multiplicative
correction factor $(1+\alpha_{\mathrm{s}}/\pi)$ for each hadronically
decaying W~boson in $\O(\alpha_{\mathrm{s}})$.  The application of
this inclusive factor to distributions is usually called naive QCD
correction. In the total cross section for W-pair production naive
QCD-correction factors cancel against the corresponding factors in the
W-boson width in the resonant W~propagators.

\subsubsection{Anomalous triple and quartic gauge-boson couplings}%
\label{subsubsec:anomalous}%

In {\RacoonWW} anomalous triple gauge-boson couplings (ATGCs) to
$\eeffff$ and anomalous quartic gauge-boson couplings (AQGCs) to
$\eeffffg$ have been implemented.  More precisely, the non-standard
couplings are introduced in the lowest-order matrix elements, which
can be dressed by initial-state radiation, while the contributions
from the remaining radiative corrections add linearly to the
non-standard effects.  The ATGC contributions are taken into account
when ${\tt SATGC}=1$ is set in the input file (see also
\citere{Denner:2001bd}).  The charged ATGCs are defined as in
\citere{Gounaris:1996rz}. The conventions for the neutral ATGCs follow
those of \citeres{Gounaris:1999kf,Gounaris:2000dn}.  The ATGC
contributions to the matrix elements are expressed in terms of five
generic functions corresponding to the $\PZ VV$ and $V\PW\PW$
couplings with $V=\gamma,\PZ$, which have been implemented in the
subroutines {\tt ADDMATGCVVZ} and {\tt ADDMATGCVWW} of {\tt
  ee4fa\_amps.f}, respectively.  These ATGC contributions are added to
the generic CC and NC helicity amplitudes in the subroutines {\tt
  M\_NC} and {\tt M\_CC}.  When ${\tt SATGC}=1$ is chosen the result
depends on the ATGC parameters $\Delta g_1^{V}$, $\Delta \kappa^{V}$,
$\lambda^{V}$, $\tilde \kappa^{V}$, $\tilde \lambda^{V}$, $g_{4,5}^V$,
$f_{4,5}^V$, $h_{1,3}^V$ with $V=\gamma,\PZ$, which are specified in
the input file (see \refse{subsec:input}).

As described in detail in \citere{Denner:2001vr}, AQGCs that involve
at least one photon have been implemented in {\RacoonWW}.  They are
taken into account when ${\tt QAQGC}=1$ is set in the input file. Only
genuine AQGCs are considered, \ie all operators that contribute also
to ATGCs, such as the quadrilinear part of the well-known operator
$F^{\mu\nu}W^{+,\rho}_\nu W^-_{\rho\mu}$, are omitted.  Imposing in
addition custodial $\SU(2)_{\mathrm{c}}$ invariance
\cite{Sikivie:1980hm} to keep the $\rho$ parameter close to 1, only
operators of dimension~6 or higher occur.  Following
\citeres{Denner:2001vr,Stirling:2000ek,Belanger:1992qh,Eboli:1994wg,Stirling:2000sj}
dimension-6 operators for genuine AQGCs are considered that respect
local $\U(1)_{\mathrm{em}}$ invariance and global custodial
$\SU(2)_{\mathrm{c}}$ invariance. These symmetries reduce the set of
such operators to a phenomenologically accessible basis.  More general
AQGCs were discussed in \citere{Belanger:2000aw}.  The AQGC
contributions are expressed in terms of three generic functions which
correspond to the $\gamma\gamma VV$ and $\gamma \PZ\PW\PW$ couplings
with $V=\PW,\PZ$, respectively.  They are added to the generic
Standard Model functions in {\tt M\_CC} and {\tt M\_NC} by calling the
subroutines {\tt ADDMAAQGCAAWW}, {\tt ADDMAAQGCAAZZ}, and {\tt
  ADDMAAQGCAZWW} in {\tt ee4fa\_amps.f}. When ${\tt QAQGC}=1$, the
result depends on the AQGC parameters $a_{0,c,n}/\Lambda^2$ and
$\tilde a_{0,n}/\Lambda^2$ specified in the input file (see
\refse{subsec:input}).
      
\section{The usage of {\RacoonWW}}
\label{sec:usage}
In the sequel we describe the input parameters and flags for the
options of {\RacoonWW}, starting with those that are set in the
input file.%
\footnote{%
  The flags {\tt SBORNG5}, {\tt SSIGEPEMG5}, {\tt SRECOMBG}, and the
  parameters {\tt PRECOMBG} are not listed here and are not discussed any
  further, since they only affect the QCD corrections beyond the naive
  approximation, which are not supported by {\RacoonWW} yet.}%
A list of all possible values of the flags of {\RacoonWW} discussed
below is given in \refta{tab:options}.
\begin{table}
\bce
\begin{tabular}{|l|l|l|c|c|c|c|c|} \hline
flag &  set in & switch for &  \multicolumn{5}{|c|}{possible values ( * means not relevant)} \\ \hline
{\tt SRC}       &input file &tree-level, IBA, & \multicolumn{2}{|c|}{${\tt SRC}=0$} & ${\tt SRC}=2$
 & ${\tt SRC}=3$ & ${\tt SRC}=1$ \\ 
          &             &and RC mode      & \multicolumn{2}{|c|}{tree-level} 
& IBA  & IBA & RC  \\ \hline
          &             &                 &$4f$ &$4f\gamma$ & $4f$ &
$4f\gamma$ & $4f, 4f\gamma$    \\ \hline
{\tt SMC}       &input file   &MC branch        & 1,2 & 1,2 &  1,2 & 1,2 & 1--3 \\
{\tt SBORN4}    &input file   &tree-level $4f$& 1--3 &  0  &   0  &  0  & 0--3 \\
{\tt SBORN5}    &input file   &tree-level $4f\gamma$ &  0  &  1  &  0  &  1  &  1  \\
{\tt SISR}      &input file   &higher-order ISR & 0,1 & 0,1 &  1  &  1  & 0,1 \\
{\tt SCOULTREE} &input file   &Coulomb sing.\ for& 0,1 & 0--2  &  *  & 0--2 &  *  \\
          &             &$\eeffff,4f\gamma$  &     &      &     &     &     \\
{\tt QNF}       &input file   &Coulomb sing.\ for&  *  &  *  &  *  &  *  & 1--3 \\
          &             &$\eeffff$ (RC)      &     &      &     &     &     \\
{\tt QREAL}     &input file   &imaginary part   &  *   &  *  &  *  &  *  & 0,1 \\
{\tt QALP}      &input file   &parameter scheme & 0--2 & 0--2 &   2  &  2  & 0--2 \\
{\tt QGW}       &input file   &W-boson width    & 0--4 & 0--4 & 0,2 & 0,2 & 0--4 \\
{\tt QPROP}     &input file   &width approach   & 0--4 & 0--4 &  1  &  1  &  1  \\
{\tt SSIGEPEM4} &input file   &sub-diagrams ($4f$)& 0--5 &  *  &  1  &  *  & 0--5 \\
{\tt SSIGEPEM5} &input file   &sub-diag.\ ($4f\gamma$) &  *  & 0--5   &  *  & 0--5 &  5  \\
{\tt QQCD}      &input file   &QCD RCs          & 0,2 & 0,2  & 0,2 & 0,2 & 0,2 \\
{\tt SQCDEPEM}  &input file   &gluon-ex.\ backg.& 0--2 & 0--2  &  *  & 0,1 & 0,1 \\
{\tt SRECOMB}   &input file   &photon recomb.   &  *  & 0--3  &  *  & 0--3 & 0--3 \\
{\tt SCUTS}     &input file   &separation cuts       & 0--11& 0--11 & 0--11& 0--11& 0--11\\
{\tt SATGC}     &input file   &ATGCs            & 0,1 &  *    & 0,1 &  *  & 0,1 \\
{\tt QAQGC}     &input file   &AQGCs            &  *  & 0,1   &  *  & 0,1 &  0  \\
{\tt QFAST}     &racoonww.f   &Legendre expans. &  *  &  *    &  *  &  *  & 0,1 \\
{\tt SON}       &racoonww.f   &on-shell projection & 1--4 &  *    &  *  &  *  & 1--4 \\
{\tt QSOFT}     &racoonww.f   &photon DPA part  &  *  &  *    &  *  &  *  & 1--3 \\
%{\tt DSDC}      &slicing.f    &technical cuts   &  *  &  *    &  *  &  *  & 0--2 \\
{\tt OPT}       &slicing.f    &optimization     & 0,1 & 0,1 & 0,1 & 0,1 & 0,1 \\
{\tt SSUB}      &subtraction.f&subtr.\ functions &  *  &  *  &  *  &  *  & 1,2 
\\ \hline
\end{tabular}
\ece
\vspace*{.5em}
\caption{The allowed choices for the options
in the different modes of {\RacoonWW}}
\label{tab:options}
\end{table}
Some more information on the options and details about the format in the input file can be found in the
file {\tt README}.
 
\subsection{Input and initialization of\/ {\RacoonWW} via the input file}
\label{subsec:input}
\begin{cpcdescription}
\item[{\bf {\tt OUTPUTFILE:}}] a character string that specifies the
name of the output file.
\item[{\bf {\tt ENERGY:}}] the total CM energy 
in GeV (${\tt ENERGY}=\sqrt{s}$). \\ 
\item[{\bf {\tt NEVENTSW:}}] number of weighted events.\\ 
\noindent
The number of weighted events must be at least ${\tt NEVENTSW}=2 \times
10^6$ in the tree-level and IBA modes and ${\tt NEVENTSW}=10^7$ in the
RC mode to guarantee that the multi-channel integration yields
reliable estimates for the MC integration error.  When unweighted
events are generated (${\tt NEVENTSUNW}>0$), {\tt NEVENTSW} is the
number of weighted events used to initialize the unweighting
procedure; these weighted events
 are used to determine the maximal weights (see
\refse{subsec:unweighting}).
\item[{\bf {\tt SMC:}}] flag for the slicing and subtraction branches 
of {\RacoonWW}:
\cpcitemtable{${\tt SMC}=1$}{%
${\tt SMC}=1$ &:& chooses the slicing branch in its default setup, \\  
${\tt SMC}=2$ &:& chooses the subtraction branch, \\
${\tt SMC}=3$ &:& chooses a slightly modified version of 
the slicing branch, which has been introduced to minimize the number of 
negative weights in the RC mode. 
}%
In the RC mode, {\RacoonWW} provides an unweighting 
procedure only for ${\tt SMC}=3$.  More details can be found in 
\refse{subsec:unweighting}. 
\item[{\bf {\tt SBORN4:}}] flag for the tree-level $\eeffff$ cross section: 
\cpcitemtable{${\tt SBORN4}=0$}{%
${\tt SBORN4}=0$ &:& the tree-level contribution to $\eeffff$ is not included, \\
${\tt SBORN4}=1$ &:& calculates the off-shell Born $\eeffff$ cross section, \\
${\tt SBORN4}=2$ &:& calculates the cross section to $\eeWWffff$ in DPA with
phase-space for off-shell W~bosons,  \\
${\tt SBORN4}=3$ &:& same as ${\tt SBORN4}=2$ but with phase-space for on-shell W~bosons. \\
}%
For ${\tt SBORN4}=1$ the selection of a subset of diagrams is governed
by the flag {\tt SSIGEPEM4} (see \refta{tab:subsets}). For ${\tt
  SBORN4}=2,3$ the Born cross section is based on the CC03 diagrams
only.  While for ${\tt SBORN4}=2$ the full phase space for off-shell W
bosons is generated, the matrix element is calculated with on-shell
momenta which result from a projection of the phase space with
off-shell W bosons onto a phase space with on-shell W bosons.  For
${\tt SBORN4}=3$ both the phase space and the matrix element squared
are calculated with on-shell kinematics. In both cases the \PZ-boson
width is set to zero. The flags ${\tt SBORN4}=2,3$ should only be used
for checks and comparisons but not for the generation of physical
results. Moreover, ${\tt SBORN4}=2,3$ is only a valid choice for CC
processes or mixed processes with ${\tt SYM}=0,3$ (see
\refta{tab:classification}).
\item[{\bf {\tt SBORN5:}}] flag for the tree-level cross section to
  $\eeffffg$:\\ 
\cpcitemtable{${\tt SBORN5}=0$}{%
${\tt SBORN5}=0$ &:& the tree-level contribution to $\eeffffg$ 
is not included,\\
${\tt SBORN5}=1$ &:& calculates the tree-level cross section
to $\eeffffg$. \\
}%
In the tree-level mode with ${\tt SBORN5}=1$, {\RacoonWW} calculates
the Born contribution to $\eeffffg$. In the RC mode (${\tt SRC}=1$)
and in the IBA mode to $\eeffffg$ (${\tt SRC}=3$), {\RacoonWW} always
sets ${\tt SBORN5}=1$ in {\tt racoonww.f} independent of the choice in
the input file. In the slicing branch (${\tt SMC}=1,3$), when
radiative corrections to $\eeWWffff$ are calculated (${\tt SRC}=1$),
${\tt SBORN5}=1$ calculates the hard-photon bremsstrahlung
contribution, \ie photon radiation away from soft and collinear
singularities.
\item[{\bf {\tt SISR:}}] flag for the inclusion of higher-order ISR:
\cpcitemtable{${\tt SISR}=0$}{%
${\tt SISR}=0$ &:& higher-order LL corrections from ISR are not included, \\
${\tt SISR}=1$ &:& higher-order LL corrections from ISR are included in the 
LL approximation up to order $\O(\alpha^3)$ for ${\tt SMC}=1,2$ and up to 
order $\O(\alpha^2)$ for ${\tt SMC}=3$.\\
}
Higher-order LL corrections from ISR are taken into account by
convoluting the tree-level cross section with the structure functions
of \citeres{Beenakker:1996kt,sf} (see also
\refse{subsec:matrixelement}).  In the RC mode the ISR beyond
$\O(\alpha)$ is switched on/off by the choice of {\tt SISR}, \ie
$\O(\alpha)$ ISR contributions are always included as part of the
$\O(\alpha)$ corrections even when ${\tt SISR}=0$.  In the RC mode
with ${\tt SISR}=0$ the full $\O(\alpha)$ ISR contribution is included
for all diagrams of the CC11 process class.  For ${\tt SRC}=1$ and
${\tt SISR}=1$ the LL contribution to the ISR is taken into account
for the full four-fermion process, \ie also for the background
diagrams, depending on the choices for {\tt SSIGEPEM4}. When ATGCs are
considered (${\tt SATGC}=1$), the complete ISR corrections in LL
approximation ($\O(\alpha)+$higher-orders) to the additional
contributions involving anomalous couplings is also switched on/off by
the choice of ${\tt SISR}$.  When calculating the IBA (${\tt
  SRC}=2,3$), or the tree-level processes $\eeffff$ or $\eeffffg$
(${\tt SRC}=0$, ${\tt SISR}=1$, ${\tt SBORN4}=1$ or ${\tt SBORN5}=1$),
the full corrections from ISR are taken into account in LL
approximation up to order $\O(\alpha^3)$.
\item[{\bf {\tt SRC:}}] flag for choosing the tree-level, IBA, or RC mode:\\
\cpcitemtable{${\tt SRC}=0$ (tree-level mode)}{%
${\tt SRC}=0$ (tree-level mode)&:& calculation of the 
tree-level processes $\eeffff$ (${\tt SBORN4}=1$) or 
$\eeffffg$ (${\tt SBORN5}=1$), \\
${\tt SRC}=1$ (RC mode) &:& $\O(\alpha)$ radiative
corrections to $\eeWWffff$ are included in DPA, \\
${\tt SRC}=2$ (IBA mode) &:& IBA for $\eeWWffff$ is 
calculated for the CC03 class of diagrams, \\
${\tt SRC}=3$ (IBA mode) &:& IBA for $\eeffffg$ is calculated 
based on the full matrix element. \\
}
\item[{\bf {\tt SCOULTREE:}}] flag for the inclusion of the Coulomb singularity
in the tree-level and IBA modes:\\
\cpcitemtable{${\tt SCOULTREE}=0$}{%
${\tt SCOULTREE}=0$ &:& the Coulomb singularity 
to $\eeffff$ and $\eeffffg$ is not included, \\
${\tt SCOULTREE}=1$ &:& the off-shell Coulomb singularity 
to $\eeffff$ in the tree-level mode and to $\eeffffg$ in the tree-level 
and IBA modes is included, \\
${\tt SCOULTREE}=2$ &:& the off-shell Coulomb singularity 
to $\eeffffg$ in the tree-level and IBA modes is included. \\
}
For ${\tt SCOULTREE}=1$ the Coulomb singularity is calculated with
invariant masses derived from the four-momenta of the final-state
fermions only. For ${\tt SCOULTREE}=2$, which is only a valid choice
for the processes $\eeffffg$, the Coulomb singularity is calculated
with the invariant masses of those $\PWp$ and $\PWm$ bosons which are
most resonant.  Note that the invariant masses of the W bosons may or
may not include the photon momentum depending whether the photon is
emitted from the initial or final state (see
\refse{subsec:matrixelement}).  The choices ${\tt SCOULTREE}>0$ are
only valid for WW mediated processes, \ie for CC processes and mixed
processes with ${\tt SYM}=0,3$.
\item[{\bf {\tt QNF:}}] flag for the inclusion for the Coulomb singularity 
in the RC mode:\\
\cpcitemtable{${\tt QNF}=1$}{%
${\tt QNF}=1$ &:& DPA Coulomb singularity, \\
${\tt QNF}=2$ &:& off-shell Coulomb singularity with DPA Born cross section, \\
${\tt QNF}=3$ &:& off-shell Coulomb singularity with (off-shell) CC03 
Born cross section.\\
}
This flag is only effective in the RC mode.  According to the choice
of ${\tt QNF}$, different implementations of the Coulomb singularity
are included in the virtual non-factorizable corrections (the real
non-factorizable corrections are implicitly contained in the full
matrix-element calculation of $\eeffffg$).  The off-shell Coulomb
singularity is implemented as defined in \citere{Denner:1998ia}.  In
\citere{Denner:2000bj} different versions of the DPA have been
investigated by modifying the implementation of the DPA.  Different
treatments of the Coulomb singularity, ${\tt QNF}=2$ (``def'') and
${\tt QNF}=1$ (``Coul''), are one of the considered options.  The
recommended setting is ${\tt QNF}=3$.
\item[{\bf {\tt QREAL:}}] flag for the inclusion of the
imaginary parts of the virtual electroweak corrections:\\
\cpcitemtable{${\tt QREAL}=0$}{%
${\tt QREAL}=0$ &:& imaginary parts of the one-loop corrections are included, \\
${\tt QREAL}=1$ &:& imaginary parts of the one-loop corrections are not included. \\
}
This flag is only effective in the RC mode. The imaginary parts of the
one-loop corrections contribute only significantly to distributions
that depend on the azimuthal decay angle of one of the W bosons. If
such distributions are considered, {\tt QREAL} has to be set to ${\tt
  QREAL}=0$ (imaginary parts are not neglected).
\item[{\bf {\tt QALP:}}] flag to specify the input-parameter scheme:
\cpcitemtable{${\tt QALP}=0$}{%
${\tt QALP}=0$ &:& $\alpha(0)$ scheme, \\
${\tt QALP}=1$ &:& $\alpha(M_Z)$ scheme, \\
${\tt QALP}=2$ &:& $G_{\mu}$ scheme.\\
}
Independent of the choice of the input-parameter scheme, the
electroweak corrections are always calculated by using the fine
structure constant $\alpha(0)$, \ie the relative (electroweak)
corrections $(\rd \sigma-\rd \sigma_{{\rm Born}})/ \rd \sigma_{{\rm
    Born}}$ are always proportional to $\alpha(0)$.  Not all input
parameters defined in the subroutine {\tt PARAMETER} in {\tt public.f}
are used in the different input-parameter schemes, e.g.\ {\tt ALPHAZ}
is only relevant for ${\tt QALP}=1$, {\tt GF} only for ${\tt QALP}=2$.
While the small fermion masses have to be non-zero for all
input-parameter schemes, their actual values matter only if ${\tt
  QALP}=0$, except for the electron mass that is used in the structure
functions.
\item[{\bf {\tt QGW:}}] flag for the W-boson width:
\cpcitemtable{${\tt QGW}=0$}{%
${\tt QGW}=0$ &:& the total W-boson width is an input 
parameter, set in the subroutine {\tt PARAMETER} in {\tt public.f}, \\
${\tt QGW}=1$ &:& the tree-level W-boson width is calculated, \\
${\tt QGW}=2$ &:& the IBA W-boson width is calculated, \\
${\tt QGW}=3$ &:& the one-loop W-boson width is calculated 
without $\O(\alpha_s)$ corrections, \\
${\tt QGW}=4$ &:& the one-loop W-boson width is calculated 
with $\O(\alpha_s)$ corrections. \\
}
When the one-loop W-boson width is calculated (${\tt QGW}=3,4$ and
${\tt SRC}=0,1$), the full electroweak $\O(\alpha)$ corrections in the
chosen input-parameter scheme (controlled by the flag {\tt QALP}) are
included.  When ${\tt QGW}=2$, the W-boson width is calculated in IBA,
\ie the tree-level W-boson width in the $G_{\mu}$ scheme (${\tt
  QALP}=2$) including $\O(\alpha_s)$ corrections.  The only allowed
options for ${\tt SRC}=2,3$ are ${\tt QGW}=0,2$.  In {\tt racoonww.f}
the partial decay widths for the decays into leptons and quarks are
also calculated using the same options as for the calculation of the
amplitude. In the output file, {\RacoonWW} provides the ``effective''
branching ratios: leptonic width/total width, hadronic width/total
width, and ($\mathrm{leptonic} + \mathrm{hadronic}$ width)/total
width.  The widths in the numerators are calculated with the same
options as the DPA matrix element for $\eeWWffff$. Since the total
W-boson width and the partial widths are not necessarily calculated
with the same options, the total branching ratio can be different from
one. The effective branching ratios are not used in {\RacoonWW} and
are only provided for information.
\item[{\bf {\tt QPROP:}}] flag for choosing a width scheme:
\cpcitemtable{${\tt QPROP}=0$}{%
${\tt QPROP}=0$ &:& gauge-boson widths are set to zero,\\
${\tt QPROP}=1$ &:& constant width is used in all propagators,\\
${\tt QPROP}=2$ &:& constant width is used only in time-like propagators, \\
${\tt QPROP}=3$ &:& running width is used (no width in space-like propagators), \\
${\tt QPROP}=4$ &:& constant width is used in all 
propagators and the weak mixing angle is calculated from the complex
W- and Z-boson masses.\\
}%
For ${\tt SRC}\ne0$, ${\tt QPROP=1}$ is automatically chosen in the
subroutine {\tt INITIALIZE} in {\tt racoonww.f}.
\item[{\bf {\tt SSIGEPEM4:}}] chooses a subset of diagrams for 
the tree-level processes $\eeffff$ (see \refta{tab:subsets}).
\item[{\bf {\tt SSIGEPEM5:}}] chooses a subset of diagrams
for the tree-level and IBA processes $\eeffffg$ (see \refta{tab:subsets}).

\begin{table}
\bce\begin{tabular}{|c|l|} \hline
\begin{tabular}{c}{\tt SSIGEPEM4} \\ {\tt SSIGEPEM5}\end{tabular} &  
subset of diagrams included \\ \hline
0 & all electroweak diagrams \\
1 & only electroweak WW signal diagrams (CC03) \\
2 & only electroweak ZZ signal diagrams (NC02) \\
3 & only electroweak ZZ, Z$\gamma$, and $\gamma\gamma$ signal diagrams  \\
4 & only electroweak WW, ZZ, Z$\gamma$, and $\gamma\gamma$ signal diagrams  \\
5 & only diagrams corresponding to the CC11 process class\\ \hline
\end{tabular}
\ece
\caption{Possible choices for subsets of Feynman diagrams as governed by
{\tt SSIGEPEM4} and {\tt SSIGEPEM5}}
\label{tab:subsets}
\end{table}

\item[{\bf {\tt QQCD:}}] flag for the inclusion of QCD corrections:

\cpcitemtable{${\tt QQCD}=0$}{%
${\tt QQCD}=0$ &:& QCD corrections are not included, \\ 
${\tt QQCD}=2$ &:& naive QCD corrections are included. \\
}
In {\RacoonWW} only so-called naive QCD corrections can be included
yet, \ie for each hadronically decaying W boson {\RacoonWW} applies a
global factor $(1+\alpha_s/ \pi)$ to the differential cross sections.
\item[{\bf {\tt FNAME(3:6)}:}] particle names for the
specification of the final-state fermions:
\cpcitemtable{${\tt FNAME = u, d, s, c, t, b, e, mu, tau, nu\_e, nu\_mu, nu\_tau}.$}{%
${\tt FNAME = u, d, s, c, t, b, e, mu, tau, nu\_e, nu\_mu, nu\_tau}.$
\\}
When specifying the final-state particles the order of
\refta{tab:classification} must be respected.
\item[{\bf {\tt PP,PM}:}] degrees of positron- and electron-beam polarization;
for unpolarized beams choose ${\tt PP}={\tt PM}=0$. 
\item[{\bf {\tt SRECOMB}:}] flag for the photon-recombination scheme:
\cpcitemtable{${\tt SRECOMB}=0$}{%
${\tt SRECOMB}=0$ &:& no recombination is performed, \\
${\tt SRECOMB}=1$ &:& TH recombination scheme is applied 
(see also \citere{Denner:2000bj}), \\
${\tt SRECOMB}=2$ &:& EXP recombination scheme is applied 
(see also \citere{Denner:2000bj}), \\
${\tt SRECOMB}=3$ &:& TH recombination scheme is applied including
photon recombination also with the beams.\\
}%
When ${\tt SRECOMB}=1,2,3$ the parameters {\tt PRECOMB(1:4)} must be
specified in the input file (see below).  For ${\tt SRECOMB}=3$ no
separation cuts are required and ${\tt PRECOMB(1)}=0$ can be chosen.
Note that if the unweighting procedure is used, the recombination
procedure and separation cuts influence the resulting unweighted
events.
\item[{\bf {\tt PRECOMB(1:4)}:}] recombination parameters:
\cpcitemtable{{\tt PRECOMB(1)}}{%
{\tt PRECOMB(1)}&:& separation cut on angle between beam and photon in 
degrees if ${\tt SRECOMB}=1,2,3$, \\
{\tt PRECOMB(2)}&:& recombination cut for photon energy in GeV
if ${\tt SRECOMB}=1,2,3$, \\
{\tt PRECOMB(3)}&:& invariant-mass recombination cut in GeV 
if ${\tt SRECOMB}=1,3$ or angular recombination cut for leptons 
in degrees if ${\tt SRECOMB}=2$, \\
{\tt  PRECOMB(4)}&:& angular recombination cut for quarks in degrees if 
${\tt SRECOMB}=2$. \\
}
The recombination scheme is defined in the subroutine {\tt
  RECOMBINATION} in {\tt public.f}. The recombination scheme for ${\tt
  SRECOMB}=1$ is described in detail in \citere{Denner:2000bj} for the
parameters ${\tt PRECOMB(1)}=5$, ${\tt PRECOMB(2)}=1$, and ${\tt
  PRECOMB(3)}=5$ or 25, called ``\bare'' or ``calo'' cuts,
respectively.  For all three photon recombination schemes (${\tt
  SRECOMB}=1,2,3$), the photon momentum is set to zero if the angle
between the photon and the beam is smaller than {\tt PRECOMB(1)}. This
cut is applied at the very beginning before photon recombination is
performed.

For ${\tt SRECOMB}=1$, the minimal invariant mass of the photon and
any charged final-state fermion, $M(\gamma,{\tt FCOMB})$, is
determined, and the photon is recombined with the final-state fermion
${\tt FCOMB}$ if $M(\gamma,{\tt FCOMB}) < {\tt PRECOMB(3)}$ or the
photon energy $E_{\gamma} < {\tt PRECOMB(2)}$.  Explicitly, this means
that the photon momentum is added to the momentum of the fermion {\tt
  FCOMB}, and the photon is discarded.

For ${\tt SRECOMB}=2$, the minimal angle between the photon and any
charged final-state fermion, $\theta(\gamma,{\tt FCOMB})$, is
determined, and the photon is recombined with the fermion ${\tt
  FCOMB}$ if $\theta(\gamma,l^\pm) < {\tt PRECOMB(3)}$ in case $l^\pm$
is charged lepton of the final state or if $\theta(\gamma,q) < {\tt
  PRECOMB(4)}$ in case $q$ is a final-state quark, or if the photon
energy $E_{\gamma} < {\tt PRECOMB(2)}$.

For ${\tt SRECOMB}=3$ the minimal invariant mass of the photon and any
charged fermion (including the incoming electron and positron),
$M(\gamma,{\tt FCOMB})$, is determined, and the photon is recombined
with the fermion ${\tt FCOMB}$ if $M(\gamma,f) < {\tt PRECOMB(3)}$,
where $f$ is one of the initial- or final-state charged fermions, or
if $E_{\gamma} < {\tt PRECOMB(2)}$.  In case ${\tt FCOMB}$ is a beam
electron or positron, the photon momentum is set to zero.

\item[{\bf {\tt SQCDEPEM:}}] flag for the inclusion of gluon-exchange
diagrams in the tree-level mode:
\cpcitemtable{${\tt SQCDEPEM}=0$}{%
${\tt SQCDEPEM}=0$ &:& gluon-exchange background diagrams are not included, \\
${\tt SQCDEPEM}=1$ &:& gluon-exchange background diagrams are included, \\
${\tt SQCDEPEM}=2$ &:& only gluon-exchange background diagrams are included. \\
}
The choice of {\tt SQCDEPEM} only affects the tree-level cross
sections to $\eeffff$ and $\eeffffg$, and for ${\tt SISR}=1$ also the
LL corrections from ISR.  This QCD contribution then consists of
gluon-exchange background diagrams to the tree-level processes
$\eeffff$ and $\eeffffg$ as described in \citere{Denner:1999gp}. In
the RC mode gluon-exchange diagrams can be taken into account only in
the tree-level matrix element to $\eeffff$.
\item[{\bf {\tt SATGC:}}] flag for the inclusion of anomalous
  triple gauge-boson couplings (ATGCs):
\cpcitemtable{${\tt SATGC}=0$}{%
${\tt SATGC}=0$ &:& ATGCs are not included, \\
${\tt SATGC}=1$ &:& ATGCs are included. \\
}
If ${\tt SATGC}=1$, the parameters {\tt GTGC(1:11,1:2)} must be
specified in the input file. They are stored in the common block {\tt
  ATGC} and correspond to the ATGCs $\Delta g_1^{V}$, $\Delta
\kappa^{V}$, $\lambda^{V}$, $\tilde \kappa^{V}$, $\tilde \lambda^{V}$,
$g_{4,5}^V$, $f_{4,5}^V$, $h_{1,3}^V$ with $V=\gamma(\equiv\mathrm
A),\PZ$.  The ATGCs affect only the tree-level cross section for
$\eeffff$, and for ${\tt SISR}=1$ also the LL corrections from ISR are
applied to the tree-level cross sections involving ATGCs.
\item[{\bf {\tt QAQGC:}}] flag for the inclusion of 
anomalous quartic gauge-boson couplings (AQGCs):
\cpcitemtable{${\tt QAQGC}=0$}{%
${\tt QAQGC}=0$ &:& AQGCs are not included, \\
${\tt QAQGC}=1$ &:& AQGCs are included. \\
}
If ${\tt QAQGC}=1$, the parameters {\tt ALAM2(0:4)} must be specified
in the input file.  They are stored in the common block {\tt AQGC} and
correspond to the AQGCs $a_{0,c,n}/\Lambda^2$ and $\tilde
a_{0,n}/\Lambda^2$.  The inclusion of AQGCs is not supported in the RC
mode and only affects the results obtained in the tree-level mode
(${\tt SRC}=0$, ${\tt SBORN5}=1$) and the IBA mode (${\tt SRC}=3$) for
$\eeffffg$.
\item[{\bf {\tt SCUTS:}}] flag for choosing the separation cuts:
\cpcitemtable{${\tt SCUTS}=10$}{%
${\tt SCUTS}=0$ &:& no separation cuts are applied, \\
${\tt SCUTS}=1$ &:& ADLO cuts are applied as defined in {\tt racoonww.f}
and in \citere{Grunewald:2000ju},\\
${\tt SCUTS}=2$ &:& LC cuts are applied as defined in {\tt racoonww.f},\\
${\tt SCUTS}=10$&:& a minimal set of cuts must be specified in the
input file, \\
${\tt SCUTS}=11$&:& cuts on the energies of all final-state particles, 
on all angles, and on invariant masses must be specified in the input file. \\
}
All energies and invariant masses are measured in GeV and all angles
in degrees.  Non-standard cuts, \ie cuts which are not covered by the
choices of {\tt SCUTS} that {\RacoonWW} provides, can be included at
the end of the subroutine {\tt CUT} in {\tt public.f}.
\end{cpcdescription}

\subsection{Options not provided in the input file}

{\RacoonWW} has a few more options that are not provided in the input
file. Changing those options is only advisable if the user is familiar
with the details of the calculation as described in
\citere{Denner:2000bj}.  If they are common to both the slicing and
subtraction branches, {\RacoonWW} sets them to their default values in
subroutine {\tt INITIALIZE} in {\tt racoonww.f} (after ``{\tt setting
  options}'').  Special options for the slicing and subtraction
branches are set in the subroutines {\tt INITSLICING} and {\tt
  INITSUBTRACTION}, respectively.

The following flags are set in subroutine {\tt INITIALIZE} in {\tt racoonww.f}:
\begin{description}\setlength{\itemsep}{2ex}
\item[{\bf {\tt SON:}}] flag for the projection of the phase space of
  the final-state particles to the one with on-shell \PW~bosons.
  {\RacoonWW} provides four different versions for the projection of
  off-shell momenta to on-shell momenta:
\cpcitemtable{${\tt SON}=0$}{%
${\tt SON}=0$ &:& directions of momenta of $\PW^+$, $f_3$, and $f_5$ are fixed (default), \\
${\tt SON}=1$ &:& directions of momenta of $\PW^+$, $\bar{f}_4$, and ${f}_5$ are fixed, \\
${\tt SON}=2$ &:& directions of momenta of $\PW^+$, $f_3$, and $\bar{f}_6$ are fixed, \\
${\tt SON}=3$ &:& directions of momenta of $\PW^+$, $\bar{f}_4$, and 
$\bar{f}_6$ are fixed \\  
\multicolumn{3}{@{}l}{for the process 
$\Pe^+ \Pe^- \to f_3 \bar{f}_4 f_5 \bar{f}_6$
with $\PWp \to f_3 \bar{f}_4$ and $\PWm \to f_5 \bar{f}_6$.}\\
}
Since the projections are performed in the CM frame of the incoming
$\Pe^+\Pe^-$, the direction of the
$\PW^-$ is opposite to the direction of the $\PW^+$.  The projection
is done in the subroutine {\tt OFTOON} in {\tt eeWW4f\_DPA.f}. In
\citere{Denner:2000bj} the accuracy of the DPA has been investigated
by modifying the implementation of the DPA and comparing the results.
The comparison of the on-shell projections, ${\tt SON}=1$ (``def'') and 
${\tt SON}=2$ (``proj''), is one of the considered options.
\item[{\bf {\tt QFAST:}}] flag for the evaluation of the virtual
corrections in DPA:   
\cpcitemtable{${\tt QFAST}=0$}{%
${\tt QFAST}=0$ &:& form factors for virtual corrections from full
formulas (slow), \\
${\tt QFAST}=1$ &:& evaluation from expansion into Legendre polynomials 
(fast); this is the default value of {\tt QFAST}. \\
}
Details of this approach can be found in \refse{subsubsec:radcor} and
in \citere{Denner:2000bj}.  
\item[{\bf {\tt QSOFT:}}] flag for the definition of the DPA part of 
radiative corrections. Terms that are subtracted from the virtual
corrections in DPA are fixed by
\cpcitemtable{${\tt QSOFT}=1$}{%
  ${\tt QSOFT}=1$ &:& endpoint part of \citere{Dittmaier:2000mb} as
  given in Eq.~(4.29) of
  \citere{Denner:2000bj} (default), \\
  ${\tt QSOFT}=2$ &:& only logarithms as defined in Eq.~(4.2.53)
  of \citere{Roth:1999kk}, \\
  ${\tt QSOFT}=3$ &:& YFS factor for virtual photons as defined in
  Eq.~(4.31) of \citere{Denner:2000bj}. \\
  }
Note that results obtained with different choices for {\tt QSOFT}
differ only by terms that are beyond DPA accuracy. In
\citere{Denner:2000bj} the accuracy of the DPA has been investigated
by modifying the implementation of the DPA and comparing the results.
The definition of the finite virtual corrections, ${\tt QSOFT}=1$
(``def'') and ${\tt QSOFT}=3$ (``eik''), is one of the considered
options.
\end{description}

{\RacoonWW} provides the possibility to use {\sc PAW/HBOOK}
\cite{Brun:1989vg,Brun:1987vv} for histogram filling and plotting.
The usage of {\sc PAW/HBOOK} is only allowed when generating weighted
events ({\tt NEVENTSUNW=0}).  This option can be activated as follows:
\begin{itemize}\setlength{\itemsep}{2ex}
\item Set ${\tt SPAW}=1$ in the subroutine {\tt INITIALIZE} in 
{\tt racoonww.f}.
\item Uncomment all lines in the if-environment starting with 
``{\tt if(spaw.eq.1$\ldots$}'' in the subroutine {\tt INITIALIZE}. 
\item Edit the {\tt makefile} and add {\tt pawgraphs.o} to the list of
  the objects {\tt OBJS}.  When using {\sc HBOOK/PAW}, {\sc CERNLIB}
  must be linked in the {\tt makefile}.
\end{itemize}
In {\tt pawgraphs.f} the histograms are defined in 
the subroutine {\tt SETUP\_GRAPHS} and filled in the subroutine 
{\tt GRAPHS}, which are called in {\tt kern.f}.  
This is the place where the user can add and modify histograms.

In the following we describe some branch-specific options of {\RacoonWW}.

\subsubsection{Options specific for the slicing branch}

All the flags and parameters specific for the slicing branch are set
in the subroutine {\tt INITSLICING} in {\tt slicing.f}:
\begin{description}\setlength{\itemsep}{2ex}

\item[{\bf {\tt OPT:}}] flag for the optimization of the a priori weights:
\cpcitemtable{${\tt OPT}=0$}{%
${\tt OPT}=0$ &:& no optimization, \\
${\tt OPT}=1$ &:& self-optimization of the a priori weights (default).\\
}
In the slicing branch the self-optimization of the a priori weights is
performed when ${\tt OPT}=1$ in slicing.f. The number of events used
for the optimization of the a priori weights and how often the
optimization is performed can be separately specified for the
calculation of the $4 f$ weights ({\tt NOPT,IOPT}) and the $4 f
\gamma$ weights ({\tt NOPTG,IOPTG}). The default setting is ${\tt
  NOPT}=10^6$, ${\tt IOPT}=10$, and ${\tt NOPTG}=25 \times 10^6$,
${\tt IOPTG}=1$.  The flag {\tt OPT} and the parameters enter in
function {\tt KERN} in {\tt kern.f}. The a priori weights are
optimized as long as the number of generated events is smaller than
{\tt NOPT} or {\tt NOPTG}.  For the $\eeffff$ part the a priori
weights are optimized after $5 \times 10^3$, $10 \times 10^3$, $15
\times 10^3$, $25 \times 10^3$ generated events and when the number of
events, {\tt NEVENTS}, is a multiple of ${\tt NOPT/IOPT}$ as long as $
{\tt NEVENTS}\le{\tt NOPT}$.  For the $\eeffffg$ part the a priori
weights are optimized after $5 \times 10^5$, $15 \times 10^5$, $30
\times 10^5$ generated events and when {\tt NEVENTS} is a multiple of
${\tt NOPTG/IOPTG}$ as long as $ {\tt NEVENTS}\le{\tt NOPTG}$.
\end{description}

\subsubsection{Options specific for the subtraction branch}

The flags and parameters specific for the subtraction branch are set
in the subroutine {\tt INITSUBTRACTION} in {\tt subtraction.f}:
\begin{description}\setlength{\itemsep}{2ex}
\item[{\bf {\tt SSUB:}}] flag for the definition of the subtraction functions:
\cpcitemtable{${\tt SSUB}=1$}{%
${\tt SSUB}=1$ &:& subtraction terms as defined in 
\citere{Dittmaier:2000mb}, \\
${\tt SSUB}=2$ &:& subtraction terms as defined in 
\citere{Roth:1999kk}. \\
}
The results obtained by the two subtraction variants differ only by
terms that are beyond the DPA accuracy. More details can be found in
\citere{Denner:2000bj}.
\item[{\bf {\tt NOPT:}}] parameters for the adaptive optimization.
\\
\noindent
{\tt NOPT} defines the numbers of events after which the adaptive
optimization is performed, \ie after which event the a priori weights
are recalculated.  The optimization steps are given by ${\tt
  NOPT(I,NG)} \times \mbox{number of channels}$, where ${\tt
  NG}=1,2,3$ denotes the phase-space generators discussed in
\refse{subsubsec:subtraction}.  The adaptive optimization stops after
the $i$-th optimization step if ${\tt NOPT(I}+1{\tt ,NG)} = 0$.  The
default settings are ${\tt NOPT(I,NG)}={\tt I} \times 100$ for ${\tt
  I}=1,\ldots,8$ and zero otherwise.
\item[{\bf {\tt ALPHAMIN:}}] the a priori weights have a minimal value which 
is fixed by setting the parameter {\tt ALPHAMIN}.                             
\end{description}

\subsection{The output of {\RacoonWW}}
\label{subsec:output}
The output consists of an output file and 27 (or more, if the user
added histograms) data files, which contain the data for the
distributions.  The name of the output file is set in the input file
and stored in the variable {\tt OUTPUTFILE}.  For ${\tt SPAW}=1$ there
is also a {\sc PAW} file {\tt pawplot.paw} with the histograms
generated with {\sc PAW/HBOOK} \cite{Brun:1989vg,Brun:1987vv}. If
${\tt IOUT} \ne 0$ is set in the subroutine {\tt INITIALIZE} in {\tt
  racoonww.f}, a file {\tt optimization.info} will be generated.  This
file contains information about the kinematical channels used in the
multi-channel integration and about the optimization of the a priori
weights as described in \refse{subsec:phasespace}.  By default ${\tt
  IOUT}=0$ is set.

The event information is stored in the common block {\tt EVENT} (see
\refse{sec:structure}).  The weights of the slicing and subtraction
branches are described in \refta{tab:wslicing1} for ${\tt SMC}=1$, in
\refta{tab:wsubtraction} for ${\tt SMC}=2$, and in
\refta{tab:wslicing2} for ${\tt SMC}=3$.
\begin{table}
\bce
\cpcframetable{{\tt WEIGHT(1)}}{
{\tt WEIGHT(1)}&:& tree-level cross section to $\eeffff$, 
vanishes for ${\tt SBORN4}=0$ or  ${\tt SISR}=1$, \\
{\tt WEIGHT(2)}&:& $\eeWWffff$ DPA part of $\O(\alpha)$ corrections, \\
{\tt WEIGHT(3)}&:& virtual+soft+final-state collinear photon part
(to avoid double counting, LL $\O(\alpha)$ corrections from ISR are
subtracted when ${\tt SISR}=1$),\\
{\tt WEIGHT(4)}&:& initial-state collinear photon part of $\O(\alpha)$ 
corrections
(to avoid double counting, LL $\O(\alpha)$ corrections from ISR are
subtracted when ${\tt SISR}=1$), collinear photon radiation off $\Pe^+$, \\
{\tt WEIGHT(5)}&:& 
same as {\tt WEIGHT(4)} but for
collinear photon radiation off $\Pe^-$, \\
{\tt WEIGHT(6)}&:& 
\cpcsubtable{${\tt SRC}=1$}{%
${\tt SRC}=1$ &:& %LL corrections from ISR up to $\O(\alpha^3)$, 
the tree-level cross section is convoluted with the ISR 
structure functions including LL corrections up to $\O(\alpha^3)$\\
${\tt SRC}=2$ &:& IBA for $\eeWWffff$, the CC03 IBA cross section is 
convoluted with the ISR structure functions,
}\\
&&vanishes for ${\tt SISR}=0$,\\
{\tt WEIGHT(7)}&:& 
\cpcsubtable{${\tt SRC}=1$}{%
${\tt SRC}=0$ &:& tree-level process $\eeffffg$, \\
${\tt SRC}=1$ &:& $\eeffffg$ bremsstrahlung contribution,\\       
${\tt SRC}=3$ &:& IBA for $\eeffffg$. \\
}
%${\tt WEIGHT(8)}$&:& is zero since QCD beyond the naive corrections is 
%not supported by {\RacoonWW} yet.\\
}%
\ece
\caption{Description of the weights {\tt WEIGHT(1:7)} of 
the slicing branch (${\tt SMC}=1$)}
\label{tab:wslicing1}
\end{table}

\begin{table}
\bce
\cpcframetable{\tt WEIGHT(8:29)}{
{\tt WEIGHT(1)}&:& tree-level cross section to $\eeffff$, \\
&& vanishes for ${\tt SBORN4}=0$ or ${\tt SISR}=1$,\\ 
{\tt WEIGHT(2)}&:& 
\cpcsubtable{${\tt SRC}=0,1$}{%
${\tt SRC}=0,1$&:& tree-level cross section for $\eeffff$, \\
${\tt SRC}=2$&:& IBA cross section for $\eeWWffff$, \\
}%
\\
&&convoluted with ISR structure functions, LL corrections from ISR up
to $\O(\alpha^3)$, collinear photon radiation off $\Pe^+$ and $\Pe^-$,\\
&&vanishes for ${\tt SISR}=0$, \\
{\tt WEIGHT(3:5)}&:& $4f$ part of radiative corrections including
virtual $\O(\alpha)$ corrections and subtraction functions for ${\tt SRC}=1$
($\O(\alpha)$ LL corrections from ISR omitted to avoid double counting), \\
&& vanishes for ${\tt SRC}=0,2,3$, \\
&&
\cpcsubtable{{\tt WEIGHT(3)}}{%
{\tt WEIGHT(3)}&:& contribution with $4f$ kinematics,           \\            
{\tt WEIGHT(4)}&:& collinear photon radiation off $\Pe^+$, \\              
{\tt WEIGHT(5)}&:& collinear photon radiation off $\Pe^-$, \\                  
}%
\\
{\tt WEIGHT(6)}&:& 
\cpcsubtable{${\tt SRC}=1$}{%
${\tt SRC}=0$&:& tree-level process $\eeffffg$ when ${\tt SBORN5}=1$,\\
${\tt SRC}=1$&:& bremsstrahlung part of $\O(\alpha)$ corrections, \\
${\tt SRC}=3$&:& IBA for $\eeffffg$,
}%
\\
{\tt WEIGHT(7)}&:&  presently not used,\\
%is zero since QCD beyond the naive corrections is 
%not supported by {\RacoonWW} yet,\\
{\tt WEIGHT(8:29)}&:& 
contributions from the subtraction functions to the $4f\ga$ part for ${\tt SRC}=1$, 
vanishes for ${\tt SRC}=0,2,3$. 
}\ece
\caption{Description of the weights {\tt WEIGHT(1:29)} of 
the subtraction branch (${\tt SMC}=2$)}
\label{tab:wsubtraction}
\end{table}

\begin{table}
\bce
\cpcframetable{{\tt WEIGHT(2:3)}}{%
{\tt WEIGHT(1)}&:& tree-level cross section to $\eeffff$, 
$\eeWWffff$ DPA part of $\O(\alpha)$ corrections, 
virtual+soft+final-state collinear photon part, and soft part 
of the $\O(\alpha^2)$ LL corrections from ISR when ${\tt SISR}=1$, \\ 
{\tt WEIGHT(2:3)}&:& vanishes, \\
{\tt WEIGHT(4)}&:& initial-state collinear photon part of $\O(\alpha)$ 
corrections, collinear part of the $\O(\alpha^2)$ LL  
corrections from ISR when ${\tt SISR}=1$, 
collinear photon radiation off $\Pe^+$, \\ 
{\tt WEIGHT(5)}&:& same as {\tt WEIGHT(4)} but for
collinear photon radiation off $\Pe^-$, \\ 
{\tt WEIGHT(6)}&:& collinear part of the $\O(\alpha^2)$ LL 
corrections from ISR when ${\tt SISR}=1$, 
collinear photon radiation off both $\Pep$ and $\Pem$ simultaneously, \\ 
{\tt WEIGHT(7)}&:& 
$\eeffffg$ bremsstrahlung contribution.       
}\ece
\caption{Description of the weights {\tt WEIGHT(1:7)} of 
the slicing branch (${\tt SMC}=3$). This branch is only supported in the 
RC mode (${\tt SRC}=1$)}
\label{tab:wslicing2}
\end{table}

\subsubsection{The output file of {\RacoonWW}}
\label{subsubsec:outputfile}
In the output file {\RacoonWW} provides the following information:
\begin{itemize}
\item
the input parameters, the effective branching ratios, and the choices
for the options,
\item
the total cross section together with the maximal values of the weights
(in the subtraction branch also the corresponding event numbers and 
channels are given), and
\item
the total cross sections for the subcontributions 
(tree-level $4f$ cross section, 
$2\to 4$ part of the corrections,
ISR up to $\O(\alpha^3)$,
and $2\to 5$ part of the corrections)
together with the maximal values of the weights.  
\end{itemize}

Finally, the number of rejected events is given. Events are discarded
owing to the applied separation cuts and in case of the slicing branch
also owing to the imposed technical cuts. {\RacoonWW} provides the
number of rejected events in the calculation of each weight
separately.

\subsubsection{The data files for distributions}
\label{subsubsec:data}
The data for the distributions are written in files {\tt
  dat.01},$\ldots$,{\tt dat.27}.  The histograms are defined in the
subroutine {\tt SETTINGS} in {\tt public.f} as described in
\refta{tab:data}. The file {\tt dat.01} contains the total cross
section for test purposes.  Note that the momenta of the initial-state
particles after ISR, ${\tt P(1:2,J,K)}$, do not correspond to
physically measurable quantities and, thus, must not be used for
defining histograms.
\begin{table}
\bce
\cpcframtable{data file}{%
   data file  &  distribution in the \\ \hline
{\tt dat.02}  &  invariant mass of $f_3$ and $\bar{f}_4$, [75:85]\,GeV
                 (in case of CC processes this corresponds to 
                 the $\PW^+$ invariant mass), \\    
{\tt dat.03}  &  invariant mass of $f_5$ and $\bar{f}_6$, [75:85]\,GeV 
                 (in case of CC processes this corresponds to 
                 the $\PW^-$ invariant mass), \\     
{\tt dat.04}  &  cosine of the angle between $V_1$ and $\bar{f}_1=\Pep$, 
                 [$-1$:1],       \\
{\tt dat.05}  &  cosine of the angle between $V_2$ and $f_2=\Pem$, 
                 [$-1$:1],       \\ 
{\tt dat.06}  &  cosine of the angle between $V_1$ and $f_3$, [$-1$:1], \\ 
{\tt dat.07}  &  cosine of the angle between $V_1$ and $\bar{f}_ 4$,
[$-1$:1], \\  
{\tt dat.08}  &  cosine of the angle between $V_2$ and $f_5$, [$-1$:1] \\ 
{\tt dat.09}  &  cosine of the angle between $V_2$ and $\bar{f}_6$,
[$-1$:1], \\  
{\tt dat.10}  &  energy of the photon, [0:50]\,GeV  
                 ($4 f$ weights are added to the first bin), \\
{\tt dat.11}  &  cosine of the angle between $\gamma$ and $\bar{f}_1=\Pep$, 
                 [$-1$:1] ($4 f$ weights are added to the last bin), \\
{\tt dat.12}  &  minimal angle between $\gamma$ and charged final-state
                 fermions, [0:180]\,degrees 
                 ($4 f$ weights are added to the first bin), \\
{\tt dat.13} &  energy of $f_3$, [0:$\sqrt{s}/2$]\,GeV,  \\
{\tt dat.14} &  energy of $\bar{f}_4$, [0:$\sqrt{s}/2$]\,GeV,  \\
{\tt dat.15} &  energy of $f_5$, [0:$\sqrt{s}/2$]\,GeV,  \\
{\tt dat.16} &  energy of $\bar{f}_6$, [0:$\sqrt{s}/2$]\,GeV,   \\
{\tt dat.17} &  cosine of the angle between $V_1$ and
                $f_3$ in the rest frame of $V_1$, [$-1$:1], \\
{\tt dat.18} &  cosine of the angle between $V_1$ and
                $\bar{f}_4$ in the rest frame of $V_1$, [$-1$:1], \\
{\tt dat.19} &  cosine of the angle between $V_2$ and
                $f_5$ in the rest frame of $V_2$, [$-1$:1], \\
{\tt dat.20} &  cosine of the angle between $V_2$ and
                $\bar{f}_6$ in the rest frame of $V_2$, [$-1$:1], \\
{\tt dat.21} &  angle between decay planes of $f_3$, $\bar{f}_4$ and 
                $f_5$, $\bar{f}_6$, [0:360]\,degrees, \\
{\tt dat.22} &  angle between decay planes of $V_1$, $\bar{f}_1=\Pep$ and 
                $V_1$, $\bar{f}_4$, [0:360]\,degrees,    \\
{\tt dat.23} &  angle between decay planes of $V_2$, $f_2=\Pem$ and 
                $V_2$, $f_5$, [0:360]\,degrees,      \\
{\tt dat.24} &  cosine of the angle between $f_3$ and $\bar{f}_4$, 
                [$-1$:1],   \\
{\tt dat.25} &  cosine of the angle between $f_5$ and $\bar{f}_6$, 
                [$-1$:1],  \\
{\tt dat.26} &  cosine of the angle between $V_1$ and $\bar{f}_1=\Pep$ 
                in the CM frame of $V_1$ and $V_2$, [$-1$:1],  \\
{\tt dat.27} &  cosine of the angle between $V_2$ and $f_2=\Pem$ 
                in the CM frame of $V_1$ and $V_2$, [$-1$:1]. 
%\end{tabular}
}
\ece
\caption{Description of the data files for the differential
cross sections to 
$\bar{f}_1 f_2 \to V_1 V_2 \to f_3 \bar{f}_4 f_5 \bar{f}_6 (\gamma)$
where 
$\bar{f}_1 = \Pe^+$, $f_2=\Pe^-$, 
$V_1 \to f_3 \bar{f}_4$, and $V_2 \to f_5 \bar{f}_6$ 
and the $\Pep$ beam goes in the positive $z$~direction. All histograms 
are defined with 50 bins.}
\label{tab:data}
\end{table}

The data files contain the observable (column 1), the differential
cross section for the chosen setup (column 2) and the corresponding
lowest-order cross section of the tree-level process $\eeffff$ (column
4). The statistical errors are provided in columns 3 and 5,
respectively. These are calculated as described in \refeq{eq:MCerror}
and below. The histograms are normalized such that the total cross
section is obtained as $\si=\sum_{\mathrm{bins}}
\mbox{bin-height}\times\mbox{bin-width}$.  The angles of the
histograms {\tt dat.21} and {\tt dat.22} are defined in Eqs.~(7.4) and
(7.3) of \citere{Denner:2000bj}, respectively.

In the subroutine {\tt SETTINGS} in {\tt public.f} the user can define
additional histograms as described at the end of {\tt SETTINGS}. 
The histograms are filled in 
\cpcsuptable{\tt MIN,MAX}{%
\multicolumn{3}{l}{{\tt SUBROUTINE HISTOGRAM(MIN,MAX,X,Y,HISTO,N,NAME,PARTS,STEPS)}}\\*
{\tt MIN,MAX}&:& minimal and maximal value of the {\tt X} range, \\
{\tt X,Y}&:& {\tt X} and {\tt Y} value, {\tt Y} is the weight of the event, \\
{\tt HISTO}&:& number of the histogram (${\tt HISTO}\le {\tt MAXH}$), \\
{\tt N}&:& total number of (weighted) events, used only for ${\tt STEPS}=4$, \\
{\tt NAME}&:& name of the data file, \\
{\tt PARTS}&:& number of bins of the histogram (${\tt PARTS}\le {\tt
  MAXP}$), 
\\
{\tt STEPS}&:& initialization, filling, and output of the 
histogram,\\
&&\cpcsubtable{${\tt STEPS}=1$}{%
${\tt STEPS}=1$&:& initialization of the histogram, \\
${\tt STEPS}=2$&:& filling the histogram with one of the weights of 
%a single subcontribution to
an event (column 2), \\
${\tt STEPS}=3$&:& evaluation of the histogram information for a complete event
(column 2,3), and filling the histogram with a single
event of the tree-level cross section to $\eeffff$ (column 4,5), \\
${\tt STEPS}=4$&:& writing data file of the histogram \\
}\\
}%
at the end of {\tt public.f}. 
The parameters ${\tt MAXH}$ and  ${\tt MAXP}$ specify the maximal
number of histograms and bins and are set to ${\tt MAXH}=50$ and
${\tt MAXP}=1000$, respectively. If more histograms or bins are
required, these parameters have to be increased.

For the subtraction branch the splitting into ${\tt STEPS}=2$ and
${\tt STEPS}=3$ is needed since the weights for the bremsstrahlung
process $\eeffffg$ and the corresponding subtraction functions are
related to different phase-space points and, hence, cannot be combined
into a single weight.  In the soft and collinear limits the momenta
corresponding to these different weights approach each other and the
%corresponding
weights approximatively cancel, leading to IR-safe observables. These
cancellations of the soft and collinear singularities occur within
single histogram bins. Therefore, the histograms must be filled in a
first step (${\tt STEPS}=2$) with all subcontributions corresponding
to one weighted event. From the resulting contributions to each
histogram bin the corresponding statistical errors can be calculated
(${\tt STEPS}=3$).

In addition, for ${\tt SPAW}=1$ the {\sc HBOOK} file {\tt pawplot.paw}
is generated. It contains the histograms for the distributions as
specified in {\tt pawgraphs.f}. In {\tt racoonww.f} the subroutine
{\tt SETUP\_GRAPHS} is called where the histograms are defined ({\tt
  HBOOK}). The subroutine {\tt GRAPHS} is called in {\tt racoonww.f}
after the weights have been calculated.  There the observables are
defined and the corresponding histograms are filled ({\tt HFILL}). In
addition three more subroutines, {\tt GRAPHS\_B}, {\tt GRAPHS\_K}, and
{\tt GRAPHS\_R}, are provided which generate histograms showing the
distributions separately for the tree-level $\eeffff$ contribution,
the whole $4 f$ part, and the $4 f \gamma$ part. As usual, the
histograms can be viewed by opening the {\tt PAW} \cite{Brun:1989vg}
main browser, and loading the file {\tt pawplot.paw}.

\subsection{The unweighting procedure and the interface with {\sc Pythia}}
\label{subsec:unweighting}

As default, {\RacoonWW} generates only weighted events.  
For ${\tt
  SRC}=1$ and ${\tt SMC}=3$, or for ${\tt SRC}=0,2,3$, {\RacoonWW} can
generate unweighted events. The unweighting procedure is activated as
follows:
\begin{itemize}
\item The number of generated unweighted events {\tt NEVENTSUNW} must
  be set in the main program in {\tt racoonww.f}. By default ${\tt
    NEVENTSUNW}=0$, and no unweighted events are generated. In the RC
  mode, to obtain a precision comparable to the ``best'' result of
  \citere{Denner:2000bj}, we recommend to use ${\tt NEVENTSUNW}=5
  \times 10^4$.  With an efficiency of one unweighted event per $10^3$
  weighted events, this corresponds to the calculation of
  approximately $5 \times 10^7$ weighted events.  In the tree-level
  and IBA modes the efficiency is about a factor 100 better and we
  recommend to use ${\tt NEVENTSUNW}=5 \times 10^5$.  Note that for
  certain processes the efficiency can strongly depend on the
  separation cuts.
\item
The number of weighted events ({\tt NEVENTSW}) that are used only to
determine the maximal weights for the hit-and-miss algorithm is set
in the input file. This should be much smaller than what is usually
used without unweighting, for instance, we recommend ${\tt NEVENTSW}=10^5$.
\end{itemize}
When including radiative corrections (${\tt SRC}=1$), only the branch 
${\tt SMC}=3$ can be used in the unweighting procedure.
 
The unweighted events are generated in the subroutine {\tt EVENTGENERATION} in
{\tt racoonww.f} from weighted events with a hit-and-miss algorithm.
More precisely, for each weighted event an unweighted event is
generated if the corresponding weight is larger than the product of a
random number {\tt RAND} ($0 \le {\tt RAND} \le 1$) and the maximal weight.
Otherwise no unweighted event is generated. The maximal weight, 
{\tt WEIGHTTOTMAX}, is calculated by multiplying the maximal weight of the
first {\tt NEVENTSW} events with a factor two. The number {\tt NEVENTSW} and 
the factor can be tuned to suppress weights that are larger than 
{\tt WEIGHTTOTMAX} in the main calculation. If a weight exceeds the
maximal weight, {\tt WEIGHTTOTMAX} is redefined by the value of the
weight and a warning is written in the output file.

The maximal weight of the bremsstrahlung contribution exceeds by far
the maximal weights of the other subcontributions. Therefore the other
weights are not calculated for each event in order to enhance the
efficiency of the unweighting procedure. They are calculated only all
{\tt NWEIGHT} times but are multiplied by {\tt NWEIGHT}. In this way
the probability of an unweighted event from the different
subcontributions remains unchanged. The integers {\tt NWEIGHT(1:29)}
are chosen in such a way that each subcontribution leads to roughly
the same maximal weight, which has to be smaller than {\tt
  WEIGHTTOTMAX}.

The information about the (unweighted) event is provided in the HEP
standard format in the common block {\tt HEPEVT} \cite{hepevt}.
%\multicolumn{3}{l}{{\tt COMMON/HEPEVT/NEVHEP,NHEP,ISTHEP,IDHEP,JMOHEP,JDAHEP,PHEP,VHEP}} \\
%{\tt PHEP(I,NMXHEP)} &:& four-momenta after possible ISR, \\
%&& \ie ${\tt phep(4,1)+phep(4,2)} \leq \sqrt{s}$ if ${\tt SISR}=1$, \\
%&& ${\tt i}=1,\ldots,5$: $x,y,z$,energy components, mass of the particle,\\
%&& ${\tt nmxhep}=0,\ldots,8$: numerates the particles 
%($7,8=\mathrm{photon}$), \\
%&& 1: beam $\Pe^+$ \\
%&& 2: beam $\Pe^-$ \\
%&& 3: final-state fermion $f_3$ \\
%&& 4: final-state fermion $\bar{f}_4$ \\
%&& 5: final-state fermion $f_5$ \\
%&& 6: final-state fermion $\bar{f}_6$ \\
%}
The parameters and flags of the common block {\tt HEPEVT} are set in
the subroutine {\tt EVENTGENERATION} in {\tt racoonww.f}. For a
detailed description we refer to \citere{Sjostrand:2001yu}, p.~63.
While the first six particle entries in {\tt HEPEVT} are the external
fermions, $\Pe^+ \Pe^- \to f_3 \bar{f}_4 f_5 \bar{f}_6$, the last up
to two entries are filled by a possible hard photon from the
bremsstrahlung process $\eeffffg$ and, if higher-order LL corrections
from ISR are included, a photon collinear to the beam.  The
four-momentum of the ISR photon collinear to the beam is calculated
from the four-momenta of the other external particles using momentum
conservation.  Note that the positron beam goes in $+z$ direction and
that cuts and photon recombination influence the resulting unweighted
events.

Although negative weights appear in the RC mode also for ${\tt
  SRC=3}$, they have only a negligible influence on the observables
that we have studied so far.  Nevertheless, {\RacoonWW} also unweights
the events with negative weights, but does not fill the {\tt HEPEVENT}
common block with these weights.  The resulting unweighted
negative-weight events can, for instance, be used to estimate the
uncertainty resulting from discarding negative weights.

In the output file, {\RacoonWW} provides the number of unweighted events,
the number of unweighted events with negative weights, and the number
of events that enforce a redefinition of the maximal weight.
Note that the result for the total cross section and for the
distributions are always calculated from the weighted events.

To include parton shower and hadronization for coloured final states,
{\RacoonWW} provides an interface with {\sc Pythia} for unweighted events 
%(${\tt NEVENTSUNW}>0$) 
by calling the {\sc Pythia} subroutine {\tt PY4FRM} (see p.~236 of
\citere{Sjostrand:2001yu}):
\cpcsuptable{\tt ATOTSQR}{%
  \multicolumn{3}{l}{{\tt SUBROUTINE PY4FRM(ATOTSQR,A1SQR,A2SQR,ISTRAT,IRAD,ITAU,ICOM)}}\\
  {\tt ATOTSQR}&:& total matrix-element squared for the event
  calculated in the subroutine {\tt EVENTGENERATION} as ${\tt
    ATOTSQR}={\tt ATOTSQ(NS)}$, where {\tt ATOTSQ(NS)} is the
  matrix element squared of the {\tt NS}-th weight {\tt WEIGHT(NS)}, \\
  {\tt A1SQR}&:& matrix-element squared for the configuration when
  $f_3$, $\bar{f}_4$ and $f_5$, $\bar{f}_6$ are colour singlets
  (${\tt A1SQR}={\tt A1SQ(NS)}$), \\
  {\tt A2SQR}&:& matrix-element squared for the configuration when
  $f_3$, $\bar{f}_6$ and $f_5$, $\bar{f}_4$
  are colour singlets (${\tt A2SQR}={\tt A2SQ(NS)}$), \\
  {\tt ISTRAT} &:& flag for the strategy to select either of both
  colour configurations; we choose ${\tt ISTRAT}=0$, \\
  {\tt IRAD}&:& flag for final-state photon radiation in {\sc Pythia}.
  As default, photon radiation is switched off (${\tt IRAD}=0$) to
  avoid double counting in the RC mode (${\tt SRC}=1$);
  in the tree-level and IBA modes ${\tt IRAD}=1$ may be chosen, \\
  {\tt ITAU} &:& flag for the handling of $\tau$-lepton decays; by
  default $\tau$ decays are not included (${\tt ITAU}=0$), since the
  final-state fermions are considered to be massless in {\RacoonWW};
  however, {\RacoonWW} provides a subroutine to generate a massive
  two-particle phase space (see below), so that
  $\tau$ decays can be taken into account (${\tt ITAU}=1$),\\
  {\tt ICOM}&:& flag for the place where the information about the
  event is stored; by default ${\tt ICOM}=0$ in order to store the
  information
  in the {\tt HEPEVT} common block. \\
  }

The amplitudes {\tt ATOTSQ(1:8)}, {\tt A1SQ(1:8)}, {\tt A1SQ(1:8)},
corresponding to the weights in \reftas{tab:wslicing1},
\ref{tab:wsubtraction}, and \ref{tab:wslicing2}, are calculated by
using {\tt ATOTSQ4}, {\tt A1SQ4}, {\tt A2SQ4} for the tree-level
processes $\eeffff$ and by {\tt ATOTSQ5}, {\tt A1SQ5}, {\tt A2SQ5} for
the tree-level processes $\eeffffg$. The colour amplitudes are
calculated in the functions {\tt M2\_EPEM\_4F} and {\tt M2\_EPEM\_4FA}
in {\tt ee4fa\_amps.f} and stored in the common blocks {\tt COLAMP}
and {\tt COLAMPA}.

As default, this interface is not active in {\RacoonWW}.  It can be
activated as follows:
\begin{itemize}
\item
Link {\sc Pythia} by adding the file with the {\sc Pythia} code, e.g.\
{\tt pythia1568.f} of version 6.1, to the list of objects {\tt OBJS} in 
the {\tt makefile}. {\sc Pythia} can be downloaded from
http://www.thep.lu.se/$\; \tilde{}$ torbjorn/Pythia.html.
\item
Uncomment the line with the call of {\tt PY4FRM} in the main program in 
{\tt racoonww.f}. 
\end{itemize}
Note that the compilation of {\sc Pythia} needs extensive memory.

As mentioned above, {\RacoonWW} provides the possibility of 
converting pairs of massless momenta to massive momenta by calling
\cpcsuptable{\tt PHEP1,PHEP2}{%
\multicolumn{3}{l}{{\tt SUBROUTINE MASSIVE(M1,M2,PHEP1,PHEP2)}}\\
{\tt M1,M2} &:& masses of the two final-state particles, \\
{\tt PHEP1,PHEP2} &:& four momenta of the two final-state particles 
with zero masses before the call of {\tt MASSIVE} and with 
masses {\tt M1,M2} after the call of {\tt MASSIVE}\\
}
in {\tt racoonww.f}.  This could be used to switch to massive
kinematics for the $\tau$ lepton before interfacing with {\sc Pythia}.
Note that the momenta of two final-state particles have to be changed
in order to maintain four-momentum conservation and the mass-shell
conditions.

\section{Summary}
\label{sec:summary}
We have described the MC generator \RacoonWW, version 1.3, for
four-fermion production, off-shell W-pair production, as well as the
radiative processes $\eeffffg$.  At tree level, the complete set of
diagrams is included.  For $\eeWWffff$, the prediction is based on the
complete $\Oa$ corrections in double-pole approximation and QED
initial-state radiation up to $\O(\al^3)$. The corresponding
theoretical uncertainty is at the level of $0.5\%$ for the total cross
section and of the order of 1\% for distributions in the energy range
between $\sim170$ and $500\GeV$. The use of a double-pole
approximation, in general, puts a restriction on the theoretical
uncertainty at the level of a few per mil away from threshold.
\RacoonWW\ also provides an improved-Born approximation with an
uncertainty of 2\%, which is also applicable in the threshold region.
More precise calculations require the inclusion of radiative
corrections beyond double-pole approximation.

\section*{Acknowledgements}
We are grateful to D.~Graudenz and R.~Pittau for useful discussions,
and to T.~Sj\"ostrand and S.~Roth for their help in implementing the
interface to {\sc Pythia}.  We thank the authors of
\citere{Beenakker:1998gr}, in particular A.~Chapovsky, and the authors
of \YFSWW, in particular, W.~P\l{}aczek, for useful numerical
comparisons.  We are indebted to R.~Chierici, F.~Cossutti, L.~Malgeri,
K.~Rabbertz, A.~Str\"assner, and M.~Thomson for valuable feedback
concerning the performance of the code and to all members of the LEP2
WW/4f working group, in particular E.~Lan\c{c}on, R.~Tanaka,
A.~Valassi, and M.~Verzocchi for many useful discussions.

This work has been partly supported by
the European Commission 5th framework contract HPRN-CT-2000-00149,
the Swiss Bundesamt f\"ur Bildung und Wissenschaft contract 99.0043, % Ansgar
and the US DOE Contract DE-FG02-91ER40685. S.D. is a Heisenberg Fellow
of the Deutsche Forschungsgemeinschaft

\appendix
\section{Sample runs and output}

Two test runs are discussed in the following. One is a run of
{\RacoonWW} in the RC mode. The second one is the calculation of the
cross section to $\eeffffg$ within the IBA mode.  Both test runs and
three additional sample runs are discussed in the file {\tt README}.

\subsection{Test run in the RC mode}

The first test run calculates the ``best'' predictions for the cross
section to the process $\Pep \Pem \to \Pu \Pdbar \mu^- \bar\nu_\mu
(\gamma)$ for a CM energy of $\sqrt{s}=200 \GeV$.  Thereby, the full
$\O(\alpha)$ corrections in DPA and leading higher-order corrections
are included, and the W-boson width is calculated.  The recombination
scheme and separation cuts of \citere{Denner:2000bj} are applied.  The
numerical integration is performed in the slicing branch with $5\times
10^7$ (weighted) events.  On a DEC/ALPHA workstation the computing
time is about 30 hours.

\subsubsection{Input file {\tt inputsli}}

{\tt\small
\begin{verbatim}
outputfile  ! name of output file
200d0       ! energy: CMF energy (in GeV)
50000000    ! neventsw: number of weighted events
1           ! smc: choice of MC branch: 1(or 3):slicing 2:subtraction
1           ! sborn4: include Born ee->4f: 0:no 1-3:yes
1           ! sborn5: include Born ee->4f+photon: 0:no 1:yes
0           ! sborng5: include Born ee->4f+gluon: 0:no 1:yes
1           ! sisr: include higher-order ISR: 0:no 1:yes
1           ! src: include radiative corrections: 0:no 1:DPA 2:IBA-4f 3:IBA-4fa
0           ! scoultree: Coulomb singularity for ee->4f,4f+ga: 0:no 1,2:yes
3           ! qnf: Coulomb singularity for ee->4f (DPA): 1,2, or 3
1           ! qreal: neglect imaginary part of virt. corr.: 0:no 1:yes
2           ! qalp: choice of input-parameter scheme: 0,1, or 2
4           ! qgw: calculate the W-boson width: 0:no 1-4:yes
1           ! qprop: choice of width scheme: 0,1,2,3 or 4 
0           ! ssigepem4: choice of diag. for Born ee->4f: 0:all 1-5:subsets
5           ! ssigepem5: choice of diag. for Born ee->4f+ga: 0:all 1-5:subsets
0           ! ssigepemg5: choice of diag. for Born ee->4f+gl: 0:all 1,5:subsets
2           ! qqcd: include QCD radiative corr.: 0:no 1:CC03 2:naive 3:CC11
0           ! sqcdepem: include gluon-exch. diag. in Born: 0:no 1:yes 2:only
u           ! fermion 3
d           ! anti-fermion 4
mu          ! fermion 5
nu_mu       ! anti-fermion 6
0d0         ! pp: degree of positron beam polarization [$-1$d0:1d0]
0d0         ! pm: degree of electron beam polarization [$-1$d0:1d0]
1           ! srecomb: recombination cuts: 0:no 1:TH 2:EXP
5d0         ! precomb(1): angular rec. cut between photon and beam
1d0         ! precomb(2): rec. cut on photon energy
5d0         ! precomb(3): inv.-mass rec.(TH) or angular rec. cut for lept.(EXP)
0d0         ! precomb(4): angular rec. cut for quarks(EXP)
0           ! srecombg: gluon recombination cuts: 0:no 1:TH 2:EXP
0d0         ! precombg(1): rec. cut on gluon energy
0d0         ! precombg(2): inv.-mass (TH) or angular (EXP) recombination cut 
0           ! satgc: anomalous triple gauge couplings (TGC): 0:no 1:yes
0d0         ! TGC Delta g_1^A
0d0         ! TGC Delta g_1^Z
0d0         ! TGC Delta kappa^A
0d0         ! TGC Delta kappa^Z
0d0         ! TGC lambda^A
0d0         ! TGC lambda^Z
0d0         ! TGC g_4^A
0d0         ! TGC g_4^Z
0d0         ! TGC g_5^A
0d0         ! TGC g_5^Z
0d0         ! TGC tilde kappa^A
0d0         ! TGC tilde kappa^Z
0d0         ! TGC tilde lambda^A
0d0         ! TGC tilde lambda^Z
0d0         ! TGC f_4^A
0d0         ! TGC f_4^Z
0d0         ! TGC f_5^A
0d0         ! TGC f_5^Z
0d0         ! TGC h_1^A
0d0         ! TGC h_1^Z
0d0         ! TGC h_3^A
0d0         ! TGC h_3^Z
0           ! qaqgc: anomalous quartic gauge couplings (QGC): 0:no 1:yes
0d0         ! QGC a_0/Lambda^2
0d0         ! QGC a_c/Lambda^2  
0d0         ! QGC a_n/Lambda^2  
0d0         ! QGC tilde a_0/Lambda^2
0d0         ! QGC tilde a_n/Lambda^2
10          ! scuts: separation cuts: 0:no 1,2:default(ADLO,LC) 10,11:input
0d0         ! photon(gluon) energy cut
0d0         ! charged-lepton energy cut
0d0         ! quark energy cut
0d0         ! quark-quark invariant mass cut
0d0         ! angular cut between photon and beam
0d0         ! angular cut between photon and charged lepton
0d0         ! angular cut between photon(gluon) and quark
0d0         ! angular cut between charged leptons
0d0         ! angular cut between quarks
0d0         ! angular cut between charged lepton and quark
10d0        ! angular cut between charged lepton and beam
10d0        ! angular cut between quark and beam
\end{verbatim}}

\subsubsection{Output file}

{\tt
\begin{verbatim}
      smc= 1: Phase-space-slicing branch of RacoonWW
              ======================================
 
 technical cutoff parameters (photon): 
  delta_s =    1.000000000000000E-003
  delta_c =    5.000000000000000E-004

 Input parameters:
 -----------------
    CMF energy = 200.00000 GeV,    Number of events =   50000000,

  alpha(0) = 1/ 137.0359895, alpha(MZ) = 1/128.88700,  alpha_s = 0.11900,
        GF = .1166370E-04,
        MW =  80.35000,             MZ =  91.18670,       MH = 150.00000,
        GW =   2.08699,             GZ =   2.49471,
        me = .51099907E-03,        mmu =   0.105658389, mtau =   1.77705,
        mu =   0.00485,             mc =   1.55000,       mt = 174.17000,
        md =   0.00485,             ms =   0.15000,       mb =   4.50000.

 Effective branching ratios: 
 leptonic BR =  0.32520, hadronic BR =  0.67480, total BR =  1.00000
  
 Process:  anti-e e -> u anti-d mu anti-nu_mu (+ photon)              
  
       pp= 0.0: degree of positron beam polarization.
       pm= 0.0: degree of electron beam polarization.
     qalp= 2: GF-parametrization scheme.
      qgw= 4: one-loop W-boson width calculated (with QCD corr.).
    qprop= 1: constant width.
  
   sborn4= 1: tree-level process ee -> 4f.
ssigepem4= 0: all electroweak diagrams included.
     qqcd= 2: naive QCD corrections included. 
  
      src= 1: virtual corrections in DPA and real corrections included.
ssigepem5= 5: real photon corr. : only CC11 class of diagrams included.
     qqcd= 2: naive QCD corrections included 
    qreal= 1: imaginary part of virtual corrections neglected.
      qnf= 3: off-shell Coulomb singularity with off-shell Born included.
     sisr= 1: initial-state radiation up to order alpha^3 included.
  
  srecomb= 1: with photon recombination:
 precomb(1) =   5.00000
 precomb(2) =   1.00000
 precomb(3) =   5.00000
 precomb(4) =   0.00000
  
    scuts=10: with separation cuts:
 angle(1,3) >  10.00000 deg
 angle(1,4) >  10.00000 deg
 angle(1,5) >  10.00000 deg
 angle(2,3) >  10.00000 deg
 angle(2,4) >  10.00000 deg
 angle(2,5) >  10.00000 deg
  
    events  :  intermediary results   :  preliminary results 
   1000000  :  592.61877 +-  2.47251  :  592.61877 +-  2.47251
   2000000  :  591.72383 +-  1.89683  :  592.17130 +-  1.55815
   3000000  :  594.65445 +-  1.89273  :  592.99902 +-  1.21535
   4000000  :  593.26888 +-  1.80112  :  593.06648 +-  1.01667
   5000000  :  587.37590 +-  1.65706  :  591.92837 +-  0.87826
   6000000  :  591.10834 +-  2.22044  :  591.79170 +-  0.82013
   7000000  :  594.42877 +-  1.76546  :  592.16842 +-  0.74684
   8000000  :  590.00165 +-  1.68242  :  591.89757 +-  0.68649
   9000000  :  589.54677 +-  1.59058  :  591.63637 +-  0.63529
  10000000  :  594.99013 +-  1.90289  :  591.97175 +-  0.60260
  11000000  :  588.48466 +-  1.82284  :  591.65474 +-  0.57233
  12000000  :  593.04183 +-  1.90019  :  591.77033 +-  0.54801
  13000000  :  592.36715 +-  1.68580  :  591.81624 +-  0.52221
  14000000  :  590.27937 +-  1.64218  :  591.70646 +-  0.49890
  15000000  :  591.71289 +-  1.79049  :  591.70689 +-  0.48070
  16000000  :  591.53899 +-  1.76907  :  591.69640 +-  0.46402
  17000000  :  587.50308 +-  1.58333  :  591.44973 +-  0.44654
  18000000  :  587.89929 +-  1.55550  :  591.25249 +-  0.43050
  19000000  :  592.93474 +-  1.73210  :  591.34103 +-  0.41790
  20000000  :  592.16324 +-  1.74407  :  591.38214 +-  0.40647
  21000000  :  592.24008 +-  1.80517  :  591.42299 +-  0.39655
  22000000  :  592.44203 +-  1.72115  :  591.46931 +-  0.38652
  23000000  :  588.26438 +-  1.68691  :  591.32997 +-  0.37692
  24000000  :  590.49483 +-  1.81216  :  591.29517 +-  0.36902
  25000000  :  591.15275 +-  1.79673  :  591.28947 +-  0.36148
  26000000  :  593.47325 +-  1.78914  :  591.37346 +-  0.35432
  27000000  :  591.22938 +-  1.63230  :  591.36813 +-  0.34651
  28000000  :  591.06261 +-  1.77909  :  591.35722 +-  0.34013
  29000000  :  590.90543 +-  1.71644  :  591.34164 +-  0.33369
  30000000  :  591.59423 +-  1.68095  :  591.35006 +-  0.32740
  31000000  :  591.88912 +-  1.70980  :  591.36745 +-  0.32160
  32000000  :  588.95475 +-  1.67046  :  591.29205 +-  0.31589
  33000000  :  589.07952 +-  1.59758  :  591.22500 +-  0.31012
  34000000  :  591.09482 +-  1.83148  :  591.22117 +-  0.30578
  35000000  :  593.56319 +-  1.67756  :  591.28809 +-  0.30089
  36000000  :  592.41598 +-  1.67903  :  591.31942 +-  0.29623
  37000000  :  593.56115 +-  1.88861  :  591.38001 +-  0.29270
  38000000  :  588.14474 +-  1.69946  :  591.29487 +-  0.28849
  39000000  :  591.59522 +-  1.65363  :  591.30257 +-  0.28427
  40000000  :  590.01906 +-  1.63315  :  591.27048 +-  0.28016
  41000000  :  590.88361 +-  1.92445  :  591.26105 +-  0.27732
  42000000  :  592.07152 +-  1.69336  :  591.28034 +-  0.27371
  43000000  :  590.52329 +-  1.63555  :  591.26274 +-  0.27003
  44000000  :  588.91748 +-  1.60674  :  591.20943 +-  0.26641
  45000000  :  590.68588 +-  1.61247  :  591.19780 +-  0.26294
  46000000  :  590.26056 +-  1.80667  :  591.17743 +-  0.26021
  47000000  :  590.89561 +-  1.66101  :  591.17143 +-  0.25711
  48000000  :  593.22822 +-  1.64885  :  591.21428 +-  0.25409
  49000000  :  591.98897 +-  1.63064  :  591.23009 +-  0.25112
  50000000  :  593.47276 +-  1.71196  :  591.27494 +-  0.24847
 
 Result:
 ------- 
 Number of weighted events   =         50000000
 Average            =   591.2749427045 fb
 Standard deviation =     0.2484667990 fb
 Maximal weight     =     0.0287650976 fb
 
 Subcontributions without naive QCD:
 -----------------------------------
 Tree-level four-fermion cross section:
 Average            =   627.6779997923 fb
 Standard deviation =     0.1075731739 fb
 Maximal weight     =     0.0003787331 fb
  
 2->4 DPA part :
 Average            =    -6.1930315855 fb
 Standard deviation =     0.0030507290 fb
 Maximal weight     =     0.0000240297 fb
  
 2->4 virtual+soft+final-state collinear photon part:
 without LL O(alpha) ISR
 Average            =   -67.4968488454 fb
 Standard deviation =     0.0117625527 fb
 Maximal weight     =     0.0000250099 fb
  
 2->4 initial-state collinear photon part (e+):
 without LL O(alpha) ISR
 Average            =   -65.5094255283 fb
 Standard deviation =     0.0131360874 fb
 Maximal weight     =     0.0000551979 fb
  
 2->4 initial-state collinear photon part (e-):
 without LL O(alpha) ISR
 Average            =   -65.4479754318 fb
 Standard deviation =     0.0131306505 fb
 Maximal weight     =     0.0000523650 fb
  
 2->4 LL initial-state radiation up to O(alpha^3):
 Average            =   -54.3690820517 fb
 Standard deviation =     0.0645784370 fb
 Maximal weight     =     0.0003228727 fb
  
 2->5 Bremsstrahlung contribution (QED):
 Average            =   201.0338802635 fb
 Standard deviation =     0.2234350653 fb
 Maximal weight     =     0.0277012770 fb
  
 2->5 Bremsstrahlung contribution (QCD):
 Average            =     0.0000000000 fb
 Standard deviation =     0.0000000000 fb
 Maximal weight     =     0.0000000000 fb
  
 % events rejected in calculation of weight( 1): 5.3%
 % events rejected in calculation of weight( 2): 5.2%
 % events rejected in calculation of weight( 3): 5.1%
 % events rejected in calculation of weight( 4): 6.1%
 % events rejected in calculation of weight( 5): 6.2%
 % events rejected in calculation of weight( 6): 6.0%
 % events rejected in calculation of weight( 7):34.6%
 % events rejected in calculation of weight( 8): 0.0%
\end{verbatim}
}

\subsection{Test run in the IBA mode for the processes $\eeffffg$}

The second test run calculates the IBA for the process $\Pep \Pem \to
\Pu \Pdbar \mu^- \bar\nu_\mu \gamma$ in the subtraction branch for a
CM energy of $\sqrt{s}=200 \GeV$. The ADLO cuts are applied, and the
IBA W-boson width is calculated.  On a DEC/ALPHA workstation the
computing time is about 2 hours.

\subsubsection{Input file {\tt inputiba4fa}}

{\tt\small
\begin{verbatim}
outputfile  ! name of output file
200d0       ! energy: CMF energy (in GeV)
10000000    ! neventsw: number of weighted events
2           ! smc: choice of MC branch: 1(or 3):slicing 2:subtraction
0           ! sborn4: include Born ee->4f: 0:no 1-3:yes
1           ! sborn5: include Born ee->4f+photon: 0:no 1:yes
0           ! sborng5: include Born ee->4f+gluon: 0:no 1:yes
1           ! sisr: include higher-order ISR: 0:no 1:yes
3           ! src: include radiative corrections: 0:no 1:DPA 2:IBA-4f 3:IBA-4fa
1           ! scoultree: Coulomb singularity for ee->4f,4f+ga: 0:no 1,2:yes
3           ! qnf: Coulomb singularity for ee->4f (DPA): 1,2, or 3
1           ! qreal: neglect imaginary part of virt. corr.: 0:no 1:yes
2           ! qalp: choice of input-parameter scheme: 0,1, or 2
2           ! qgw: calculate the W-boson width: 0:no 1-4:yes
1           ! qprop: choice of width scheme: 0,1,2,3 or 4 
0           ! ssigepem4: choice of diag. for Born ee->4f: 0:all 1-5:subsets
0           ! ssigepem5: choice of diag. for Born ee->4f+ga: 0:all 1-5:subsets
0           ! ssigepemg5: choice of diag. for Born ee->4f+gl: 0:all 1,5:subsets
2           ! qqcd: include QCD radiative corr.: 0:no 1:CC03 2:naive 3:CC11
0           ! sqcdepem: include gluon-exch. diag. in Born: 0:no 1:yes 2:only
u           ! fermion 3
d           ! anti-fermion 4
mu          ! fermion 5
nu_mu       ! anti-fermion 6
0d0         ! pp: degree of positron beam polarization [$-1$d0:1d0]
0d0         ! pm: degree of electron beam polarization [$-1$d0:1d0]
0           ! srecomb: recombination cuts: 0:no 1:TH 2:EXP
0d0         ! precomb(1): angular rec. cut between photon and beam
0d0         ! precomb(2): rec. cut on photon energy
0d0         ! precomb(3): inv.-mass rec.(TH) or angular rec. cut for lept.(EXP)
0d0         ! precomb(4): angular rec. cut for quarks(EXP)
0           ! srecombg: gluon recombination cuts: 0:no 1:TH 2:EXP
0d0         ! precombg(1): rec. cut on gluon energy
0d0         ! precombg(2): inv.-mass (TH) or angular (EXP) recombination cut 
0           ! satgc: anomalous triple gauge couplings (TGC): 0:no 1:yes
0d0         ! TGC Delta g_1^A
0d0         ! TGC Delta g_1^Z
0d0         ! TGC Delta kappa^A
0d0         ! TGC Delta kappa^Z
0d0         ! TGC lambda^A
0d0         ! TGC lambda^Z
0d0         ! TGC g_4^A
0d0         ! TGC g_4^Z
0d0         ! TGC g_5^A
0d0         ! TGC g_5^Z
0d0         ! TGC tilde kappa^A
0d0         ! TGC tilde kappa^Z
0d0         ! TGC tilde lambda^A
0d0         ! TGC tilde lambda^Z
0d0         ! TGC f_4^A
0d0         ! TGC f_4^Z
0d0         ! TGC f_5^A
0d0         ! TGC f_5^Z
0d0         ! TGC h_1^A
0d0         ! TGC h_1^Z
0d0         ! TGC h_3^A
0d0         ! TGC h_3^Z
0           ! qaqgc: anomalous quartic gauge couplings (QGC): 0:no 1:yes
0d0         ! QGC a_0/Lambda^2
0d0         ! QGC a_c/Lambda^2  
0d0         ! QGC a_n/Lambda^2  
0d0         ! QGC tilde a_0/Lambda^2
0d0         ! QGC tilde a_n/Lambda^2
1           ! scuts: separation cuts: 0:no 1,2:default(ADLO,LC) 10,11:input
\end{verbatim}
}

\subsubsection{Output file}

{\tt\small
\begin{verbatim}
  
      smc= 2: Subtraction-method branch of RacoonWW
              =====================================

 Input parameters:
 -----------------
    CMF energy = 200.00000 GeV,    Number of events =   10000000,

  alpha(0) = 1/ 137.0359895, alpha(MZ) = 1/128.88700,  alpha_s = 0.11900,
        GF = .1166370E-04,
        MW =  80.35000,             MZ =  91.18670,       MH = 150.00000,
        GW =   2.09436,             GZ =   2.49471,
        me = .51099907E-03,        mmu =   0.105658389, mtau =   1.77705,
        mu =   0.00485,             mc =   1.55000        mt = 174.17000,
        md =   0.00485,             ms =   0.15000,       mb =   4.50000.

 Effective branching ratios: 
 leptonic BR =  0.32512, hadronic BR =  0.67488, total BR =  1.00000
  
 Process:  anti-e e -> u anti-d mu anti-nu_mu + photon                
  
       pp= 0.0: degree of positron beam polarization.
       pm= 0.0: degree of electron beam polarization.
     qalp= 2: GF-parametrization scheme.
      qgw= 2: IBA W-boson width calculated.
    qprop= 1: constant width.
  
   sborn4= 0: tree-level process ee -> 4f NOT included.
      src= 3: improved Born approximation ee->4f+photon.
scoultree= 1: Coulomb sing. to ee->4f+photon included.
ssigepem5= 0: all electroweak diagrams included.
     qqcd= 2: naive QCD corrections included. 
     sisr= 1: initial-state radiation up to order alpha^3 included.
  
    scuts= 1: with separation cuts:
 energy(3)  >   3.00000 GeV
 energy(4)  >   3.00000 GeV
 energy(5)  >   1.00000 GeV
 energy(7)  >   0.10000 GeV
 angle(1,5) >  10.00000 deg
 angle(1,7) >   1.00000 deg
 angle(2,5) >  10.00000 deg
 angle(2,7) >   1.00000 deg
 angle(3,5) >   5.00000 deg
 angle(3,7) >   5.00000 deg
 angle(4,5) >   5.00000 deg
 angle(4,7) >   5.00000 deg
 angle(5,7) >   5.00000 deg
 mass(3,4)  >   5.00000 GeV
  
    events  :  intermediary results   :  preliminary results 
   1000000  :  185.52994 +-  1.33873  :  185.52994 +-  1.33873
   2000000  :  184.94817 +-  1.21671  :  185.23906 +-  0.90451
   3000000  :  187.67723 +-  1.22280  :  186.05178 +-  0.72784
   4000000  :  188.12389 +-  1.29820  :  186.56981 +-  0.63508
   5000000  :  186.24278 +-  1.21187  :  186.50440 +-  0.56291
   6000000  :  184.67305 +-  1.27906  :  186.19918 +-  0.51526
   7000000  :  186.27700 +-  1.29118  :  186.21029 +-  0.47862
   8000000  :  188.42698 +-  1.29575  :  186.48738 +-  0.44903
   9000000  :  188.10039 +-  1.28908  :  186.66660 +-  0.42406
  10000000  :  187.29950 +-  1.27811  :  186.72989 +-  0.40248
 
 Result:
 ------- 
 Number of weighted events   =         10000000
 Average            =   186.7298922322 fb
 Standard deviation =     0.4024824609 fb
 Maximal weight     =     0.0452822218 fb
 
 Subcontributions without naive QCD:
 -----------------------------------
 Tree-level four-fermion cross section:
 Average            =     0.0000000000 fb
 Standard deviation =     0.0000000000 fb
 Maximal weight     =     0.0000000000 fb
 (event =         0, channel =   0)
  
 2->4 part of radiative corrections without O(alpha) LL ISR:
 Average            =     0.0000000000 fb
 Standard deviation =     0.0000000000 fb
 Maximal weight     =     0.0000000000 fb
 (event =         0, channel =   0)
  
 Initial-state radiation up to O(alpha^3):
 Average            =     0.0000000000 fb
 Standard deviation =     0.0000000000 fb
 Maximal weight     =     0.0000000000 fb
 (event =         0, channel =   0)
  
 Improved Born Approximation ee->4f+photon:
 Average            =   179.9149173069 fb
 Standard deviation =     0.3877932869 fb
 Maximal weight     =     0.0436295823 fb
 (event =   5915698, channel =  50)
  
 % events rejected in calculation of weight( 1): 0.0%
 % events rejected in calculation of weight( 2): 0.0%
 % events rejected in calculation of weight( 3): 0.0%
 % events rejected in calculation of weight( 4): 0.0%
 % events rejected in calculation of weight( 5): 0.0%
 % events rejected in calculation of weight( 6):10.2%
 % events rejected in calculation of weight( 7): 0.0%
\end{verbatim}
% events rejected in calculation of weight( 8): 0.0%
% events rejected in calculation of weight( 9): 0.0%
% events rejected in calculation of weight(10): 0.0%
% events rejected in calculation of weight(11): 0.0%
% events rejected in calculation of weight(12): 0.0%
% events rejected in calculation of weight(13): 0.0%
% events rejected in calculation of weight(14): 0.0%
% events rejected in calculation of weight(15): 0.0%
% events rejected in calculation of weight(16): 0.0%
% events rejected in calculation of weight(17): 0.0%
% events rejected in calculation of weight(18): 0.0%
% events rejected in calculation of weight(19): 0.0%
% events rejected in calculation of weight(20): 0.0%
% events rejected in calculation of weight(21): 0.0%
% events rejected in calculation of weight(22): 0.0%
% events rejected in calculation of weight(23): 0.0%
% events rejected in calculation of weight(24): 0.0%
% events rejected in calculation of weight(25): 0.0%
% events rejected in calculation of weight(26): 0.0%
% events rejected in calculation of weight(27): 0.0%
% events rejected in calculation of weight(28): 0.0%
%\qquad\qquad \vspace{-1ex}\vdots\vspace{-1ex} 
\qquad\qquad\vdots
\begin{verbatim}
 % events rejected in calculation of weight(29): 0.0%
\end{verbatim}
}

\end{document}